\documentclass[12pt]{article}

\pdfoutput=1

\usepackage[top=80pt,bottom=85pt,left=85pt,right=85pt]{geometry}
\usepackage{amssymb}
\usepackage{amsmath}
\usepackage{setspace}
\usepackage{sectsty}
\usepackage{cite}
\usepackage{bbm}
\usepackage[utf8]{inputenc}
\usepackage[T1]{fontenc}
\usepackage{comment}
\usepackage[english]{babel}
\usepackage{afterpage}
\usepackage{bbold}
\usepackage{mathdots}
\usepackage{physics}
\usepackage{makecell,booktabs}
\usepackage{multirow}
\usepackage[dvipsnames]{xcolor}
\usepackage{lscape}
\usepackage{mathtools}
\usepackage{dashrule}

\usepackage{graphicx,subcaption}
\usepackage{setspace}
\usepackage{float}
\usepackage{booktabs}
\usepackage[debug,pageanchor=false]{hyperref}
\definecolor{link}{rgb}{.8,.15,.1}
\definecolor{pigment}{rgb}{0.36, 0.54, 0.66}
\definecolor{pigment2}{rgb}{0.19, 0.55, 0.91}
\definecolor{pigment3}{rgb}{0.2, 0.2, 0.6}
\definecolor{light-gray}{gray}{0.75}
\hypersetup{colorlinks=true,linkcolor=link,citecolor=link,urlcolor=link,linktocpage}

\usepackage{array,multirow,booktabs,longtable}
\usepackage{mathrsfs}
\usepackage{tikz-cd} 
\usetikzlibrary{backgrounds, arrows,calc,shapes,decorations.pathreplacing, decorations.markings, automata,positioning}
\tikzset{%
  >={Latex[width=2mm,length=2mm]},
            base/.style = {rectangle, rounded corners, draw=black,
                           minimum width=4cm, minimum heigwht=1cm,
                           text centered, font=\sffamily},
  activityStarts/.style = {base, fill=orange!15},
       startstop/.style = {base, fill=orange!15},
    activityRuns/.style = {base, fill=orange!15},
         process/.style = {base, minimum width=2.5cm, fill=orange!15,
                           font=\ttfamily},
}

\newcommand{\red}[1]{}

\tikzset{
        cvertex/.style={circle,draw=black,inner sep=1pt,outer sep=3pt},
        vertex/.style={circle,fill=black,inner sep=1pt,outer sep=3pt},
        star/.style={circle,fill=yellow,inner sep=0.75pt,outer sep=0.75pt},
        tvertex/.style={inner sep=1pt,font=\scriptsize},
        gap/.style={inner sep=0.5pt,fill=white}}

\tikzstyle{mybox} = [draw=black, fill=blue!10, very thick,
    rectangle, rounded corners, inner sep=10pt, inner ysep=20pt]
\tikzstyle{boxtitle} =[fill=blue!50, text=white,rectangle,rounded corners]

\newcommand{\cc}{\mathbb{C}}

\newcommand\scalemath[2]{\scalebox{#1}{\mbox{\ensuremath{\displaystyle #2}}}}

\newcolumntype{C}{>{\hfil$}p{3cm}<{$\hfil}}
\newcolumntype{P}{>{\hfil$}p{7.7cm}<{$\hfil}}
\newcolumntype{L}{>{\hfil$}p{2.8cm}<{$\hfil}}
\newcolumntype{S}{>{\hfil$}p{1.8cm}<{$\hfil}}
\newcolumntype{R}{>{\hfil$}p{5.2cm}<{$\hfil}}
\newcolumntype{U}{>{\hfil$}p{4.2cm}<{$\hfil}}
\newcolumntype{Q}{>{\hfil$}p{6.4cm}<{$\hfil}}
\newcolumntype{T}{>{\hfil$}p{1.9cm}<{$\hfil}}
\newcolumntype{V}{>{\hfil$}p{5.8cm}<{$\hfil}}
\newcolumntype{H}{>{\hfil$}p{1.8cm}<{$\hfil}}
\newcolumntype{W}{>{\hfil$}p{3.7cm}<{$\hfil}}
\newcolumntype{A}{>{\hfil$}p{6cm}<{$\hfil}}
\newcolumntype{B}{>{\hfil$}p{2cm}<{$\hfil}}
\newcolumntype{D}{>{\hfil$}p{7.4cm}<{$\hfil}}

\newcommand{\todo}[1]{}
\renewcommand{\todo}[1]{{\color{red} TODO: {#1}}}
\renewcommand{\red}[1]{{\color{red} {#1}}}

\newcommand{\be}{\begin{equation}}  
\newcommand{\ee}{\end{equation}}  
\newcommand{\bea}{\begin{align}}
\newcommand{\eea}{\end{align}}
\newcommand{\bp}{\begin{bmatrix*}[r]}  
\newcommand{\bpp}{\begin{bmatrix}}  
\newcommand{\epp}{\end{bmatrix}}  
\newcommand{\bcd}{\begin{center}
\begin{tikzcd}}
\newcommand{\ecd}{\end{tikzcd} \end{center}}
\newcommand{\bm}{\begin{pmatrix}}  
\newcommand{\eem}{\end{pmatrix}}

\makeatletter
\@addtoreset{equation}{section}
\makeatother

\begin{document}

\begin{titlepage}

\begin{center}

\vskip .3in \noindent

{\Large \bf{5d Higgs Branches from M-theory on \\  \vspace{.5cm} quasi-homogeneous cDV threefold singularities}}

\bigskip\bigskip\bigskip

Mario De Marco$^{a}$, Andrea Sangiovanni$^{b}$ and Roberto Valandro$^{b}$ \\

\bigskip


\bigskip
{\footnotesize
 \it

$^a$ SISSA and INFN, Via Bonomea 265, I-34136 Trieste, Italy\\
\vspace{.25cm}
$^b$ Dipartimento di Fisica, Universit\`a di Trieste, Strada Costiera 11, I-34151 Trieste, Italy \\
and INFN, Sezione di Trieste, Via Valerio 2, I-34127 Trieste, Italy	
}

\vskip .5cm
{\scriptsize \tt mdemarco at sissa dot it \hspace{1cm} andrea dot sangiovanni at phd dot units dot it} \\
{\scriptsize \tt    roberto dot valandro at ts dot infn dot it}

\vskip 2cm
     	{\bf Abstract }
\vskip .1in
\end{center}

We classify rank zero 5d SCFTs geometrically engineered from M-theory on quasi-homogeneous compound Du Val isolated threefold singularities. For \textit{all} such theories, we characterize the Higgs Branch, by computing the dimension, the continuous and discrete symmetry groups, as well as more refined details such as the charges of the hypermultiplets under these groups. We derive these data by means of a  gauge-theoretic method, that we have recently introduced, based on establishing a correspondence between an adjoint Higgs field and the M-theory geometry. As a byproduct, this further allows us to construct  several T-brane backgrounds, that yield inequivalent 5d spectra but are associated with the same geometry.

\noindent

\vfill
\eject

\end{titlepage}

\tableofcontents
\section{Introduction}
The dynamics of five-dimensional Superconformal Field Theories (SCFTs) has
been attracting, in the course of the recent years, a great deal of attention. 
One of the main tools for studying SCFTs in
various dimensions is to regard them as fixed points of supersymmetric gauge
theories. In five dimensions, however, the gauge theory analysis fails to
capture relevant features of the SCFTs dynamics due to the presence of
non-perturbative degrees of freedom becoming massless at the infinite
coupling point.
A different perspective is to
rely on string theory inspired techniques. Among the huge variety of approaches, in this paper we adopt the so-called geometric engineering point of view.
In particular, we study the 5d
dynamics of M-theory on an isolated, non-compact Calabi-Yau (CY) threefold
singularity $X$ (the subject was initiated in \cite{Seiberg:1996bd,Intriligator:1997pq}\footnote{
In the past years, the subject has seen a revival, with systematic studies   of such theories and various methods for constructing them, producing a vast literature; some key works include 
\cite{DelZotto:2017pti,Xie:2017pfl,Jefferson:2017ahm,Jefferson:2018irk,Apruzzi:2018nre,Closset:2018bjz,Bhardwaj:2018yhy,Apruzzi:2019opn,Apruzzi:2019enx,Bhardwaj:2019jtr,Closset:2019juk,
Saxena:2020ltf,
Closset:2020scj,Closset:2020afy,Apruzzi:2021vcu,Closset:2021lwy,DelZotto:2022fnw,DelZotto:2022joo,Cvetic:2022imb}.}).
More specifically, we will concentrate on isolated
hypersurface singularities of $\mathbb C^4$, admitting a quasi-homogeneous
$\mathbb C^*$-action. Furthermore, we will focus exclusively on singularities
leading to rank-zero theories with an empty Coulomb Branch (CB). This reflects, from the geometric point of
view, the fact that either $X$ does not admit any crepant
K\"ahler resolution, or that it admits a small crepant resolution $\pi: \hat{X}
\twoheadrightarrow X$. This can be equivalently formulated requiring that $X$ only possesses an isolated terminal singularity\footnote{In this
context, ``small'' means that the exceptional locus of the
resolution $\pi$ has $\mathbb{C}$-dimension one, or, equivalently, is a
collection of rigid curves, contracted on the singular point by $\pi$.}. In
the framework of M-theory geometric engineering, the degrees of freedom of
the five-dimensional SCFTs corresponding to such CY threefolds $X$ descend from the M-theory M2-branes, wrapped on the
curves contracted by $\pi$. The theory is effectively five-dimensional, as a consequence of the fact that the eleven-dimensional profile describing the
M2-brane state is peaked on the point where the curves have been
contracted, namely on the singular point of $X$. 

It is a known mathematical fact that all Gorenstein isolated terminal CY threefold singularities can be described,
locally \cite{Katz:1992aa}, as \textit{compound Du Val (cDV) singularities}: these are  families of ALE spaces over a complex line $\mathbb C_w$,
developing an ADE singularity at the origin $w = 0$. The singular point of
the ALE fiber, over $w=0$, corresponds to the singular point of
$X$. In particular, in the course of this work we focus on the subclass of \textit{quasi-homogeneous} cDV singularities, that have been completely classified in \cite{Wang:2015mra,Yau:2003uk}.\\
\indent The simplest example
of this setup is the conifold singularity, that can be thought of as a family of
deformed $A_1$ singularities over a complex parameter $w$:
\begin{equation}
  \label{conifold}
  x^2+ y^2 + z^2 + w^2=0, \quad (x,y,w,z) \in \mathbb C^4.
\end{equation}
The crepant resolution of the conifold blows-up a single $\mathbb P^1$,
hence being a small resolution, and supports a single M2-brane state that yields a
five-dimensional hypermultiplet. The divisor dual to the inflated $\mathbb P^1$
is associated with a non-normalizable two-form, that, upon reduction of the
M-theory three-form $C_3$, gives the Cartan  of the $Sp(2)$ flavor symmetry associated
with a single five-dimensional hypermultiplet.\\

In this paper we complete the study, initiated in \cite{DeMarco:2021try}, of M-theory dynamics on \textit{all} the isolated,
quasi-homogeneous, cDV hypersurface singularities. In particular, we compute the Higgs Branches (HBs) of the 5d $\mathcal{N}=1$ SCFTs arising from M-theory on all the quasi-homogeneous cDV singularities. 
To achieve this result,
we use the techniques outlined in \cite{Collinucci:2021wty,Collinucci:2021ofd,Collinucci:2022rii}, regarding
$X$ as a deformation of a trivial fibration of $\mathcal G$-type ADE
singularities, with $\mathcal{G} \in A,D,E$. M-theory on a trivial ADE fibration over $\mathbb C_w$ gives
origin to a seven-dimensional $\mathcal N=1$ gauge theory on $\mathbb{R}^5\times \mathbb{C}_w$, with
gauge algebra $\mathcal G$.
We now break half of the supercharges of the seven-dimensional theory as follows:
We re-organize the three real scalars into a complex adjoint scalar $\Phi=\phi_1+i\phi_2$ and a real adjoint scalar $\phi_3$.
\emph{We then switch on a vev for $\Phi$ that  depends holomorphically on the complex coordinate~$w$.}
As a result we obtain:
\begin{itemize}
\item  The 7d gauge algebra $\mathcal{G}$ is broken to the commutant $\mathcal{H}$ of $\Phi$. The 7d vector boson resides now in 5d background vector multiplets that support the 5d flavor group. There can also be a discrete part of the 7d gauge group that survives the Higgsing: this leads to 5d discrete gauging.
\item The zero modes of $\Phi$ are deformations in $\mathcal{G}$ that cannot be gauge fixed to zero; in particular one obtains zero modes that are localized at $w=0$, i.e. they are 5d modes. These organize in 5d hypermultiplets and correspond, in M-theory, to the M2 brane states. The total number of hypermultiplets gives the dimension of the HB (as there is no continuous 5d gauge group).
\item With our method, one can easily derive the charges of the hypermultiplets with respect to the continuous flavor group and the discrete symmetry.\footnote{This allows, in principle, to compute refined quantities, such as the Hilbert series of the Higgs branch.}
\end{itemize}
One can physically interpret the seven-dimensional theory, and the fields $\Phi, \phi_3$, as describing the dynamics of type IIA brane systems. In the case of classical Lie algebras, this is simply the type IIA limit of M-theory on $\mathbb{C}^*$-fibered threefolds. The $E_6,E_7,E_8$ singularities are instead elliptically fibered: we can then consider F-theory on them, reducing to type IIB with seven-branes, and T-dualize the system to produce the seven-dimensional gauge theory with $E_6,E_7,E_8$ gauge algebra.

From a geometric point of view, the Casimir invariants of the vevs $\Phi$ and $\phi_3$ (for fixed $w$) control,
respectively, the complex structure and the K\"ahler moduli of the ALE fiber
over the point $w$. In other words, introducing a $w$-dependence on the
vev of $\Phi$ deforms the trivial ADE fibration to a non-trivially 
fibered threefold $X$. In this fashion we can realize all the cDV singular threefolds.

Some of 5d Higgs Branches studied in this paper were already discussed in previous works using the following different method \cite{Closset:2020scj,Closset:2020afy,Closset:2021lwy}: the 5d HB is computed in the 3d theory coming from reducing M-theory on $X\times T^2$; this theory is found starting from type IIB string theory on $X\times S^1$, applying 3d mirror symmetry and finally ungauging some $U(1)$ symmetries. In \cite{Carta:2021whq,Carta:2022spy} the authors analyze type IIB reduced on a large subclass of quasi-homogeneous cDV singularities (among many other threefolds that we do not consider here); applying the method of \cite{Closset:2020scj,Closset:2020afy,Closset:2021lwy} that we have just outlined, one finds complete agreement with our results in the overlapping cases.\footnote{We thank F. Carta, S. Giacomelli, N. Mekareeya and A. Mininno  for intense exchanges on this point.}

The main technical problem we had to solved in order to carry this program to completion is finding the explicit Higgs background $\Phi$ corresponding to a given quasi-homogeneous cDV singularity: indeed, once it is in our hands, we can compute all the relevant quantities needed to characterize the 5d Higgs Branch. In this paper we introduce a novel method to directly identify the Higgs backgrounds, solely by looking at the quasi-homogeneous cDV threefold equations and relying on the homogeneity of the coefficients of the versal deformation of the ADE singularities. 

As one could expect, there is an ambiguity: several different $\Phi$'s (leading to different 5d symmetries and modes) can lead to the same CY equation. This means that the geometry is not able to capture all the information of the 5d theory; one needs to add more, and our claim is that the field $\Phi$, that we specify, does the job. This phenomenon is common in the F-theory literature, in the context of T-brane backgrounds \cite{Cecotti:2010bp,Anderson:2013rka,Collinucci:2014qfa,Collinucci:2014taa,Collinucci:2016hpz,Collinucci:2017bwv,Anderson:2017rpr}.

The work is structured as follows: in Section~\ref{sec: cDV} we recall
the main features of compound Du Val threefolds, focusing on how to interpret them as
non-trivial ADE fibrations over $\mathbb C_w$. In Section~\ref{Sec:MthonX} we describe the dynamics of M-theory on these singularities, and how it
is related to the physics of the seven-dimensional theory associated with
M-theory on the trivial ADE fibration. In Section~\ref{sec: higgs data} and Section~\ref{section higgs vev} we  build the
dictionary between the physics of the seven-dimensional gauge theory, and
M-theory on quasi-homogeneous cDV singularities:  in Section~\ref{sec: higgs data} we show how starting from the 7d vev $\Phi$ one can derive the associated CY equation, count the zero modes, and pinpoint the 5d symmetry group; in Section~\ref{section higgs vev} we invert the procedure, namely we start from the
quasi-homogeneous cDV singularity and exhibit a method to swiftly obtain an Higgs vev $\Phi$ capturing the
physics of M-theory on the considered singularity.
In Section~\ref{section cDV HB} we provide explicit examples, and sum up the results that we obtained for all the
quasi-homogeneous cDV singularities. In Section~\ref{Sec:T-branes} we describe the physics of the so-called T-branes states. In Section~\ref{conclusion} we draw our conclusions.

In Appendix~\ref{appendix C} we perform an explicit computation of the five-dimensional modes for the $(A_2,D_4)$ singularity. In Appendix~\ref{appendix D} we focus on some mathematical aspects related to the main text. In Appendix~\ref{appendix E} we list the expressions of the deformed $E$ singularities as functions of the Casimirs of the respective algebras. Finally, in Appendix~\ref{Appendix B} we present the ancillary Mathematica code that we used to compute the Higgs branches.

\section{cDV 3-folds as families of deformed ADE surfaces}\label{sec: cDV}

\subsection{Isolated quasi-homogeneous 3-fold singularities of cDV type }

The rational Gorenstein singularities in dimension two are called Du Val singularities and take the form 
\begin{equation}\label{ADEsingSurf}
x^2 + P_{\mathcal G}(y,z)=0\:,
\end{equation}
with $P_{\mathcal G}(y,z)$ following the ADE classification:
\begin{equation}\label{ADE singularities}
\begin{split}
& P_{A_n}=y^2+z^{n+1}\\
& P_{D_n}=z y^2+z^{n-1}\\
& P_{E_6}=y^3+z^4 \\
& P_{E_7}=y^3+yz^3 \\
& P_{E_8}=y^3+z^5 \\
\end{split}
\end{equation}

Now, let us consider the class of threefolds singularities called \emph{compound Du Val (cDV)}. They are of the form
\begin{equation}\label{Eq:cDVsing}
   x^2 + P_{\mathcal G}(y,z) + w \, g(x,y,z,w) =0\:.
\end{equation}
The isolated quasi-homogeneous 3-fold singularities of cDV type are precisely the ones listed in Table~\ref{Xie table}, computed in \cite{Wang:2015mra}.
A famous sub-class of cDV singularities are, for example, the threefolds of type $(A,{\mathcal G})$, with the following threefold hypersurfaces\footnote{It is known that reducing type IIB string theory on such threefolds, one obtains Argyres-Douglas theory of type $(A,\mathcal{G})$ \cite{Cecotti:2010fi,Xie:2015rpa}.}~in~$\mathbb{C}^4$:
\begin{equation}\label{3foldFamilyGw}
x^2 + P_{\mathcal G}(y,z) + w^{N} = 0 \:.
\end{equation}

\subsection{ADE families and cDV threefolds}
In all the threefold equations in Table~\ref{Xie table} the first three monomials reconstruct the ADE singularity of type ${\mathcal G}$ in \eqref{ADEsingSurf}, while the last term can be interpreted as a deformation of this singularity. This is manifest in  the equations \eqref{Eq:cDVsing} and  \eqref{3foldFamilyGw}. Hence the equations in Table~\ref{Xie table} describe one-parameter families of deformed ${\mathcal G}$-singularities, fibered over a complex plane $\mathbb{C}_w$.

\begin{table}[t]\centering
\begin{equation}
\begin{array}{|c|c|c|c|}
\hline 
 \text{ADE} & \text { Label }  & \text { Singularity } & \begin{array}{c}
\makecell{\text{Non-vanishing} \\
\text {deformation parameter}}
\end{array} \\
\Xhline{4\arrayrulewidth} A &  (A_{N-1},A_{k-1}) & x^{2}+y^{2}+z^{k}+w^{N}=0 & \mu_{k}=w^{N}  \\
\hline& A_{k-1}^{(k-1)}[N]  & x^{2}+y^{2}+z^{k}+w^{N} z=0 & \mu_{k-1}=w^{N} \\
\Xhline{4\arrayrulewidth}  D& (A_{N-1},D_{k}) & x^{2}+z y^{2}+z^{k-1}+w^{N}=0 & \mu_{2 k-2}=w^{N} \\
\hline & D_{k}^{(k)}[N]& x^{2}+z y^{2}+z^{k-1}+w^{N} y=0 & \tilde{\mu}_{k}=w^{N}  \\
\Xhline{4\arrayrulewidth}  E_{6}& (A_{N-1},E_6) & x^{2}+y^{3}+z^{4}+w^{N}=0 & \mu_{12}=w^{N} \\
\hline& E_{6}^{(9)}[N]  & x^{2}+y^{3}+z^{4}+w^{N} z=0 & \mu_{9}=w^{N} \\
\hline & E_{6}^{(8)}[N]& x^{2}+y^{3}+z^{4}+w^{N} y=0 & \mu_{8}=w^{N}  \\
\Xhline{4\arrayrulewidth} E_{7}& (A_{N-1},E_7) & x^{2}+y^{3}+y z^{3}+w^{N}=0 & \mu_{18}=w^{N}  \\
\hline & E_{7}^{(14)}[N] & x^{2}+y^{3}+y z^{3}+w^{N} z=0 & \mu_{14}=w^{N} \\
\Xhline{4\arrayrulewidth} E_{8} &(A_{N-1},E_8)& x^{2}+y^{3}+z^{5}+w^{N}=0 & \mu_{30}=w^{N}  \\
\hline& E_{8}^{(24)}[N]  & x^{2}+y^{3}+z^{5}+w^{N} z=0 & \mu_{24}=w^{N} \\
\hline  & E_{8}^{(20)}[N] & x^{2}+y^{3}+z^{5}+w^{N} y=0 & \mu_{20}=w^{N} \\
\hline
\end{array}\nonumber
\end{equation}
\caption{Quasi-homogeneous cDV singularities as ADE families.}\label{Xie table}
\end{table}

Threefolds that are families of ADE surfaces have been extensively studied in the past and techniques have been developed to extract their geometric data \cite{Katz:1992aa,Cachazo:2001gh,Karmazyn:2017aa,BrownWemyss,Collinucci:2018aho,Collinucci:2019fnh}. In particular, in this paper we apply the methods recently developed in \cite{Collinucci:2021ofd,Collinucci:2022rii} to work out the dynamics of M-theory on all the quasi-homogeneous compound Du Val 3-fold singularities, adding data that were not present before.

To tackle this task, we first look more closely on such threefolds, built as families of deformed ADE surfaces.
We start from the generic versal deformation of an ADE singularity of type $\mathcal{G}$ (an ADE Lie algebra of rank $r$):
\begin{equation}\label{VersalDefADE}
x^2 + P_{\mathcal G}(y,z)+\sum_{i=1}^r \mu_ig_i = 0,
\end{equation}
where the monomials $g_i$ belong to the ring $
R = \frac{\mathbb{C}[x,y,z]}{\left(f,\frac{\partial f}{\partial x},\frac{\partial f}{\partial y},\frac{\partial f}{\partial z}\right)}
$, where $f = x^2 + P_{\mathcal G}(y,z)$.
The equation \eqref{VersalDefADE} describes the total space of a family of deformed ADE singularities of type ${\mathcal G}$. This is a fibration over the space $\mathcal{B}_\mu$ of deformations parametrized by the coefficients $\mu_i$ of the monomials $g_i$ in \eqref{VersalDefADE}. The generic fiber is a deformation of the corresponding ADE singularity.

The base $\mathcal{B}_\mu$ is isomorphic to the $r$-dimensional space
$\mathfrak{t}/\mathcal{W}$, where $\mathfrak{t}$ is the  Cartan subalgebra and $\mathcal{W}$ the Weyl group of the underlying Lie algebra.
At the origin of $\mathcal{B}_\mu$, the surface develops a singularity of type ${\mathcal G}$. 
At a generic point of $\mathcal{B}_\mu$, the singularity is deformed: the surface admits $r$ non-vanishing $S^2$ intersecting in the same pattern as the nodes of the Dynkin diagram and whose volume is measured by the holomorphic (2,0)-form. 
In the deformed ALE surface, the holomorphic (2,0)-form is along an element ${\bf t}$  of the Cartan subalgebra, such that ${\rm vol}_{\alpha_i}=\int_{\alpha_i}\Omega_{2,0}= \alpha_i[{\bf t}]$.
We choose a set of coordinates $t_i$ of $\mathfrak{t}$ such that the volumes of the simple roots are given by
\begin{equation}\label{volumes}
\begin{split}
& A_{r}:\quad {\rm vol}_{\alpha_i} = t_i-t_{i+1} \qquad i=1,...,r
\\
& D_{r}:\quad {\rm vol}_{\alpha_i} = \left\{\begin{array}{l}
 t_i-t_{i+1} \qquad i=1,...,r-1 \\  t_{r-1}+t_{r} \qquad i=r
\end{array}\right.
 \\
& E_{r}: \quad  {\rm vol}_{\alpha_i} = \left\{\begin{array}{l}
 t_i-t_{i+1} \qquad i=1,...,r-1 \\  -t_{1}-t_{2}-t_{3} \qquad i=r
\end{array}\right.
\end{split}
\end{equation}
The coordinates $\mu_i$'s can be written as \emph{homogeneous} functions of the $t_i$'s that are invariant under the action of the Weyl group $\mathcal{W}$. If one uses these functions to make a \emph{base change}, one obtains a family of deformed ${\mathcal G}$-singularities fibered this time over the space $\mathfrak{t}$. This family is a singular space. Resolving the singularity at $t_1=...=t_r=0$ blows up all the $\mathbb{P}^1$'s corresponding to the simple roots of ${\mathcal G}$ in the central fiber. This is called a \emph{simultaneous resolution}.

One can also choose a different base change in which $\mathfrak{t}/\mathcal{W'}$ is mapped to $\mathfrak{t}/\mathcal{W}$, where $\mathcal{W}'\subset \mathcal{W}$. In this case, the resolution of the family blows up the roots that are left invariant by $\mathcal{W}'$, in the central fiber. The base of the fibration $\mathcal{B}_\varrho\cong \mathfrak{t}/\mathcal{W'}$ is now parametrized by the $r$ $\mathcal{W}'$ invariants, that we call~$\varrho_i$~($i=1,...,r$). The $\varrho_i$'s can also be associated with $\mathcal{W}'$-invariant \emph{homogeneous} functions of coordinates $t_i$ on $\mathfrak{t}$.

In Figure~\ref{Fig:baseChanges}, we summarize the possible base changes that can be done at the level of the $(r+2)$-dimensional family.\\
\begin{figure}
\begin{equation}\label{base changes}
   \renewcommand{\arraystretch}{2.2}
    \begin{array}{l|crcccc}
&& \boldsymbol{ \mathfrak{t} }  \quad\quad\quad\quad&  \boldsymbol{\longrightarrow} &\quad\quad\quad\quad  \boldsymbol{\mathcal{B}_\varrho} \quad\quad\quad\quad&  \boldsymbol{\longrightarrow} & \quad\quad\quad\quad  \boldsymbol{\mathcal{B}_\mu} \\
 \vspace{-1.0cm} & & & & \\
  \hline 
 1) && t_i \quad\quad\quad\quad& \multicolumn{3}{c}{\xmapsto{\hspace{9cm}}} &\quad\quad\quad\quad\mu_i(t_j) \\
  \vspace{-1.1cm} & & & & \\

  2) && t_i \quad\quad\quad\quad& \xmapsto{\hspace{2cm}} &\quad\quad\quad\quad  \varrho_i(t_j) \quad\quad\quad\quad& & \\
    \vspace{-1.1cm} & & & & \\

3) &&  & & \varrho_i & \xmapsto{\hspace{2cm}} & \quad\quad\quad\quad\mu_i(\varrho_j) \\    
    \end{array}\nonumber
\end{equation}
\caption{Three possible base changes: 1) from $\mathfrak{t}$ to $\mathcal{B}_{\mu} = \mathfrak{t}/\mathcal{W}$, 2) from $\mathfrak{t}$ to $\mathcal{B}_{\rho} = \mathfrak{t}/\mathcal{W}'$, 3)
from $\mathcal{B}_{\rho} = \mathfrak{t}/\mathcal{W}'$ to $\mathcal{B}_{\mu} = \mathfrak{t}/\mathcal{W}$.}\label{Fig:baseChanges}
\end{figure}


In the cases we are interested in this paper, we restrict to families where we set $\mu_i=0$, for all $i$ except one (see the last column of Table~\ref{Xie table}). 
Let us first keep only the constant term, i.e.\ we take the family
\begin{equation}\label{VersalDefADE3fold}
x^2 + P_{\mathcal G}(y,z) + \mu = 0,
\end{equation}
where $\mu$ does not depend on $x,y,z$. This is a non-singular threefold hypersurface in $\mathbb{C}^4[x,y,z,\mu]$.
We can recover the $(A,\mathcal{G})$ space \eqref{3foldFamilyGw}, by making the \emph{base change} $\mu=w^{N}$. Now, the threefold is singular at $x=y=z=w=0$. 

We want to describe this base change in the following way: 
\begin{enumerate}\label{Twostepsproced}
\item we first go from $\mathcal{B}_\varrho$ to $\mathcal{B}_\mu$, by putting $\mu_i=\mu_i(\varrho_1,...,\varrho_r)$;
\item we then allow a holomorphic dependence $\varrho_i=\varrho_i(w)$ that makes all $\mu_i=0$, except the constant deformation that must take the form $\mu(\varrho_i(w))=w^{N}$.
\end{enumerate} 
Of course this is not unique: several choices of $\mathcal{W}'\subset\mathcal{W}$ produce the same threefold \eqref{3foldFamilyGw}, by taking the proper expressions for $\varrho_i(w)$.  
As we will see in detail below, the different choices of $\mathcal{W}'$
correspond to different T-brane backgrounds associated with the same singular threefold in M-theory. The presence of a T-brane can obstruct the resolution of some roots \cite{Cecotti:2010bp,Anderson:2013rka}, enlarging the subgroup $\mathcal{W}'$. 
In order to have a geometry without T-branes, we will make the following choice: we consider the smallest choice of $\mathcal{W}'$ that reproduces the equation \eqref{3foldFamilyGw} (that is not necessarily the trivial subgroup). This corresponds to a family over $\mathcal{B}_\varrho$ with the biggest possible number of resolved simple roots in the simultaneous resolution.

The same considerations can be done for the other compound Du Val threefold singularities in Table~\ref{Xie table} that do not belong to the $(A,{\mathcal G})$ class.

\section{M-theory on quasi-homogeneous cDV threefold singularities} \label{Sec:MthonX}

\subsection{5d from M-theory on quasi-homogeneous CY threefolds}

The threefolds $X$ in Table~\ref{Xie table} have isolated singularities, whose exceptional locus contains only  dim$_\mathbb{C}=1$  loci: these are the $\mathbb{P}^1$'s blown up in the simultaneous resolution of the singularity. 
Since there are \textit{no exceptional divisors}, M-theory on $X$ gives a \textit{rank-zero} 5d $\mathcal{N}=1$ SCFT. The theory then has no Coulomb branch.

Each exceptional $\mathbb{P}^1$ comes with a dual non-compact divisor\footnote{The non-Cartier divisor of the singularity associated with the small resolution.} that, upon reducing $C_3$, gives a background  vector multiplet corresponding to a $U(1)$ flavor symmetry under which the hypermultiplets are charged.
Hence, if the (partial) simultaneous resolution blows up $\ell \leq r$ spheres, the flavor group acting on the Higgs Branch is 
\begin{equation} \label{5dFlavGr}
G_F = U(1)^\ell.
\end{equation}
In particular, in the resolved threefold, the size of the exceptional $\mathbb{P}^1$'s are controlled by the vev's of the scalars in the background vector multiplets associated with $G_F$.

These theories have a bunch of massless hypermultiplets localized at the singularity that come from BPS M2-branes wrapping the exceptional $\mathbb{P}^1$'s. Their vev's live in the Higgs branch of the theory, that is believed to be a discrete gauging of a product of~$\mathbb{H}$~\cite{Closset:2020scj}.

\subsection{The field $\Phi$ and 7d to 5d Higgsing}\label{Sec:Phi7d5d}

Due to the fibration structure of the threefolds $X$ under study, we can apply the following point of view to compute the data of the 5d SCFT \cite{Collinucci:2021ofd,DeMarco:2021try,Collinucci:2022rii} . 

M-theory on a ADE singularity \eqref{ADEsingSurf} of type ${\mathcal G}$ gives rise to a 7d $\mathcal{N}=1$ field theory with a vector multiplet in the adjoint representation of ${\mathcal G}$. This supermultiplet contains three real scalars $\phi_1,\phi_2,\phi_3$. 

We split $\mathbb{R}^7=\mathbb{R}^5\times \mathbb{C}_w$, where $\mathbb{C}_w$ is parametrized by a complex coordinate $w$.
We give a vev to the complex adjoint scalar $\Phi=\phi_1+i\phi_2$, that is holomorphic in $w$. 
The  vev for $\Phi$ breaks the 7d Poincar\'e group to the 5d one and preserves half of the supersymmetries. We consider only cases when the 5d symmetry is abelian, say $U(1)^\ell$.
The zero modes around such a background organize then in 5d supermultiplets:
\begin{itemize}
\item The zero modes of $A_\mu$ ($\mu=0,...,4$) and $\phi_3$ propagate in 7d and are collected into $U(1)$ background vector multiplets, giving rise to the $U(1)^\ell$ flavor group.
\item There are 7d zero modes of $\Phi$ that are collected together with zero modes of $A_5+iA_6$ into background hypermultiplets,  that are neutral under the flavor $U(1)$'s. 
\item There are 5d zero modes of $\Phi$ that are localized at $w=0$. 
They are collected into 5d massless hypermultiplets that are charged under the $U(1)^\ell$ flavor group.
\end{itemize}

From the geometric point of view, a vev for $\Phi$ means that the ADE surface is deformed, with deformation parameters depending on its Casimir invariants.
Switching on a $w$-dependent vev gives then a threefold $X$ that is a family of deformed ADE singularities fibered over $\mathbb{C}_w$. The fact that the  algebra preserved by $\langle\Phi\rangle$ is abelian means that the surface is smooth on top of generic points of $\mathbb{C}_w$. If the preserved algebra contained a simple summand of type $\mathcal G'$, the threefold $X$ would have a non-isolated singularity of type $\mathcal G'$ on top of each point of the base.

Furthermore, at the level of the ALE fiber a vev for $\phi_3$, with $[\langle\phi_3\rangle,\langle\Phi\rangle]=0$, means resolving some simple roots, i.e. switching on K\"ahler deformations along some Cartan generators. 
Given a vev for $\Phi$, then, its (abelian) commutant $\mathcal{H}$
tells us what are the blown up roots in the simultaneous resolution.
Given a basis $\alpha_1^*,...\alpha_r^*$ of $\mathfrak{t}$ that is dual to the basis of simple roots,  the subalgebra $\mathcal{H}$ can be written as
\begin{equation}\label{Eq:calH}
\mathcal{H}=\langle\alpha_1^*,...\alpha_\ell^*\rangle\:.
\end{equation} 
the  blown up roots  are then $\alpha_1,...,\alpha_\ell$. 

The Higgs field\footnote{From now on, when we write $\Phi$, we mean the vev.} $\Phi$ breaks $\mathcal{G}$ to the abelian subalgebra $\mathcal{H}$ in \eqref{Eq:calH}, then it must live in the commutant of $\mathcal{H}$ in $\mathcal{G}$. 
A subalgebra $\mathcal{L}$ that is the commutant of an abelian subalgebra is called a Levi subalgebra. The Levi subalgebras of a Lie algebra ${\mathcal G}$ are in one-to-one correspondence with the choices of a set of simple roots (see \cite{Collingwood}).
$\mathcal{L}$ will be a direct sum  of $\mathcal{H}$ with a semi-simple Lie algebra, i.e. $\mathcal{L}=\mathcal{L}^{\rm semi-simp} \oplus \mathcal{H}$. In order to break to $\mathcal{H}$, it is enough that $\Phi$ belongs to 
$\mathcal{M}=\mathcal{M}^{\rm semi-simp} \oplus \mathcal{H}$, where $\mathcal{M}^{\rm semi-simp}$ is a maximal subalgebra of $\mathcal{L}^{\rm semi-simp}$ of maximal rank\footnote{This is true because the Cartan subalgebra of $\mathcal{L}^{\rm semi-simp}$ coincides with the Cartan subalgebra of $\mathcal{M}^{\rm semi-simp}$, as it is a maximal subalgebra of maximal rank.}.
We are assuming $\Phi \rvert_{\mathcal{M}^{\rm semi-simp}}$ to be a generic element of $\mathcal{M}^{\rm semi-simp}$, i.e. it is not contained in any proper subalgebra of $\mathcal{M}^{\rm semi-simp}$. We can then write
\begin{equation}\label{MaxSubAlgLevi}
\Phi\in\mathcal{M}\equiv\bigoplus_h\mathcal{M}_h\oplus \mathcal{H}\:, 
\end{equation}
where $\mathcal{M}_h$ are simple Lie algebras.

To sum up, we have the relations
\begin{equation}
 \mbox{simult.resol. :  } \alpha_1,...,\alpha_\ell \quad \leftrightarrow \quad \mathcal{H}=\langle\alpha_1^*,...\alpha_\ell^*\rangle \quad \leftrightarrow \quad \mathcal{L} \:.
\end{equation}
We can summarize these data in the Dynkin diagram of $\mathcal{G}$: we color in black the nodes corresponding to roots belonging to $\mathcal{H}$ (namely, they are the nodes that get blown-up by the simultaneous resolution). Then the semi-simple part $\mathcal{L}^{\rm semi-simp}$ of the corresponding Levi subalgebra is given by the Dynkin diagram colored in white. 
Hence, a coloring of the nodes of the Dynkin diagram completely and univocally fixes a Levi subalgebra $\mathcal{L}=\mathcal{H}\oplus \mathcal{L}^{\rm semi-simp}$. Given the Levi $\mathcal{L}$, one can look for maximal subalgebras of the form \eqref{MaxSubAlgLevi}, employing the usual technique based on extended Dynkin diagrams. A $\Phi$ producing a fibration with the given simultaneous resolution must belong to one of these maximal subalgebras.

\begin{figure}[t]
    \centering
    \includegraphics[scale=0.13]{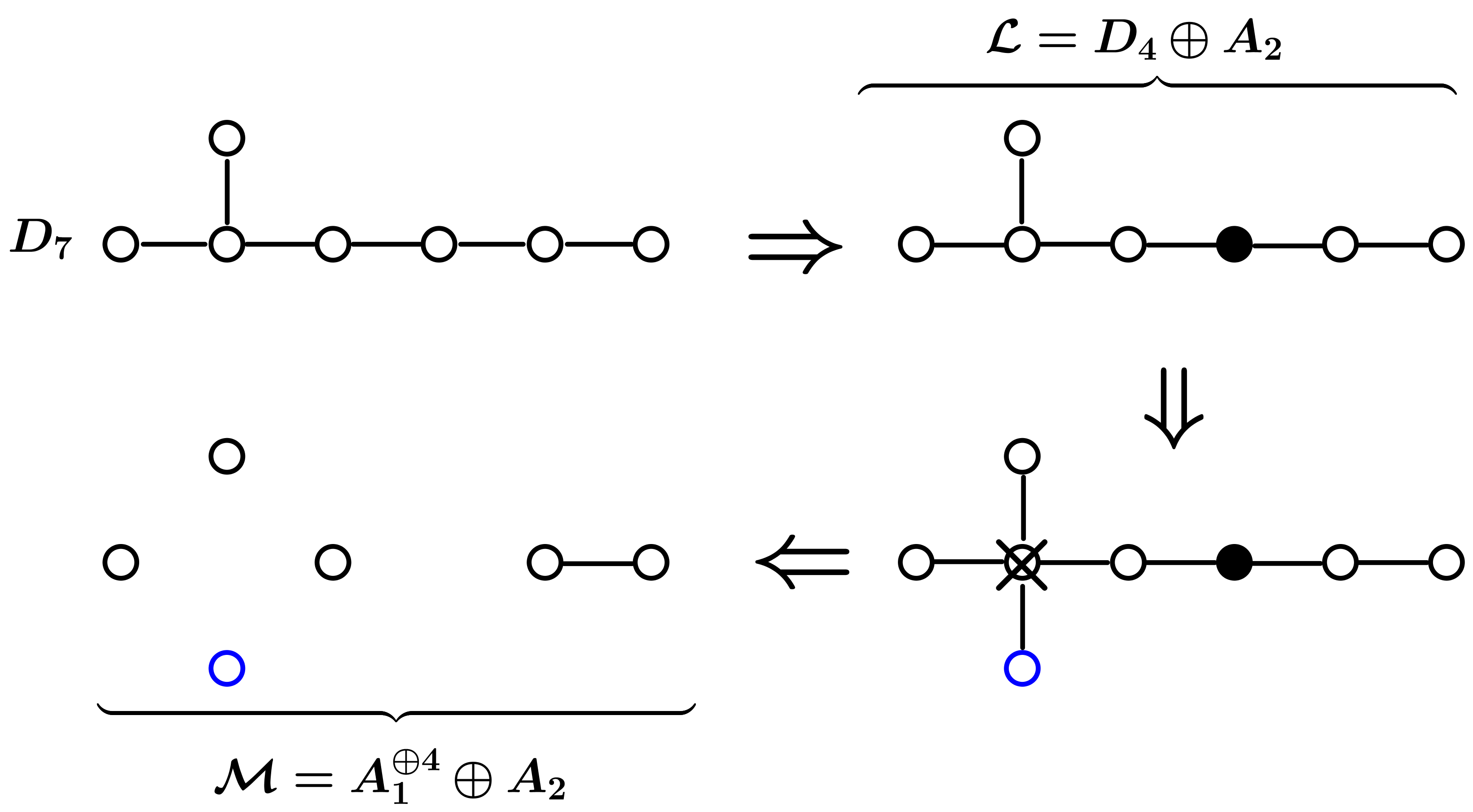}
    \caption{$A_1^4 \oplus A_2 $ subalgebra of $D_7$.}
    \label{D7discrete}
\end{figure}

Let us see a simple example of this framework: take the Dynkin diagram of the Lie algebra $D_7$ and color one node $\alpha_{\text{res}}$ in black as in Figure~\ref{D7discrete}. We immediately read  $\mathcal{L}=D_4 \oplus A_2 \oplus\langle \alpha_{\text{res}}^* \rangle$ from the white nodes. $D_4$ has a maximal subalgebra $A_1^{\oplus 4}$, that we can extract pictorially as in Figure~\ref{D7discrete}. If we want $\Phi$ to produce a threefold with a simultaneous resolution of only the root $\alpha_{\text{res}}$, then either $\Phi\in \mathcal{M}=\mathcal{L}$ or $\Phi\in\mathcal{M} = A_1^4 \oplus A_2 \oplus \langle \alpha_{\text{res}}^* \rangle$.

\section{Data from the Higgs field}\label{sec: higgs data}

In this paper we want to analyze the 5d theories coming from M-theory on specific threefolds $X$, namely the quasi-homogeneous cDV singularities in Table~\ref{Xie table}. We first derive which vev of $\Phi$ produces such threefold fibrations. Afterwards, we derive the data of 5d theory by studying the Higgsing of the corresponding 7d theory. In the following, we describe the necessary techniques to attain such results.

\subsection{The threefold equation from  $\langle\Phi\rangle$}\label{Sec:3foldEqFromPhi}

Once we define a Higgs field $\Phi(w)$, we have determined the threefold fibrations. 
The key point is the geometric correspondence between the vev of $\Phi$ and the deformed ADE singularity, or better the volumes of the spheres that get non-zero size in the deformation:
At a generic $w\in\mathbb{C}_w$, $\Phi$ is  diagonalizable. After diagonalizing $\Phi$, one obtains an element $\Phi_d$ in the Cartan subalgebra $\mathfrak{t}$. The volumes of the finite size spheres in the deformed ADE surface are now given by
\begin{equation}
   \mbox{vol}_{\alpha_i}=\alpha_i[\Phi_d] \:.
\end{equation}
Using the definitions \eqref{volumes}, we obtain the values of the $t_i$-coordinates in $\mathfrak{t}$ corresponding to a given $\Phi_d$. Given $\Phi(w)$, we can then compute $t_i=t_i(w)$.

The versal deformations in terms of the $t_i$'s are known \cite{Katz:1992aa}, and they specify which monomials are turned on when a particular sphere takes non-zero volume. One has:
\begin{equation}\label{deformed ADE singularities}
\begin{split}
& A_{r}:\quad x^{2}+y^{2}+\prod_{i=1}^{r+1}\left(z+t_{i}\right)=0 \quad\quad \sum_{i=1}^{r+1} t_{i}=0 \\
& D_{r}:\quad x^{2}+ zy^{2}+\frac{\prod_{i=1}^{r}\left(z+t_{i}^{2}\right)-\prod_{i=1}^{r} t_{i}^{2}}{z}+2 \prod_{i=1}^{r} t_{i} y=0 \\
& E_{6}: \quad x^{2}+z^{4}+y^{3}+\mu_{2} y z^{2}+\mu_{5} y z+\mu_{6} z^{2}+\mu_{8} y+\mu_{9} z+\mu_{12} =0\\
& E_{7}: \quad x^{2}+y^{3}+y z^{3}+\mu_{2} y^{2} z+\mu_{6} y^{2}+\mu_{8} y z+\mu_{10} z^{2}+\mu_{12} y+\mu_{14} z+\mu_{18}=0 \\
& E_{8}: \quad x^{2}+y^{3}+z^{5}+\mu_{2} y z^{3}+\mu_{8} y z^{2}+\mu_{12} z^{3}+\mu_{14} y z+\mu_{18} z^{2}+\mu_{20} y+\mu_{24} z+\mu_{30}=0,
\end{split}
\end{equation}
where the $\mu_i$ are known functions of $t_i \in \mathfrak{t}$ (see \cite{Katz:1992aa} for the explicit expressions of $\mu_i$ in $E_6,E_7$ and an algorithm to compute them for $E_8$). For us, the important information is that $\mu_i(t)$ is a homogeneous polynomial in the $t_i$'s of degree $i$.

In principle, in order to obtain the equation of the hypersurface related to a given Higgs field $\Phi(w)$, one hence diagonalizes  $\Phi(w)$ and then inserts the corresponding $t_i(w)$'s in \eqref{deformed ADE singularities}, obtaining  $\mu_i(w)$ and the equation of the threefold hypersurface.
Even if it is not obvious, this produces $\mu_i(w)$ that are holomorphic in $w$.
There are several ways to implement this involved procedure, that becomes more and more complicated as the rank of the ADE algebra increases \cite{Katz:1992aa}.

It is most useful for us to take the approach given in \cite{Collinucci:2022rii}. It allows to obtain directly the form of $\mu_i$'s in terms of the Casimir invariants of $\Phi\in\mathfrak{g}$. These can easily be computed by taking $\Phi(w)$  in a given representation and computing traces of powers of it.
For families of $A$- and $D$-type one obtains the hypersurface equations
\begin{equation}\label{threefolds}
\renewcommand{\arraystretch}{1.5}
\begin{array}{cl}
\boldsymbol{A_r}:\, & x^2+y^2+\text{det}(z\mathbb{1}-\Phi)=0\\
\boldsymbol{D_r}:\, & x^2+zy^2-\frac{\sqrt{\text{det}(z\mathbb{1}+\Phi^2)}-\text{Pfaff}^2(\Phi)}{z}+2y\hspace{0.1cm}\text{Pfaff}(\Phi)=0\\
 \end{array},
\end{equation}
that manifestly depends on  the Casimirs invariants of $\Phi$, i.e. 
\begin{equation}\label{Casimirs}
\begin{array}{c|cc}
\boldsymbol{A_r} &k_i=\text{Tr}(\Phi^{i}) & \text{for }i=2,\ldots,r+1\\
\hline
\boldsymbol{D_r} & \makecell{\tilde{k}_i=\text{Tr}(\Phi^{i}) \\
  \hat{k}_r=\text{Pfaff}(\Phi)}& \text{for }i=2,4,\ldots,2(r-1)\\
\end{array}.
\end{equation}
with $\Phi$ in the fundamental representation.
One can show \cite{Collinucci:2022rii} analogous formulae for the exceptional cases, where one can write the deformation parameters $\mu_i$ in terms of the Casimir invariants of $E_r$, that can easily be computed once one has the explicit form of $\Phi$. We report the formulae in Appendix \ref{appendix E}.

After we have the relations between the coefficients of the versal deformation and the Casimir invariants of $\Phi$, we can plug in a given choice of $\Phi(w)$ and easily obtain the corresponding threefold.

Let us make a simple example: take a Higgs field in $\mathcal{M}= A_1^{(1)}\oplus A_1^{(3)} \oplus \langle\alpha_2^*\rangle\subset A_3$, given by
\begin{equation}
\Phi|_{A_1^{(1)}}= \begin{pmatrix}
0 & 1 \\ w & 0 \\
\end{pmatrix}\,,\qquad \Phi|_{A_1^{(3)}}= \begin{pmatrix}
0 & 1 \\ -w & 0 \\
\end{pmatrix}\,\:
\end{equation}
and with zero coefficient along $\alpha_2^*$. 
Computing the characteristic polynomial of such $\Phi$ one immediately obtains that all the Casimirs are zero except the one of degree $4$ that is equal to $w^2$. 
If we plug this $\Phi$ into \eqref{threefolds}, we in fact obtain
\begin{equation}\label{A1A3 threefold}
    x^2+\underbrace{y^2+z^4}_{P_{A_3(y,z)}}+w^2=0,
\end{equation}
i.e. the $(A_{1},{\mathcal G})$ threefold  with ${\mathcal G}=A_3$.

\subsection{5d zero modes computation}
\label{5dmodes}

Given a Higgs field $\Phi$, the zero modes are the holomorphic deformations of the Higgs field up to (linearized) gauge transformations, i.e. $\varphi \in \mathcal{G}$ such that
\be\label{LinGaugeTr}
\partial \varphi = 0 \qquad \varphi \sim \varphi + [\langle \Phi \rangle, g]\,,
\ee
with $g \in \mathcal{G}$.
To study the zero modes, we then have to work out which components of the deformation $\varphi$ can be set to zero by a gauge transformation \eqref{LinGaugeTr}. One then tries to solve the equation
\begin{equation}\label{eqZeroModes}
  \varphi + \delta_{g} \varphi = 0 \,, \qquad\mbox{with}\qquad \delta_{g}\varphi = [ \Phi(w) , g] 
\end{equation}
with unknown $g \in \mathcal{G}$. 
Each component of $\varphi$ is a holomorphic polynomial in $w$. There will be gauge transformations that cancel the full polynomial and gauge transformations that allow to cancel only some powers in $w$. In the first case, that component does not support any zero-mode. In the second case, the gauge fixed mode may belong to $\mathbb{C}[w]/(w)$; this means that at $w\neq 0$ the mode can be gauge fixed to zero, but at $w=0$ we still have some freedom: the result is one 5d zero mode localized at $w=0$. The gauge fixing may also produce modes in $\mathbb{C}[w]/(w^k)$; in this case we have $k$ zero modes localized at $w=0$.
Finally, there are components that are not touched by the gauge fixing procedure: they host a 7d mode that extends in all $\mathbb{C}_w$.

Let us make an example: we consider the conifold singularity, that in the language of this paper is the $(A_1,A_1)$ threefold. We let $\Phi$ belong to the Cartan subalgebra of $A_1$; this gives a $U(1)$ flavor group generated by $\mathcal{H}=\langle\alpha^*\rangle$. The Higgs field corresponding to the threefold equation (using \eqref{threefolds}) is
\begin{equation}
\Phi=\begin{pmatrix}
w & 0 \\ 0 & - w \\
\end{pmatrix} \,,\qquad\qquad x^2+y^2+z^2-w^2 = 0\:.
\end{equation}

Let us count the zero modes. We parameterize both the fluctuation and gauge parameter as follows:
\be
\varphi= \bm \varphi_0 & \varphi_+\\ \varphi_- & -\varphi_0 \eem\,, \qquad g = \tfrac{1}{2}\,\bm g_0& g_+\\g_-& -g_0 \eem\,.
\ee
We then have
\be
\varphi \sim \varphi + w \bm 0&g_+\\-g_-&0 \eem\,.
\ee
This tells us a few things. The entry $\varphi_0$ is not gauge fixed; this then gives a 7d mode. Moreover, the fluctuations $\varphi_\pm$ are defined up to any multiple of $w$, i.e. $\varphi_\pm \in \cc[w]/(w) \cong \cc$. This means that they are localized on $w=0$ and are therefore genuinely 5d dynamical fields. These are charged under the $U(1)$ flavor group.
The pair $(\varphi_+, \varphi_-)$ forms a free hypermultiplet, as expected for the conifold. 

In a more complicated model, it is worth using the fact that 
the Higgs field lives in the maximal subalgebra $\mathcal M$ of the Levi subalgebra $\mathcal{L}$. We can then branch the algebra $\mathcal{G}$ w.r.t.\ $\mathcal{M}$:
\begin{equation}\label{BranchingGM}
 \mathcal{G} = \bigoplus_p R_p^{\mathcal{M}} \:.
\end{equation}
A representation $R^\mathcal{M}$ of $\mathcal{M}$ can be written more explicitly as
\begin{equation}
  R^{\mathcal{M}}=  (\,R^{\mathcal{M}_1}\,,\,  R^{\mathcal{M}_2}\,,\, ... \,)_{q_1,...,q_\ell}\:,
\end{equation}
where $(q_1,...,q_\ell)$ are the charges under the $U(1)^\ell$ flavor group generated by $\mathcal{H}$.
Since $\Phi\in\mathcal{M}$, if we take $g\in R^\mathcal{M}$, then also the commutator in \eqref{LinGaugeTr}  lives in $R^\mathcal{M}$.
We can then solve the equation \eqref{eqZeroModes} in each representation in \eqref{BranchingGM} individually, i.e.\ we
 consider the deformations $\varphi$ in each $R^\mathcal{M}$ and check how much of it can be fixed by \eqref{LinGaugeTr}. All the modes living in a given representation $R^\mathcal{M}$ have the same charges $(q_1,...,q_\ell)$ under the $U(1)^\ell$ flavor group generated by $\mathcal{H}$. We provide an example of zero modes computation in a given representation in Appendix~\ref{appendix C}.

In \cite{Collinucci:2022rii} we have worked out an algorithm that allows to compute the zero modes in all cases, together with their charges under the 5d symmetries. We will use that algorithm also to get the results in the following sections. In Appendix~\ref{Appendix B}, we describe how to use the Mathematica code that implements the algorithm (and that is uploaded on the same \emph{arXiv} page of the present paper).

\

We conclude by considering a case that we will recurrently encounter in the following. Consider two Higgs fields $\Phi$ and $\tilde{\Phi}$ related as
\begin{equation}
\tilde\Phi = w^j\Phi\:,
\end{equation}
and with $\Phi(0)\neq 0$, while $\tilde{\Phi}$ has a zero of order $j$ at $w=0$.

We can compute the zero modes of $\tilde{\Phi}$, knowing the zero modes of $\Phi$:
The components of the deformation $\varphi$ that were gauge fixed to zero by $\Phi$, now host zero modes in $\mathbb{C}[w]/(w^j)$. Components that hosted localized modes in $\mathbb{C}[w]/(w^k)$, now support zero modes in $\mathbb{C}[w]/(w^{j+k})$.
We further note that the Casimir invariants of $\Phi$ and $\tilde{\Phi}$ are related by
$  \text{Tr}\left((\tilde\Phi)^i\right) =
  \text{Tr}\left((w^j\Phi)^i\right)=
  w^{i \cdot j}\text{Tr}\left((\Phi)^i\right)$.

These simple facts will permit us to reproduce the Higgs fields of all the
quasi-homogeneous cDV, first identifying a finite set of Higgs field
profiles, and then producing all the other
Higgs fields multiplying them by an appropriate power of $w$.

\subsection{The symmetry group}
\label{sec:symmetry-group}

The 7d theory has gauge group $G$, whose Lie algebra is $\mathcal{G}$. Since all  fields are in the adjoint representation of $\mathcal{G}$, the non-trivial acting group is the quotient of the simply connected group associated with $\mathcal G$ modulo its center\footnote{Actually there is an ambiguity in choosing the global group of the 7d theory \cite{Freed:fx,Freed:zh,Garcia-Etxebarria:rw,Morrison:2020ool,Albertini:hc}. Taking the minimal choice, as we are doing, one captures the non-trivial discrete symmetries that come solely from Higgsing. Different choices would enlarge the discrete symmetries with elements of the center of the group.}. We take  such a quotient as our 7d group $G$.

Switching on the vev for $\Phi(w)$ on one side breaks $G$ and on the other side generates  zero modes localized at $w=0$, that are charged under the preserved symmetry group.
Such a  symmetry group is $\text{Stab}_G(\Phi) \subset 
G$, with
\begin{equation}
  \label{stabilizers definition}
  \text{Stab}_{ G}(\Phi) \equiv \left\{U \in  G \text{ s.t. } U \Phi
    U^{-1} = \Phi\right\}\:.
\end{equation}

Our Higgs field $\Phi$ engineers a threefold family that (simultaneously) resolves the roots $\alpha_1,...,\alpha_\ell$. This is realized by letting $\mathcal{H}$ (defined in \eqref{Eq:calH}) commute with $\Phi$. The commutant of $\mathcal{H}$ is the Levi subalgebra $\mathcal{L}$ associated with the choice of the roots $\alpha_1,...,\alpha_\ell$. If the Higgs field $\Phi$ is a generic element of $\mathcal{L}$, then $\text{Stab}_G(\Phi)=U(1)^\ell$ (generated by $\mathcal{H}$).

Such $U(1)^{\ell}$ group, namely the symmetry preserved by $\Phi \in \mathcal{L}$, is nothing but the five-dimensional flavor group, acting via its adjoint representation on the hypermultiplets coming from the deformation $\varphi$. The explicit flavor charges of the hypermultiplets can be readily computed employing the irrep decomposition \eqref{BranchingGM}, that naturally regroups hypers of the same charge into the same irrep.
In general, if we only have one $U(1)$ factor, associated with a simple root $\alpha_i$, then the flavor charges can acquire values only up to the dual Coxeter label of the node $\alpha_i$ in the Dynkin diagram of the considered 7d algebra \cite{Collinucci:2022rii}. If, instead, $\text{Stab}_G(\Phi)=U(1)^\ell$ with $\ell > 1$, this is not valid anymore.

As we have said, generically we have
\begin{equation}
  \label{maximal sub. issue}
  \Phi \in \mathcal M ,  \qquad\mbox{with}\qquad  \mathcal M=\bigoplus_h
\mathcal M_h   \oplus \mathcal H
\end{equation}
where $\mathcal{M}$ is a maximal subalgebra of  $\mathcal{L}$. If $\mathcal{M}\subset \mathcal{L}$, the preserved group will be bigger than $U(1)^\ell$ and it will develop a discrete group part.

\begin{figure}[H]
    \centering
    \includegraphics[scale=0.22]{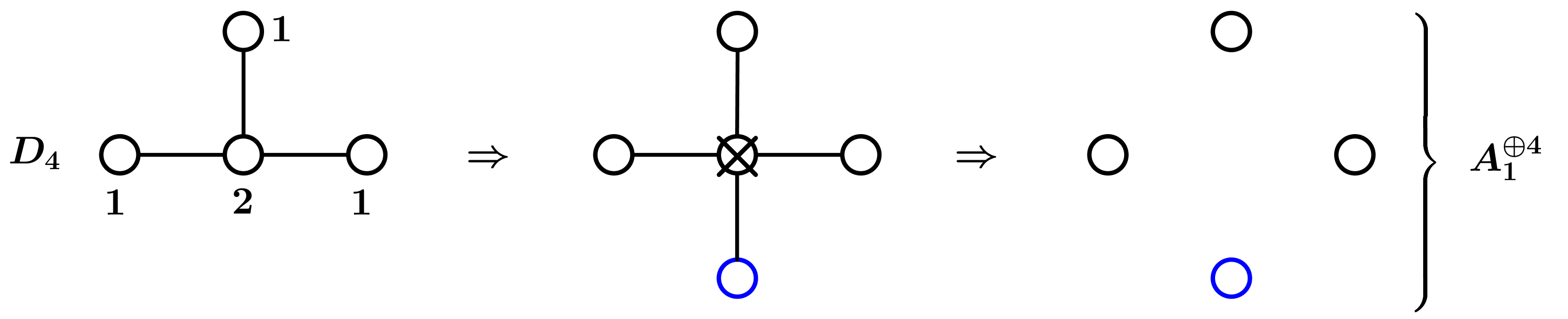}
    \caption{$A_1^{\oplus 4}$ subalgebra of $D_4$.}
    \label{D4coxeter}
\end{figure}

To explain how this works, we consider a simple example (that will appear often in the threefolds studied in the following). We take 
$$\mathcal{L}=D_4 \quad\mbox{ and }\quad \mathcal{M}=A_1^{\oplus 4}.$$ 
The Dynkin diagram of $D_4$ with its dual Coxeter labels, along with its $A_1^{\oplus 4}$ subalgebra, is depicted in Figure \ref{D4coxeter}.  The $A_1^{\oplus 4}$ maximal subalgebra is generated by adding the external node of the extended $D_4$ Dynkin diagram and removing the central one.

There are transformations of\footnote{Given a subalgebra $\mathcal{L}\subset \mathcal{G}$, we call $G_\mathcal{L}$ the subgroup of $G$, whose Lie algebra is $\mathcal{L}$.
} $G_\mathcal{L}$ that preserve all the elements of $\mathcal{M}=A_1^{\oplus 4}$ (while they break $\mathcal{L}=D_4$). In this case there is one such element: it is generated by the Cartan $\alpha_2^*$, i.e.\ the dual of the root that should be removed from the $D_4$ extended Dynkin diagram to obtain the Dynkin diagram of $A_1^{\oplus 4}$.\footnote{In Heterotic string theory on $T^3$, this element is known as a discrete Wilson line.} The element that is in the stabilizer of $\Phi\in\mathcal{M}=A_1^{\oplus 4}$ is 
\begin{equation}\label{gammaZ2}
  \gamma = \exp\left[\frac{2\pi i}{\mathfrak{q}_{\alpha_2}} \alpha_2^*\right]\:,
\end{equation}
where  $\mathfrak{q}_{\alpha_i}$ is the dual Coxeter label of the simple root $\alpha_i$, and where $\gamma\in G$ acts on the adjoint representation. In our case, we read $\mathfrak{q}_{\alpha_2}=2$ (see Figure~\ref{D4coxeter}). In particular, we have 
\begin{equation}
\gamma\cdot e_{\alpha_i} = \mathsf{e}^{\frac{2\pi i}{2} 0} e_{\alpha_i}  = e_{\alpha_i} \quad\mbox{for }i=1,3,4,\qquad\qquad
\gamma\cdot e_{\alpha_2} = \mathsf{e}^{\frac{2\pi i}{2} 1} e_{\alpha_2}  = - e_{\alpha_2}\:,
\end{equation}
and 
\begin{equation}
\gamma\cdot e_{\alpha_\theta} = \mathsf{e}^{\frac{2\pi i}{2} (-2)} e_{\alpha_\theta}  = e_{\alpha_\theta}\:,
\end{equation}
where $\alpha_\theta$ is the (minus the) highest root corresponding to the extended node. Note that the Lie algebra element $e_{\alpha_2}$ is not preserved by $\gamma$.

We see that it is crucial for preserving a maximal subalgebra that the coefficient in front of $\alpha_2^*$ in $\gamma$ is $\frac{2\pi i}{\mathfrak{q}_{\alpha_2}}$ and not any other number.
The discrete group generated by $\gamma$ in \eqref{gammaZ2} is isomorphic to $\mathbb{Z}_2$.

\begin{figure}[H]
    \centering
    \includegraphics[scale=0.16]{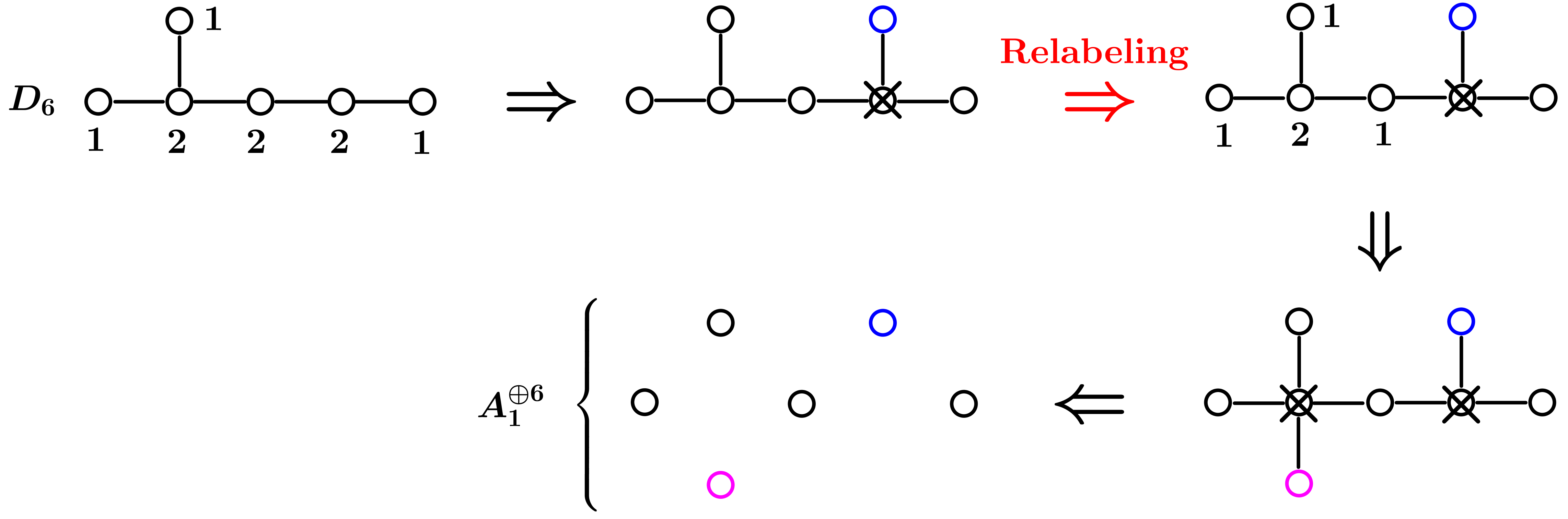}
    \caption{$A_1^{\oplus 6}$ subalgebra of $D_6$.}
    \label{D6discrete}
\end{figure}

Let us generalize this to an example that is a bit more involved, i.e.\ 
$$\mathcal{L}=D_6\quad\mbox{ and } \quad \mathcal{M}=A_1^{\oplus 6}.$$ 
In this case we proceed by steps, following the inclusions $D_6\supset D_4\oplus A_1^{\oplus 2} \supset A_1^{\oplus 4}\oplus  A_1^{\oplus 2} = A_1^{\oplus 6}$, depicted in Figure \ref{D6discrete}. In the first step, we remove a node with dual Coxeter label equal to $2$. We are then left with the final step in which we embed $A_1^{\oplus 4}$ into $D_4$: again we remove a node of $D_4$ Dynkin diagram with label equal to $2$. We conclude that the discrete group is $\mathbb{Z}_2^2$.

It is then easy to generalize to a generic case. Say that a simple summand of $\mathcal{L}$ has a maximal subalgebra, obtained by subsequently removing nodes with dual Coxeter labels $\mathfrak{q}_{\alpha_{\iota_1}},...,\mathfrak{q}_{\alpha_{\iota_k}}$ . Then the stabilizer of $\Phi$ will include the discrete group 
$$\mathbb{Z}_{\mathfrak{q}_{\alpha_{\iota_1}}} \times ...\times \mathbb{Z}_{\mathfrak{q}_{\alpha_{\iota_k}}}.$$ 
Doing this for all simple summands of $\mathcal{L}$, we obtain the full discrete symmetry $\Gamma_\Phi$. The full symmetry group is then
\begin{equation}\label{StabPhiTot}
\text{Stab}_{ G}(\Phi) = U(1)^\ell \times \Gamma_\Phi\:.
\end{equation}
Since we know how the generators of this group act on the Lie algebra $\mathcal{G}$, we can easily derive the charges under $\text{Stab}_{ G}(\Phi)$ of the deformations $\varphi$ in $R^\mathcal{M}$, i.e.\ of the 5d hypermultiplets.

The symmetry group \eqref{StabPhiTot}
is the 7d gauge group that survives the Higgsing. In order to deduce the 5d flavor and gauge symmetries we can proceed as in \cite{Collinucci:2021ofd}: we consider the 7d space as a decompactification limit from 5d times a 2-torus. Before the limit, \eqref{StabPhiTot} is a 5d gauge group; the decompactification limit will ungauge the continuous factor as its gauge coupling vanishes. The discrete part, having no gauge coupling, remain gauged in 5d.

\subsubsection*{Explicit example: $(A_2,D_4)$ singularity and discrete groups}
\label{sec:expl-exampl-a_2}

Let us visualize how it works in an explicit example. We can consider the
$(A_2,D_4)$ singularity:
\begin{equation}
  \label{A2D4}
  x^2 + z y^2 + z^3 + w^3=0, \quad (x,y,w,z) \in \mathbb C^4.
\end{equation}
The threefold can be described as a family of $D_4$ ADE singularities deformed by the
parameter $w$. The Higgs field is taken in the maximal subalgebra of $D_4$, i.e.  $\mathcal M = D_2 \oplus D_2 \cong A_1^4$.

From what we said above, it is immediate to find out
\begin{equation}
  \label{eq:183}
   \text{Stab}(\Phi)_{ G} = \mathbb Z_2 \:.
  \end{equation}
We now see how this discrete group acts on the 5d hypermultiplets.
We first branch $\mathcal G = D_4$ under $\mathcal M$:
\begin{equation}
  \label{branching D4 w.r.t. D2+D2}
  D_4 = A_1^{(I)}
\oplus  A_1^{(II)} \oplus  A_1^{(III)} \oplus  A_1^{(IV)} \oplus
  \text{(\textbf{2},\textbf{2},\textbf{2},\textbf{2})} = \mathcal M
  \oplus \text{(\textbf{2},\textbf{2},\textbf{2},\textbf{2})}.
\end{equation}
We then see how $\gamma$ in \eqref{gammaZ2} acts on the elements of $\text{(\textbf{2},\textbf{2},\textbf{2},\textbf{2})}$. The generators of $D_4$ appearing in this representation of $\mathcal{M}$ are related to roots that are linear combination of the simple roots where $\alpha_2$ appears with coefficient $1$. This immediately tells us that all elements of $\text{(\textbf{2},\textbf{2},\textbf{2},\textbf{2})}$ get a $-1$ factor when we act with $\gamma$.

This can be easily generalized to any choice of $\Phi\in\mathcal{M}\subset \mathcal{G}$ with $\mathcal{G}=A,D,E$.

\section{The Higgs vev from the threefold equation}\label{section higgs vev}

Our question is now: given a CY equation like \eqref{3foldFamilyGw}, what is the Higgs field that can generate it? The answer to this question is crucial in order to tackle the dynamics of M-theory on the quasi-homogeneous cDV singularities in Table~\ref{Xie table}.

 With this objective in mind, we will consider the two steps at page \pageref{Twostepsproced}, implemented at the Higgs level. 
From this perspective, the threefold is naturally embedded into the family over $\mathcal{B}_\varrho=\mathfrak{t}/\mathcal{W}'$ by choosing a one-dimensional subspace parametrized by $\mathbb{C}_{w}$. This means that the threefold will inherit the partial simultaneous resolution associated with $\mathcal{W}'$: both in the family and in the threefold the blown up roots will be, say $\alpha_1,...,\alpha_\ell$. This immediately tells us that the commutant of $\Phi$ is $\mathcal{H}$ in \eqref{Eq:calH}.
The choice of $\mathcal{W}'$ selects a 
maximal subalgebra $\mathcal{M}$ of the commutant $\mathcal{L}$ of $\mathcal{H}$ (see \eqref{MaxSubAlgLevi}), whose Casimirs are invariant under $\mathcal{W}'$ and are then good coordinates on $\mathcal{B}_\varrho=\mathfrak{t}/\mathcal{W}'$.
An element $\Phi\in \mathcal{M}\subseteq\mathcal{L}$ can be written as
\begin{equation}\label{PhiInMsevpieces}
  \Phi = \sum_h \Phi_h + \sum_{a=1}^\ell \varrho_1^a \alpha_a^*
\end{equation}
where $\Phi_h$ is an element of $\mathcal{M}_h$. Collecting the degree-$j$ Casimir invariants $\varrho_j^h$ of $\Phi_h$ in $\mathcal{M}_h$, together with the coefficients $\varrho_1^a$ , one obtains the invariant coordinates $\varrho_i$ on the base $\mathcal{B}_\varrho$.

\subsection{From the threefold equation to the partial Casimirs $\varrho_i(w)$}\label{section from the threefold}

Now, we will proceed as follows: We start from the equation of a threefold in Table~\ref{Xie table}. We will derive what is the minimal $\mathcal{W}'$ such that the partial Casimirs $\varrho_i$ can be taken as holomorphic (homogeneous) functions of $w$, in a way that produces the CY equation by taking $\mu_i=\mu_i(\varrho(w))$. This will tell us what is the $w$-dependence of the Casimirs $\varrho_j^h$ of each $\Phi_h$ and the $w$-dependence of the coefficients $\varrho_a$. Finally, we will look for Higgs fields $\Phi(w)\in\mathcal{M}$, holomorphic in $w$, that have the given $w$-dependence for their partial Casimirs.\footnote{Fixing the $w$-dependence of the partial Casimir invariants does not give a unique choice for a holomorphic element of $\mathcal{M}$.}

In particular, to reproduce the threefolds in Table~\ref{Xie table}, we want to determine which holomorphic functions $\varrho^I_j(w)$, with $I=(h,a)$ make all deformation parameters vanish except one of degree $M$, that is 
\begin{equation}
\label{hom mu}
\mu_M(\varrho(w))=w^N\:.
\end{equation}
We stress that $\mu_M(\varrho(w))$ is a \emph{homogeneous} polynomial in $w$ of degree $N$.

Both the $\mu_M$ and the $\varrho_j^I$ can be written as homogeneous polynomials in the $t_i \in \mathfrak{t}$ of degree, respectively, $M$ and $j$.
This implies that $\mu_M(\varrho)$ will be a weighted \emph{homogeneous} polynomial in the coordinates $\varrho^I_j$'s of degree $M$, where the coordinate $\varrho_j^I$ has weight $j$. This, together with \eqref{hom mu}, implies that $\varrho^I_j(w)$ is a homogeneous function of $w$ with degree $\frac{j\,N}{M}$, i.e.
\begin{equation}
   \varrho^I_j(w)=c_j^I \,w^{\frac{jN}{M}} \:.
\end{equation}
Now:
\begin{itemize}
\item Since we require that $\varrho^I_j(w)$ is holomorphic, the partial Casimirs that give a non-zero contribution (i.e. $c_j^I\neq 0$) are those with $j$ such that
\begin{equation} \label{MdivNj}
\frac{j\,N}{M}\,\,\in\,\,\mathbb{Z}^{>0}\:.
\end{equation} 
\item Moreover, we want to pick the smallest $\mathcal{W}'$ that allows holomorphic functions $\varrho^I_j(w)$ compatible with \eqref{hom mu}. Small $\mathcal{W}'$ correspond to subalgebras $\mathcal{M}$ with several simple summands with small rank. This subalgebra then yields the smallest degree partial Casimirs that realize \eqref{MdivNj}, for given $M,N$.

\end{itemize}

%
Choosing the threefold in Table~\ref{Xie table} determines $M$ (see the last column of the table). 
For each value of $N$, we look for the minimal value of $j$ that satisfies \eqref{MdivNj}.
Say that $M$ has $n_M$ divisors $q_1,...,q_{n_M}$, where $q_1=1$ and $q_{n_M}=M$.
Then $N$ can always be written in a unique way as
\begin{equation}\label{Nofqalpha}
 N = \frac{p}{q_\alpha}M\,\,\, \mbox{ mod }\,\,\,M \,, 
\end{equation}
with $q_\alpha$ a divisor of $M$, $p<q_\alpha$ and $(p,q_\alpha)$ coprime. 
The condition \eqref{MdivNj} becomes then
\begin{equation} \label{MdivNjbis}
\frac{j\,p}{q_\alpha}\,\,\in\,\,\mathbb{Z}^{>0}\:,
\end{equation} 
and the minimal value of $j$ fulfilling it is $j=q_\alpha$. 

Given $N$, only $\varrho^I_j$ with $j$ a multiple of $q_\alpha$ can be non zero. In other words, $c_j^I=0$ when $j\neq m\,q_\alpha$ with $m\in\mathbb{Z}$. Because of homogeneity, this implies also that $\mu_i(\varrho)=0$ with $i\neq m\,q_\alpha$. We are then left with the following equations with unknown $c_j^I$ ($j=mq_\alpha$):
\begin{equation}\label{sysLinEqmu}
\left\{\begin{array}{lcl}
\mu_{m\, q_\alpha}(c)=0 & & m\,q_\alpha<M\\
\mu_{M}(c)=1 & & 
\end{array}\right.
\end{equation}
(where we have factored out the powers in $w$).
In order to have a non-trivial solution, one requires that all $c_j^I$ with $j=m\,q_\alpha$ be non-zero\footnote{Otherwise the system of homogeneous equations in the first row of \eqref{sysLinEqmu} will force all  $c_j^I$'s to vanish. We notice that the number of holomorphic $\rho_j^I$ has to be equal to the number of all the $\mu_{m\, q_\alpha}$, $\mu_{M}$. If that was not the case, the system \eqref{sysLinEqmu} would be overconstrained, and a solution would not be guaranteed to exist.
}.

\

Let us see how we can use this information to extract the subalgebra $\mathcal{M}$ corresponding to a given choice of $(A_{N-1},G)$. We describe this in a simple example, i.e. $(A_{N-1},D_4)$. The $D_4$ algebra has four Casimirs: $\mu_2$, $\mu_4$, $\tilde{\mu}_4$ and $\mu_6$. Hence $M=6$. There are four divisors of $6$: 
$$q_\alpha\in\{1,2,3,6\}.$$
We now see which (minimal) degree can take the partial Casimirs and then what is the choice of the minimal subalgebra  $\mathcal{M}$ (minimal $\mathcal{W}'$).
Let us vary $N$:
\begin{description}
\item {\bf For $N=0$ mod $6$} ($q_\alpha=1$), the minimal degree is $j=1$. We look for a subalgebra $\mathcal{M}$ with all four partial Casimirs of degree $1$. This is the smallest possible choice, i.e. the Cartan subalgebra of $D_4$. In this case, all four roots of $D_4$ are blown up in the simultaneous resolution.
\item {\bf For $N=3$ mod $6$} ($q_\alpha=2$), the minimal degree is $j=2$. There is actually  a  subalgebra of $D_4$ with four partial Casimirs of degree $2$, i.e. $\mathcal{M}=A_1^{\oplus 4}$.\footnote{Notice that all $c_2^I$'s must be non-zero; otherwise, if one vanished,  the equations $\mu_2=\mu_4=\tilde{\mu}_4=0$ would force all the others $c_2^I$'s to be zero as well as $\mu_6$.} $\mathcal{M}$ is now a maximal subalgebra of $D_4$; correspondingly, there is no resolution at the origin of the family, hence the singularity is terminal.
\item {\bf For $N=2,4$ mod $6$} ($q_\alpha=3$), the minimal degree for the non-zero partial Casimir is $j=3$. 
In any subalgebra of $D_4$, we can have at most one partial Casimir of degree $3$. Moreover, $\mu_2$ must depend on partial Casimirs of degree lower than $3$, that must vanish identically
 (otherwise they would be non-holomorphic, due to \eqref{MdivNj}). We have $\mathcal{M}=A_2\oplus\langle\alpha_3^*,\alpha_4^*\rangle$. Only the partial Casimirs of the semi-simple part of $\mathcal{M}$, that is $A_2$, are non-vanishing.  In this case, the roots $\alpha_3$ and $\alpha_4$ of $D_4$ are blown up in the partial simultaneous resolution.
\item {\bf For $N=1,5$ mod $6$} ($q_\alpha=6$), the minimal degree for a non-vanishing partial Casimir is $j=6$, hence in this case $\mathcal{M}=D_4$ with all Casimirs equal to zero, except the maximal degree one. For $N=1$ the manifold is non-singular, while for $N=5$ there is a terminal singularity at the origin of the family.
\end{description}

As one can note in the presented example, the simple algebras $\mathcal{M}_h$ in $\mathcal{M}$ are all of the same type for a given value of $N$. This actually happens for all the cases we study in this paper. The reason is the following: we look for partial Casimirs with the lowest possible degree, realizing $\mu_M=w^N$. If one degree is allowed, we take as many partial Casimirs with that degree as we are allowed. Small degree partial Casimirs correspond to small subalgebras $\mathcal{M}_h$, hence we finish with as many summands of a given small algebra as we can.

\subsection{From the partial Casimirs $\varrho_i(w)$ to the Higgs field $\Phi(w)$}\label{section to the higgs field}

Now that we have the $w$-dependence of the $\varrho_j^I$'s, we need to take a Higgs field in $\mathcal{M}$, whose partial Casimirs have that dependence. In general, there are several choices for $\Phi_h(w)$ (see \eqref{PhiInMsevpieces}) with given $\varrho_j^h(w)$. Each choice produces a different number of zero modes. We decide to look for the Higgs field $\Phi$ that localizes the maximal number of zero modes and breaks the 7d gauge symmetry in the least disruptive way, and we interpret the others as T-brane deformations of $\Phi$, i.e.\ deformations that kill a number of modes, or destroy a preserved symmetry, without touching the threefold singularity (we come to this point in Section~\ref{Sec:T-branes}). With this choice, we pick up 
the Higgs field that leads to the same number of zero modes that are counted by the normalized complex structure deformations of the CY\footnote{In a nutshell, the procedure goes as follows \cite{Xie:2015rpa}: first we write down a basis, as $\mathbb C$-vector space, of the Jacobian ring $R = \frac{\mathbb C[x,y,w,z]}{(F,\frac{\partial F}{\partial x},\frac{\partial F}{\partial y},\frac{\partial F}{\partial w},\frac{\partial F}{\partial z})}$, with $F$ the polynomial defining $X$. It is a mathematical fact that the elements of the basis can be chosen to be monomials, and hence have a well-defined scaling w.r.t.\ the quasi-homogeneous action on $X$. It turns out that we can pair, looking at these scaling weights, a number $2 n_{\text{paired}}$ of monomials of the basis, while leaving other $n_{\text{unpaired}}$ unpaired. The expected Higgs branch quaternionic dimension (that equals the number of 5d hypers), then, is 
\begin{equation*}
    \label{paired and unpaired}
    d_{\mathbb H} = n_{\text{paired}} + n_{\text{unpaired}}.
\end{equation*}
$d_{\mathbb H}$ also coincides with the number of normalizable (and log-normalizable) complex structure deformations \cite{Shapere:1999xr}.
}.

Let us first describe what is the structure of the Higgs field. At $w=0$ the fiber of the one-parameter $\mathcal{G}$-family must develop a full $\mathcal{G}$-type singularity. This means that $\Phi(0)$ must be a nilpotent element of $\mathcal{M}$ (as all its Casimirs should vanish), that we take in its canonical  form (e.g.\ for $A_r$ it is the Jordan form; for general ADE singularities, we refer to \cite{Collingwood}).  Now, $\Phi(w)$ must be a deformation of the nilpotent element $\Phi(0)$, with deformation proportional to $w$ and belonging to $\mathcal{M}$. The way to do it in a way that goes into a transverse direction to the nilpotent orbit (that includes $\Phi(0)$) is dictated by taking $\Phi_h$ in the Slodowy slice in $\mathcal{M}_h$ passing through $\Phi_h(0)$. We give the proper definition in Appendix \ref{appendix D}. What is important here is that this allows to have canonical forms for the Higgs field in $\mathcal{M}$, that are not equivalent by gauge transformations. The Higgs field will then be given as the sum of some simple root generators of $\mathcal{G}$ multiplied by $1$ and of other generators (in $\mathcal{M}$) multiplied by powers of $w$.

To pick up the Higgs field that localizes the maximal number of modes, we need to properly choose the nilpotent orbit, which $\Phi(0)$ belongs to.
Let us consider $\Phi,\Phi'\in\mathcal{M}$ with the same expressions for $\varrho_j^I$, but such that $\Phi(0)$ and $\Phi'(0)$  belonging to two different nilpotent orbits. Then, they produce a different number of zero modes: the one whose nilpotent orbit at the origin is smaller has a bigger number of zero modes. Roughly speaking, if at the origin the orbit is bigger, one has a larger number of `$1$'s in the canonical form of the Higgs; these gauge fix to zero a bigger number of Lie algebra components in the deformation $\varphi$. A more detailed explanation of these aspects, complemented by explicit examples, can be found in Appendix \ref{appendix D}, where we lay down the complete recipe to connect the partial Casimirs to the Higgs background.

If the power of $w$ in the partial Casimirs $\varrho_j^I$ is high, the minimal orbit at the origin reproducing the required $w$-dependence will be the trivial one. In these cases, the Higgs field that leads to the maximum number of zero modes is such that
\begin{equation}
\Phi= w^{k}\hat\Phi \,,
\end{equation}
with $\hat\Phi(0)$ a non-trivial nilpotent element of $\mathcal{M}$. Knowing the zero modes of $\hat\Phi$, one is able to find the zero modes of $\Phi$.

\

Let us illustrate how we pick the right choice of $\Phi$ with given $\varrho_j^I(w)$, by using the $(A_{N-1},D_4)$ example.
\begin{description}
\item {\bf For $N=1$}, $\mathcal{M}=D_4$, $\rho_6=\mu_6=w$. $\Phi(0)$ is in the maximal nilpotent orbit of $D_4$ and its expression at generic $w$ is dictated by the $w$-dependence of the Casimir:
\begin{equation}
\Phi= e_{\alpha_1} +e_{\alpha_2}+e_{\alpha_3}+e_{\alpha_4}+\frac{w}{4}e_{-\alpha_1-2\alpha_2-\alpha_3-\alpha_4}.
\end{equation}
\item {\bf For $N=2$},  $\mathcal{M}=A_2\oplus\langle\alpha_3^*,\alpha_4^*\rangle$. The only non-zero partial Casimir is the degree 3 Casimir of $A_2$: $\varrho_3=w$. The unique (up to gauge transformations) holomorphic  Higgs field compatible with that is now
\begin{equation}
 	\Phi =\Phi_{A_2}\quad\mbox{with}\quad \Phi_{A_2} = \begin{pmatrix}
 	0 & 1 & 0 \\ 0 & 0 & 1 \\ w & 0 & 0 \\ 
 	\end{pmatrix}=  e_{\alpha_1} + e_{\alpha_2} + w\, e_{-\alpha_1-\alpha_2}  \:.
\end{equation}
\item {\bf For $N=3$}, $\mathcal{M}=A_1^{\oplus 4}$, $\varrho_2^h=c^hw$ ($h=1,...,4$), with $c^h$ solving \eqref{sysLinEqmu}. The form of the Higgs field with these partial Casimirs is again unique:
  \begin{equation}
    \label{A2D4 higgs first}
 	\Phi = \sum_{h=1}^4 \Phi_h\quad\mbox{with}\quad \Phi_h = \begin{pmatrix}
 	0 & 1 \\ c^h \,w & 0 \\ 
 	\end{pmatrix}
 	=  e_{\alpha^h} + c^hw\, e_{-\alpha^h}  \:,
\end{equation}
where $\alpha^h$ is the root of the subalgebra $A_1^h$.
\item {\bf For $N=4$},  $\mathcal{M}=A_2\oplus\langle\alpha_3^*,\alpha_4^*\rangle$. Now, differently from the $N=2$ case, the only non-zero partial Casimir of degree 3 is quadratic in $w$: $\varrho_3=w^2$. In this case we have two possible Higgs fields that are consistent with this, i.e. $\Phi=\Phi_{A_2}$ with
\begin{equation}
 {\rm either}\quad \Phi_{A_2} = \begin{pmatrix}
 	0 & 1 & 0 \\ 0 & 0 & 1 \\ w^2 & 0 & 0 \\ 
 	\end{pmatrix}\quad{\rm or}\quad  \Phi_{A_2} = \begin{pmatrix}
 	0 & 1 & 0 \\ 0 & 0 & w \\ w & 0 & 0 \\ 
 	\end{pmatrix}\:.
\end{equation}
At the origin $w=0$, the left one is in the maximal nilpotent orbit while the right one is in the minimal one. Hence we expect that choosing the right one will give us the bigger number of zero modes. Indeed this happens, as it can be easily verified by an explicit computation.
\item {\bf For $N=5$}, $\mathcal{M}=D_4$, $\varrho_6=\mu_6=w^5$, the Higgs field is of the same shape as the $N=1$ case, with some coefficients proportional to $w$:
\begin{equation}
\Phi= e_{\alpha_1} +w\left(e_{\alpha_2}+e_{\alpha_3}+e_{\alpha_4}+\frac{1}{4}e_{-\alpha_1-2\alpha_2-\alpha_3-\alpha_4}\right).
\end{equation}
\item {\bf For $N=6$}, $\mathcal{M}=\mathcal{H}$, $\varrho_1^a=c^a w$ ($a=1,...,4$).
$\Phi$ is forced to be of the form 
\begin{equation}
 	\Phi = c^1 w \,\alpha_1^* + c^2 w \,\alpha_2^* + c^3 w \,\alpha_3^* + c^4 w \,\alpha_4^*\:.
\end{equation}

\end{description}
Let us see some cases where we go up with the power $N$ of $w$ in $\mu_6$:
\begin{description}
\item {\bf For $N=8$}, we obtain the same algebra as for $N=2$, i.e.  $\mathcal{M}=A_2\oplus\langle\alpha_3^*,\alpha_4^*\rangle$. Now, the only non-zero partial Casimir of degree 3 of $A_2$ takes the following $w$-dependence $\varrho_3=w^4$. The minimal nilpotent orbit at the origin compatible with this partial Casimir is now the trivial one. The Higgs field giving the maximal number of zero modes is
\begin{equation}
 	\Phi =\Phi_{A_2}\quad\mbox{with}\quad \Phi_{A_2} = w \begin{pmatrix}
 	0 & 1 & 0 \\ 0 & 0 & 1 \\ w & 0 & 0 \\ 
 	\end{pmatrix}=  w\,e_{\alpha_1} + w\,e_{\alpha_2} + w^2\, e_{-\alpha_1-\alpha_2}  \:.
\end{equation}
\item {\bf For $N=9$}, we obtain the same algebra as for $N=3$, i.e.  $\mathcal{M}=A_1^{\oplus 4}$. The Higgs field giving the maximal number of zero modes is
\begin{equation}
\Phi = \sum_{h=1}^4 \Phi_h\quad\mbox{with}\quad \Phi_h = w\begin{pmatrix}
 	0 & 1 \\ c^h \,w & 0 \\ 
 	\end{pmatrix}
 	= w\, e_{\alpha^h} + c^hw^2\, e_{-\alpha^h}  \:.
\end{equation}
\end{description}

The same can be done for the cases $N=7,10,11,12$, where the Higgs contributing most to the zero modes is the one with $N-6$ multiplied by $w$. In general, the Higgs fields given above for $N=1,...,6$ are enough to write the Higgs field for any $N$:
If $N=n+6k$, with $n\in\{1,...,6\}$, the Higgs field is $\Phi=w^k\Phi^{(n)}$, where $\Phi^{(n)}$ is the Higgs field for $N=n$.

This is actually true for all the cDV singularities in Table~\ref{Xie table}: 
\begin{center}\label{periodicity}
\emph{Given $M$ and $N$ as above, one needs to find the Higgs fields $\Phi^{(n)}$ for $N=n$, with $n\in\{1,...,M\}$. The Higgs field for $N=n+kM$ is  then $\Phi=w^k\Phi^{(n)}$.}
\end{center}
This is remarkably convenient also from the physical point of view, as the Higgs background $\Phi$ encodes all the 5d physics, meaning the localized hypers and their charges under the flavor and discrete symmetries. What the statement in italics is telling us is that, given a quasi-homogeneous cDV singularity built as an ADE singularity with a $\mu_M=w^N$ deformation term, we need to know only the Higgs backgrounds for $N$ up to $M$: all the rest can be obtained simply by multiplying these Higgs backgrounds by some power of $w$. The 5d mode counting changes as explained at the end of Section \ref{5dmodes}, the symmetries act in the same way on the (now possibly increased) modes, and the Higgs branch content varies accordingly, so that no new computation must be performed.

\section{5d Higgs Branches from quasi-homogeneous cDV singularities}\label{section cDV HB}
In this section we exhibit the complete classification of the 5d theories arising from M-theory on quasi-homogeneous cDV singularities.\\
\indent First, given a quasi-homogeneous cDV singularity, we must find the minimal subalgebra $\mathcal{M}$ in which a Higgs background $\Phi$ can reside, compatibly with the threefold equation (see Section \ref{section from the threefold}). Then, we find the Higgs field that produces the maximal number of modes following Section \ref{section to the higgs field} (checking that it is consistent with the HB dimension given by the normalizable complex structure deformations). Once we have the Higgs field $\Phi$, we can compute the 5d continuous flavor group, the discrete gauge group and the charges of the hypermultiplets under these groups.

We proceed methodically through all the cases in Table \ref{Xie table}. We will be brief when dealing with A- and D-families, as most of the work has already been done in \cite{DeMarco:2021try}. We treat the new exceptional cases in more detail.

\subsection{Quasi-homogeneous cDV singularities of $A$ type}
\begin{figure}[H]
     \centering
     \includegraphics[scale=0.10]{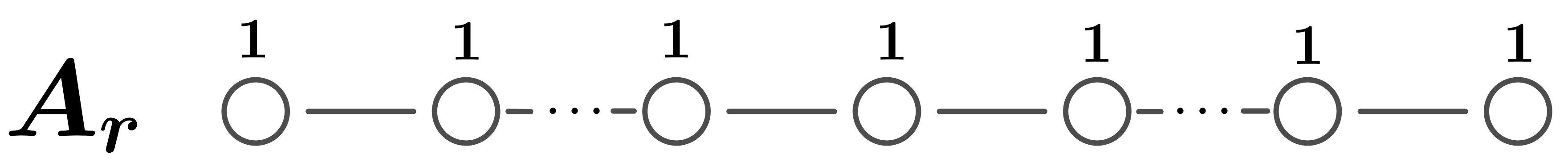}
     \caption{Dual Coxeter labels for the $A$ series.}
     \label{Coxeter A}
 \end{figure}

Two quasi-homogeneous cDV singularities of $A$ type exist: the $(A_{N-1},A_{M-1})$ and the $A^{(M)}_{M}[N]$. Their defining equations are
\begin{equation}\label{A Asing}
    \boldsymbol{(A_{M-1},A_{N-1})}: \quad\quad x^2+y^2+z^{M}+w^{N}=0,
    \end{equation}
\begin{equation}\label{A sing Xie}
    \boldsymbol{A^{(M)}_{M}[N]}: \quad\quad x^2+y^2+z\cdot( z^{M}+w^{N})=0.
\end{equation}
The non-vanishing deformation parameters are, in both cases, $\mu_M(w)=w^N$.
The equation \eqref{A Asing} is a $A_{M-1}$ family, while \eqref{A sing Xie} is a $A_{M}$ family.
It is however easy to see, adopting the technique fleshed out in section \ref{section higgs vev} (or equivalently the Type IIA approach employed in \cite{DeMarco:2021try}), that
the analysis of the $A^{(M)}_{M}[N]$ singularities can be fully traced back to the $(A_{M-1},A_{N-1})$ singularities: in particular one can see that
 the Higgs field in the $A_M$ family is living in a $A_{M-1}$ subalgebra and that both spaces are produced by the same choice of $\Phi\in A_{M-1}$.  The Higgs fields for the $(A,A)$ threefolds have already been considered in \cite{DeMarco:2021try} and can be used also for the $A^{(M)}_{M}[N]$ singularities. In general (and for some suitable choice of basis for the Cartan subalgebra), we find hypers of charge at most 1, as the dual Coxeter labels of the nodes of the $A$ Dynkin diagrams are all equal to 1, see Figure~\ref{Coxeter A}.
In Table \ref{table modes A}, we report the results for both the $(A_{k-1},A_{N-1})$ and the $A^{(k)}_{k}[N]$ singularities, rewriting them in full generality as $(A_{mp-1},A_{mq-1} )$ and $A_{mp}^{(mp)}[mq] $ singularities, respectively, and with $p$ and $q$ coprime, $p\geq q$. We give the resolution pattern, the corresponding flavor group, the number of charged hypers and the number of uncharged ones. The last ones are a signal of a non-resolvable singularity. The flavor groups are respectively $U(1)^{m-1}$ and $U(1)^m = U(1) \times U(1)^{m-1}$, where in the latter case the factor $U(1)^{m-1}$ is contained in a $A_{mp-1}$ subalgebra, as we have mentioned above. The flavor charges can be succintly understood as follows, in some basis of the Cartan subalgebra\footnote{For further details, we refer to the much more in-depth analysis of \cite{DeMarco:2021try}.}: for the $(A_{mp-1},A_{mq-1})$ cases, writing $U(1)^{m-1} \cong \frac{U(1)^{m}}{U_{\text{cm}}(1)}$ (where $U_{\text{cm}}(1)$ is the decoupled diagonal center of mass $U(1)$) there are $pq$ hypers charged in the bifundamental of every possible pair of $U(1)$'s in $U(1)^m$, as well as $m \frac{(p-1)(q-1)}{2}$ uncharged hypers. For the $A_{mp}^{(mp)}[mq] $  cases, there are $pq$ hypers charged in the bifundamental of every possible pair of $U(1)$'s in the numerator of the flavor group contained in the $A_{mp-1}$ subalgebra (regarded again as $\frac{U(1)^{m}}{U_{\text{cm}}(1)} \cong U(1)^{m-1}$), $q$ hypers charged bifundamentally under every possible pair formed by the $U(1)$ \textit{outside} the $A_{mp-1}$ subalgebra and a $U(1)$ in the numerator of $\frac{U(1)^{m}}{U_{\text{cm}}(1)}$, and finally there are $m \frac{(p-1)(q-1)}{2}$ uncharged hypers. 

\vspace{2cm}
\begin{table}[H]
$\scalemath{0.7}{
\renewcommand{\arraystretch}{2.5}
\begin{array}{|C||P|S|R|R|}
\Xhline{4\arrayrulewidth}
\large\textbf{Singularity} & \large\textbf{Resolution pattern} & \makecell{\large\textbf{Flavor}\\\large\textbf{group}\\ } & \large\textbf{Hypers} &  \large\textbf{Total hypers} \\
\hline
\hline
  (A_{mp-1},A_{mq-1} )  & \makecell{\includegraphics[scale=0.31]{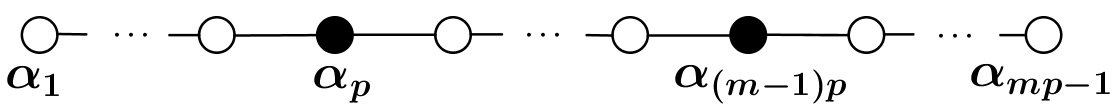} \\}& U(1)^{m-1} &\makecell{\text{Charged: } p q\frac{m(m-1)}{2} \\\text{Uncharged: } m \frac{(p-1)(q-1)}{2}\\} & \frac{1}{2} m (p (m q-1)-q+1) \\
\Xhline{4\arrayrulewidth}
\makecell{A_{mp}^{(mp)}[mq] \\}& \makecell{\includegraphics[scale=0.31]{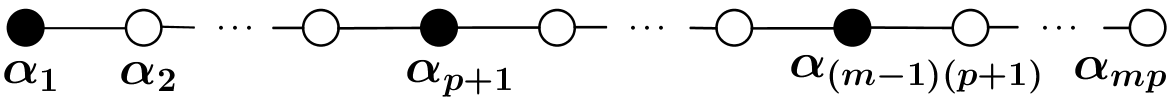} \\} &U(1)^{m} &\makecell{\text{Charged: }p q\frac{m(m-1)}{2}\boldsymbol{+mq} \\ \text{Uncharged: }m \frac{(p-1)(q-1)}{2}\\} & \frac{1}{2} m (p (m q-1)-q+1)\boldsymbol{+mq}\\
\Xhline{4\arrayrulewidth}
\end{array}
}
$
\caption{Higgs Branch data for quasi-homogeneous cDV singularities of $A$ type.}
\label{table modes A}
\end{table}

\newpage

\subsection{Quasi-homogeneous cDV singularities of $D$ type}
There exist two quasi-homogeneous cDV singularities arising from one-parameter deformations of $D$ singularities: the $(A_{N-1},D_{m+1})$ and the $D^{(m)}_{m}[N]$. Their defining equations read
\begin{equation}
    \boldsymbol{(A_{N-1},D_{m+1})}: \quad\quad x^2+zy^2+z^{m}+w^{N}=0,
    \end{equation}
\begin{equation}
    \boldsymbol{D^{(m)}_{m}[N]}: \quad\quad x^2+zy^2+z^{m-1}+yw^{N}=0.
\end{equation}
In the two cases, the non-vanishing deformation parameter is $\mu_M=w^N$, that is the maximal degree one for the first case ($M=2m$), while for the second case it is the always present $r$-degree deformation parameter of $D_r$ ($M=m$).

\begin{figure}[H]
     \centering
     \includegraphics[scale=0.10]{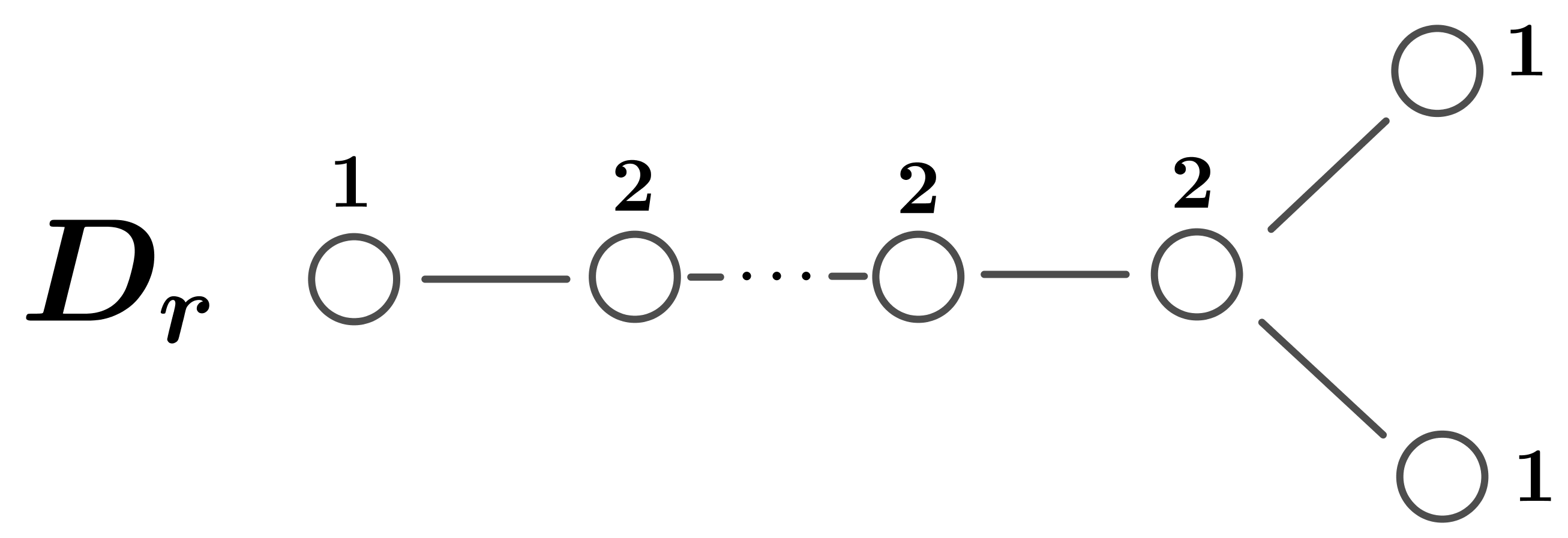}
     \caption{Dual Coxeter labels for the $D$ series.}
     \label{Coxeter D}
 \end{figure}

The 5d theories from M-theory on  $(A,D)$ singularities have been worked out in \cite{DeMarco:2021try}. We refer to that paper for the results. We have applied our method to work out also the $D^{(m)}_{m}[N]$ singularities.
As they are useful to identify the flavor charges of the hypermultiplets whenever a single node is resolved, in Figure~\ref{Coxeter D} we report the dual Coxeter labels of the nodes of the Dynkin diagrams in the $D$ series.

We notice that, in full generality, all the $(A_{2km-1},D_{m+1})$ and the $D_{m}^{(m)}[km]$ are completely resolvable, because in that case $N=kM$; this means, following Section \ref{section higgs vev}, that $q_\alpha=1$ and the minimal degree for the partial Casimirs is $j=1$, i.e.\ $\mathcal{M}$ is the Cartan subalgebra of $\mathcal{G}$.

In Table~\ref{table modes AD} and Table~\ref{table modes D} we report the results for the Higgs branch data, respectively, of the $(A_{N-1},D_4)$, $(A_{N-1},D_7)$ and $D_4^{(4)}[N],D_5^{(5)}[N],D_6^{(6)}[N]$ cases, specifying the flavor and discrete charges of the hypermultiplets. Other deformed $D_r$ examples can be treated analogously. 

\newgeometry{top=80pt,bottom=85pt,left=40pt,right=35pt}
\begin{table}[t]
\centering
$
\scalemath{0.7}{
\renewcommand{\arraystretch}{2.5}
\begin{array}{|L||V|L|W|A|L|}
\Xhline{4\arrayrulewidth}
\large\textbf{Singularity} & \large\textbf{Resolution pattern} & \boldsymbol{\mathcal{M}} & \makecell{\large\textbf{Symmetry}\\\large\textbf{group}\\ } & \large\textbf{Hypers} &  \makecell{\large\textbf{Total}\\\large\textbf{hypers}} \\
\hline
\hline
  \multirow{3}{*}{$(A_{N-1},D_4)$}& \makecell{N=6n: \includegraphics[scale=0.065]{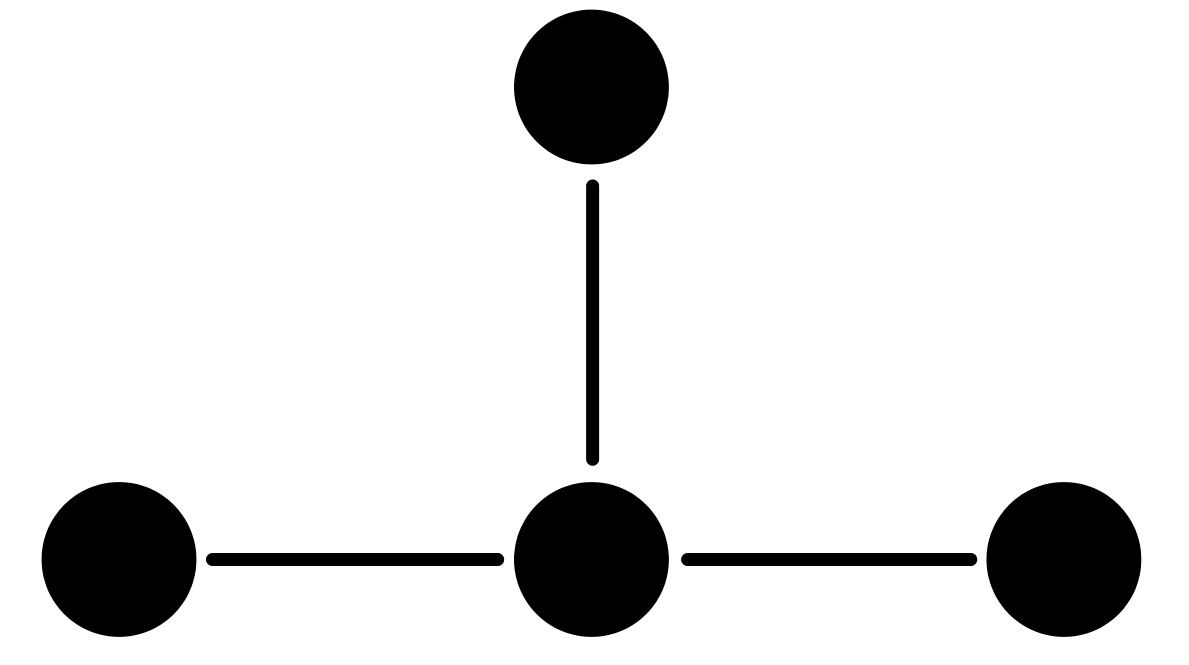}\\} &\mathfrak{t} & U(1)^4 & \renewcommand{\arraystretch}{1.1}\begin{array}{c}
  12n \\
 \text{Charges: \textbf{root system} of } D_4\end{array} &2N \\
 \cline{2-6}
 & \makecell{
 \begin{array}{c}
 N=2n \\
 n \neq 3j \\
 \end{array}:\includegraphics[scale=0.25]{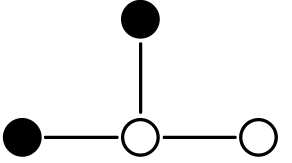}}& A_2\oplus\langle \alpha_1^*,\alpha_4^*\rangle & U(1)_a \times U(1)_b &\renewcommand{\arraystretch}{1.1}\begin{array}{c}
(q_a,q_b) = (2,0): \boldsymbol{n} \\
(q_a,q_b) = (1,1): \boldsymbol{2n} \\
(q_a,q_b) = (0,0): \boldsymbol{n-1} \\
 \end{array} & 2N-1\\
 \cline{2-6}
 & \makecell{ \begin{array}{c}
 N=3n \\
 n \neq 2j \\
 \end{array}:\includegraphics[scale=0.065]{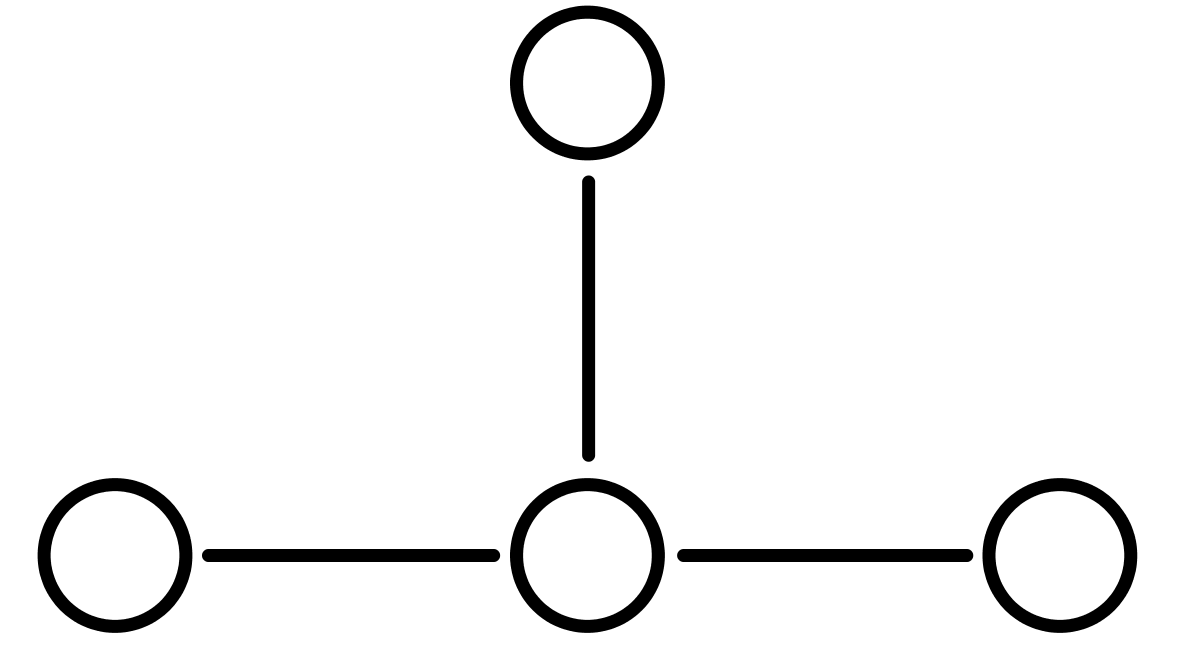}\\}& A_1^{\oplus 4} & \mathbb{Z}_2 &\renewcommand{\arraystretch}{1.1}\begin{array}{c}
q_{\mathbb{Z}_2}\neq 0: \boldsymbol{4n} \\
 q_{\mathbb{Z}_2}=0: \boldsymbol{2(n-1)}
 \end{array} & 2(N-1)\\
 \cline{2-6}
 & \makecell{N \neq 2n,3n:\includegraphics[scale=0.065]{D4nores.png}\\}& D_4  & \emptyset  & 2(N-1)& 2(N-1) \\ 
 \hline
\hline
 \multirow{3}{*}{$ (A_{N-1},D_7)$}   &\makecell{N=12n: \includegraphics[scale=0.2]{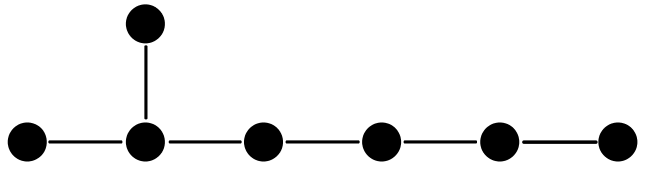} \\} &\mathfrak{t} & U(1)^7 & \makecell{42n \\
 \text{Charges: \textbf{root system} of } D_7} &\frac{7N}{2}\\
 \cline{2-6}
&\makecell{ \begin{array}{c}
 N=6n \\
 n\neq 2j \\
 \end{array}:\includegraphics[scale=0.2]{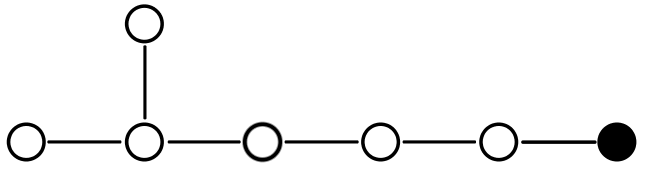} \\} & A_1^{\oplus 6}\oplus \langle \alpha_6^*\rangle & U(1)_a \times \mathbb{Z}_2^{(b)}\times \mathbb{Z}_2^{(c)} &
\renewcommand{\arraystretch}{1.1}\begin{array}{c}
(q_a,q_b,q_c) = (1,0,0): \boldsymbol{2n} \\
(q_a,q_b,q_c) = (1,0,1): \boldsymbol{2n} \\
(q_a,q_b,q_c) = (1,1,0): \boldsymbol{2n} \\
(q_a,q_b,q_c) = (0,1,1): \boldsymbol{n-1} \\
(q_a,q_b,q_c) = (0,1,0): \boldsymbol{n-1} \\
(q_a,q_b,q_c) = (0,0,1): \boldsymbol{n-1} \\
(q_a,q_b,q_c) = (0,0,0): \boldsymbol{12n} \\
\end{array} & \frac{7 N}{2}-3\\
 \cline{2-6}
 &\makecell{ \begin{array}{c}
 N=3n \\
 n \neq 2j \\
 \end{array}:\includegraphics[scale=0.2]{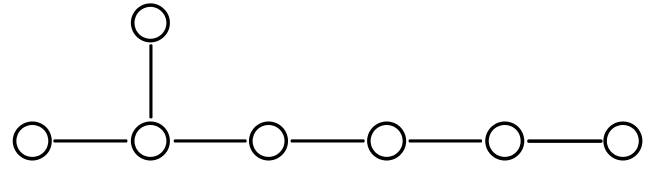} \\}&  D_4  \oplus A_3 & \mathbb{Z}_2 &
\renewcommand{\arraystretch}{1.1}\begin{array}{c}
q_{\mathbb{Z}_2}\neq 0: \boldsymbol{6n} \\
q_{\mathbb{Z}_2}= 0: \boldsymbol{\frac{9n-7}{2}}\\
\end{array} &  \frac{7(N-1)}{2}\\
 \cline{2-6}
 &\makecell{ \begin{array}{c}
 N=4n \\
 n \neq 3j \\
 \end{array}: \includegraphics[scale=0.2]{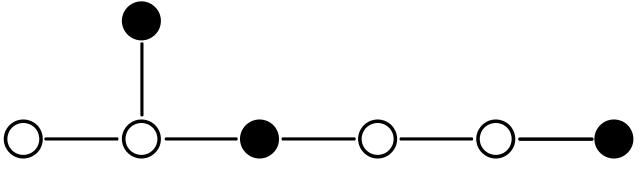}\\}& \renewcommand{\arraystretch}{1.1}\begin{array}{c} A_2 \oplus A_2 \oplus\\
 \langle \alpha_3^*,\alpha_6^*,\alpha_7^*\rangle \\
 \end{array}& U(1)_a \times U(1)_b \times U(1)_c & \renewcommand{\arraystretch}{1.1}\begin{array}{c}
(q_a,q_b,q_c) = (0,2,0): \boldsymbol{n} \\
(q_a,q_b,q_c) = (0,0,2): \boldsymbol{n} \\
(q_a,q_b,q_c) = (0,1,1): \boldsymbol{6n} \\
(q_a,q_b,q_c) = (1,1,0): \boldsymbol{2n} \\
(q_a,q_b,q_c) = (1,0,1): \boldsymbol{2n} \\
(q_a,q_b,q_c) = (0,0,0): \boldsymbol{2(n-1)} \\
\end{array} &  \frac{7N}{2}-2 \\
  \cline{2-6}
  &\makecell{ \begin{array}{c}
 N=2n \\
 n \neq 2j,3j \\
 \end{array}: \includegraphics[scale=0.2]{D7res1.png}\\}& D_6\oplus \langle\alpha_6^*\rangle & U(1) & \renewcommand{\arraystretch}{1.1}\begin{array}{c}
q = 1: \boldsymbol{5n-3} \\
q = 0: \boldsymbol{2n} \\
\end{array} & \frac{7N}{2}-3\\
 \cline{2-6}
   &\makecell{N \neq 2n,3n:\includegraphics[scale=0.2]{D7nores.png}\\}& D_7 & \emptyset & \frac{7(N-1)}{2} & \frac{7(N-1)}{2} \\
 \hline
 \end{array}
  }
  $
  \caption{Higgs branch data for quasi-homogeneous cDV singularities of $(A_{N-1},D_4)$ and $(A_{N-1},D_7)$ type.}
  \label{table modes AD}
\end{table}

\begin{table}[t]
\centering
$
\scalemath{0.7}{
\renewcommand{\arraystretch}{2.5}
\begin{array}{|L||V|L|L|A|B|}
\Xhline{4\arrayrulewidth}
\large\textbf{Singularity} & \large\textbf{Resolution pattern} & \boldsymbol{\mathcal{M}} & \makecell{\large\textbf{Symmetry}\\\large\textbf{group}\\ } & \large\textbf{Hypers} &  \makecell{\large\textbf{Total}\\\large\textbf{hypers}} \\
\hline
\hline
  \multirow{3}{*}{$D_4^{(4)}[N]$}& \makecell{N=4n: \includegraphics[scale=0.065]{D4completeres.png}\\} &\mathfrak{t} & U(1)^4 & \renewcommand{\arraystretch}{1.1}\begin{array}{c}
  12n \\
 \text{\hspace{-0.15cm}Charges: \textbf{root system} of } D_4\end{array} &3N \\
 \cline{2-6}
 & \makecell{ 
 N=2(2n-1):\includegraphics[scale=0.065]{D4nores.png}\\}& A_1^{\oplus 4}& \mathbb{Z}_2 &\renewcommand{\arraystretch}{1.1}\begin{array}{c}
q_{\mathbb{Z}_2}\neq 0: \boldsymbol{4(2n-1)} \\
 q_{\mathbb{Z}_2}=0: \boldsymbol{4(n-1)}
 \end{array} & 3N-2\\
 \cline{2-6}
 & \makecell{N \neq 4n,4n-2:\includegraphics[scale=0.065]{D4nores.png}\\}& D_4  & \emptyset  & 3N-2 & 3N-2 \\ 
\hline
\hline
 \multirow{3}{*}{$ D_5^{(5)}[N]$}   &\makecell{N=5n: \includegraphics[scale=0.12]{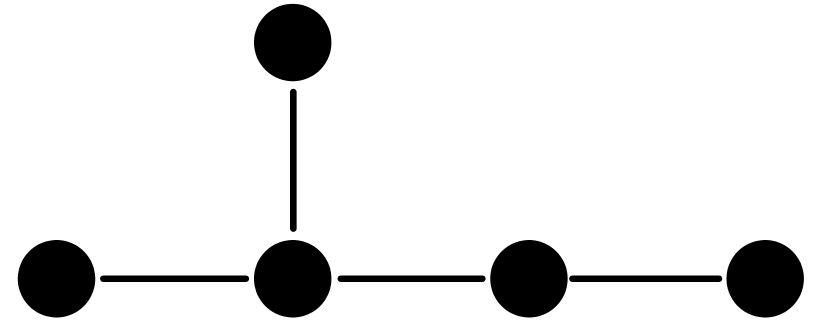} \\}& \mathfrak{t}& U(1)^5 & \makecell{20n \\
 \text{Charges: \textbf{root system} of } D_5} &4N \\
 \cline{2-6}
&\makecell{N \neq 5n:\includegraphics[scale=0.12]{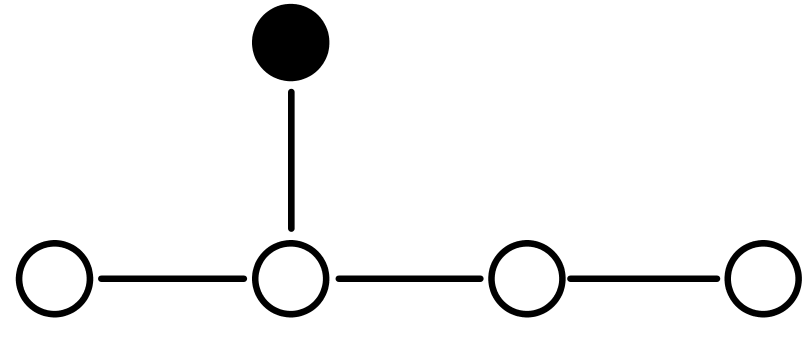} \\}& A_4\oplus \langle\alpha_5^* \rangle  & U(1) &
\renewcommand{\arraystretch}{1.1}\begin{array}{c}
q=1: \boldsymbol{2N} \\
q = 0: \boldsymbol{2(N-1)} \\
\end{array} & 2(2N-1)\\
 \hline
\hline
 \multirow{3}{*}{$ D_6^{(6)}[N]$}   &\makecell{N=6n: \includegraphics[scale=0.12]{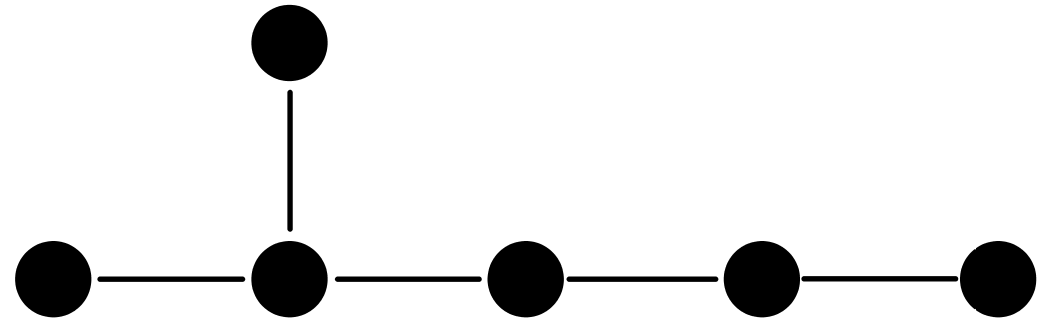} \\} &\mathfrak{t} & U(1)^6 & \makecell{30n \\
 \text{Charges: \textbf{root system} of } D_6} &5N \\
 \cline{2-6}
&\makecell{ \begin{array}{c}
 N=2n \\
 n\neq 3j \\
 \end{array}:\includegraphics[scale=0.12]{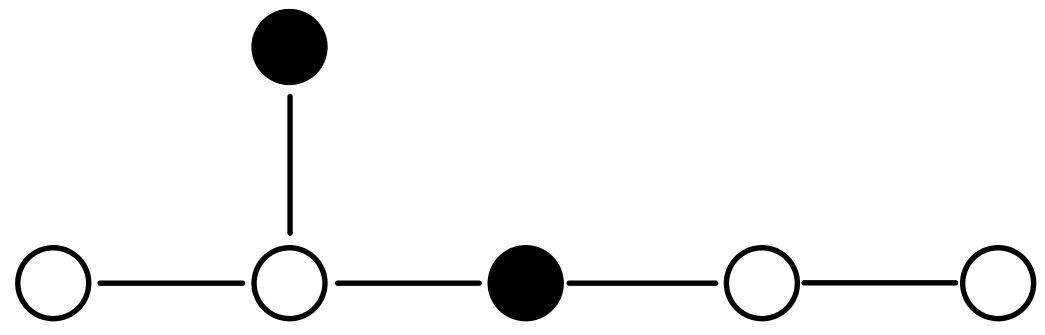} \\} & \renewcommand{\arraystretch}{1.1}\begin{array}{c} A_2 \oplus A_2 \oplus\\
 \langle\alpha_3^* ,\alpha_6^*\rangle\\
 \end{array} & U(1)_a \times U(1)_b &
\renewcommand{\arraystretch}{1.1}\begin{array}{c}
(q_a,q_b)=(2,0): \boldsymbol{n} \\
(q_a,q_b)=(0,2): \boldsymbol{n} \\
(q_a,q_b)=(1,1): \boldsymbol{6n}\\
(q_a,q_b) = (0,0): \boldsymbol{2(n-1)} \\
\end{array} & 5N-2\\
 \cline{2-6}
 &\makecell{N=6n-3:\includegraphics[scale=0.12]{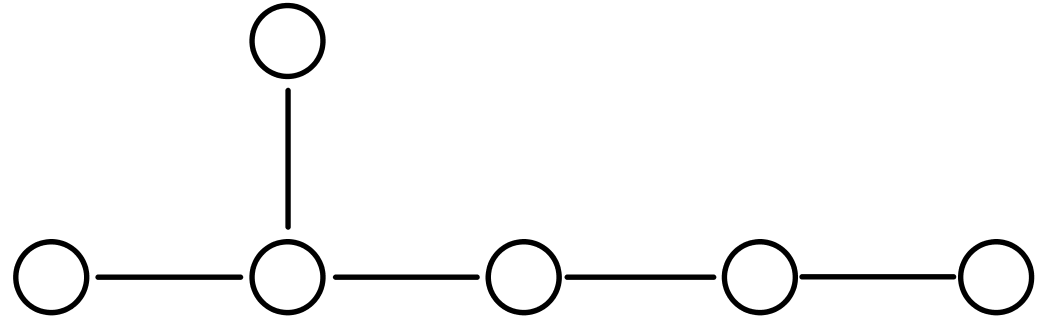} \\}& A_1^{\oplus 6} & \mathbb{Z}_2^2 &
\renewcommand{\arraystretch}{1.1}\begin{array}{c}
q_{\mathbb{Z}_2}\neq 0: \boldsymbol{12(2n-1)} \\
q_{\mathbb{Z}_2}= 0: \boldsymbol{6(n-1)}\\
\end{array} & 5N-3\\
 \cline{2-6}
 &\makecell{N \neq 2n, 6n-3:\includegraphics[scale=0.12]{D6nores.png}\\}& D_6 &\emptyset & 5N-3 & 5N-3 \\
 \hline
 \end{array}
  }
  $
  \caption{Higgs branch data for quasi-homogeneous cDV singularities of $D_4^{(4)}[N],D_5^{(5)}[N],D_6^{(6)}[N]$ type.}
  \label{table modes D}
\end{table}

\restoregeometry

\subsection{Quasi-homogeneous cDV singularities of $E_6,E_7,E_8$ type}
In this section, we focus on the deformed $E_6,E_7,E_8$ cases, looking for the minimal subalgebras containing the Higgs backgrounds reproducing a given quasi-homogeneous cDV singularity of $E_6,E_7,E_8$ type. 

\begin{figure}[H]
     \centering
     \includegraphics[scale=0.08]{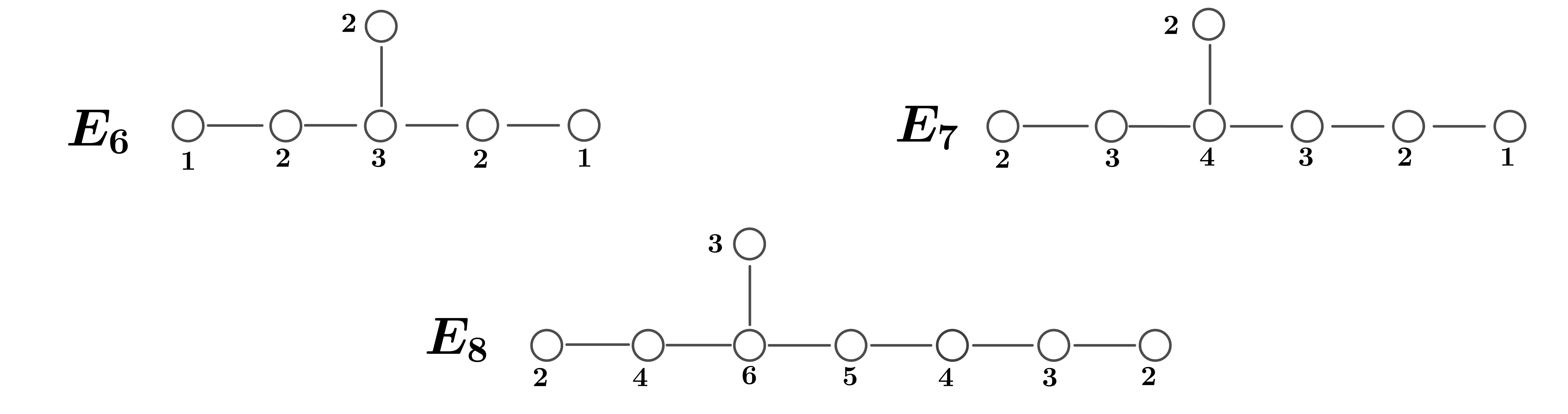}
     \caption{Dual Coxeter labels for the $E$ series.}
     \label{Coxeter E}
 \end{figure}

As they are useful to identify the flavor charges, we report the dual Coxeter labels for the $E_6,E_7,E_8$ Dynkin diagrams in Figure~\ref{Coxeter E}.

To illustrate how we get our results, we explicitly go through the $(A_{N-1},E_6)$ and the $E_7^{(14)}[N]$ cases. We sum up the results for all the cases in Tables \ref{table modes 1}, \ref{table modes 2}, \ref{table modes 3}, \ref{table discrete 1}, \ref{table discrete 2}, \ref{table discrete 3}.

\subsubsection*{$\boldsymbol{(A,E_6)}$ singularities}
\indent Let us start by showing how this works in the $(A_{N-1},E_6)$ class, employing the techniques of Section~\ref{section higgs vev}. The $(A_{N-1},E_6)$ threefolds are expressed as:
\begin{equation}
    \underbrace{x^2+y^3+z^4}_{E_6 \text{ sing}}+\underbrace{w^{N}}_{\text{def}}=0.
\end{equation}
Notice that the only non-vanishing deformation parameter is:
\begin{equation}\label{mu12E6}
    \mu_{12}(w) = w^{N}.
\end{equation}
The other (vanishing) deformation parameters are $\mu_{2},\mu_{5},\mu_{6},\mu_{8},\mu_{9}$.
Eq.\ \eqref{mu12E6} tells us that $M=12$, according to the notation of Section \ref{section higgs vev}. There are six divisors of 12:
\begin{equation}
    q_{\alpha} \in \{1,2,3,4,6,12\}.
\end{equation}
Now, we must look for the minimal degrees that the candidate partial Casimirs can acquire, thus forecasting the minimal subalgebra in which $\Phi$ can be contained. As $M=12$, the minimal subalgebras will recur with periodicity 12, namely the minimal subalgebra corresponding to the Higgs describing the $(A_k,E_6)$ singularity coincides with the one of $(A_{k+12},E_6)$. Let us proceed case by case:

\begin{description}
\item {\bf For $N=5,7,10,11$ mod $12$} ($q_{\alpha}= 12$), the minimal degree is $j=12$. This means that $\mathcal{M} = E_6$, with all Casimirs equal to zero, except the maximal degree one.  This implies that no resolution is possible.

\item {\bf For $N=2$ mod $12$} ($q_{\alpha}= 6$), the candidate minimal degree is $j=6$. This tells us that the only $c$'s that can be non-vanishing are $c^I_{6}$ and $c^I_{12}$, according to the notation of Section~\ref{section higgs vev}. 
To solve the system \eqref{sysLinEqmu} where only  $\mu_6,\mu_{12}$ appear, we need at least two Casimirs of degree 6, but this is not possible because of the rank of $E_6$\footnote{For example, one could have two degree 6 Casimirs using $\mathcal{M} = A_5 \oplus A_5$, or $\mathcal{M} = D_6$, but these cannot be embedded into $E_6$.}. This implies that no resolution is possible, and that the correct minimal subalgebra is $\mathcal{M} = E_6$ with all Casimirs equal to zero, except the maximal degree one.

\item {\bf For $N=3,9$ mod $12$} ($q_{\alpha}= 4$), the minimal degree for the non-vanishing partial Casimirs is $j=4$. To solve system \eqref{sysLinEqmu}, we have to set two parameters ($\mu_8$ and $\mu_{12}$), and thus we need at least two partial Casimirs of degree 4. They are provided by $\mathcal{M} = D_4$. This implies that the two external nodes of the $E_6$ Dynkin diagram get inflated, as can be seen in Figure~\ref{fig:D4bbranching}.
\begin{figure}[H]
    \centering
    \includegraphics[scale=0.25]{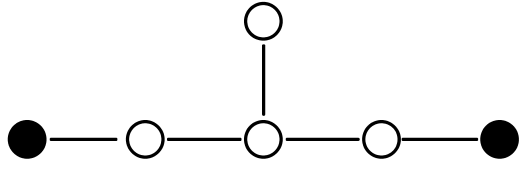}
    \caption{$D_4$ subalgebra in the $N=3,9$ case.}
    \label{fig:D4bbranching}
\end{figure}
This yields 5d hypers with charge 1 under the flavor groups corresponding to the resolved nodes, as they have dual Coxeter label equal to 1, as well as uncharged hypers.

\item {\bf For $N=4,8$ mod $12$} ($q_{\alpha}= 3$), the minimal degree for the non-zero partial Casimirs is $j=3$. System \eqref{sysLinEqmu} tells us that we need at least three Casimirs of degree 3 to extract a solution and fix the deformation parameters $\mu_6,\mu_9,\mu_{12}$. Indeed, the subalgebra $\mathcal{M} = A_2 \oplus A_2 \oplus A_2$ gives us the correct partial Casimirs. This choice produces no simultaneous resolution of the deformed family. Furthermore, the fact that $\Phi \in \mathcal{M} =A_2 \oplus A_2 \oplus A_2$ signals that in this case we have a non trivial $\text{Stab}_{\mathcal G}(\Phi) = \mathbb Z_3$, that reflects in a discrete-gauging of the hypermultiplets of the five-dimensional SCFT.
The actual discrete group $\mathbb{Z}_3$ comes because the maximal subalgebra $A_2^{\oplus 3}$ of $E_6$ is obtained removing the trivalent node from the extended Dynkin diagram of $E_6$, that has dual Coxeter number equal to $3$ (see Section~\ref{sec:symmetry-group}), as depicted in Figure \ref{fig:E6branchingA3}.

\begin{figure}[H]
    \centering
    \includegraphics[scale=0.14]{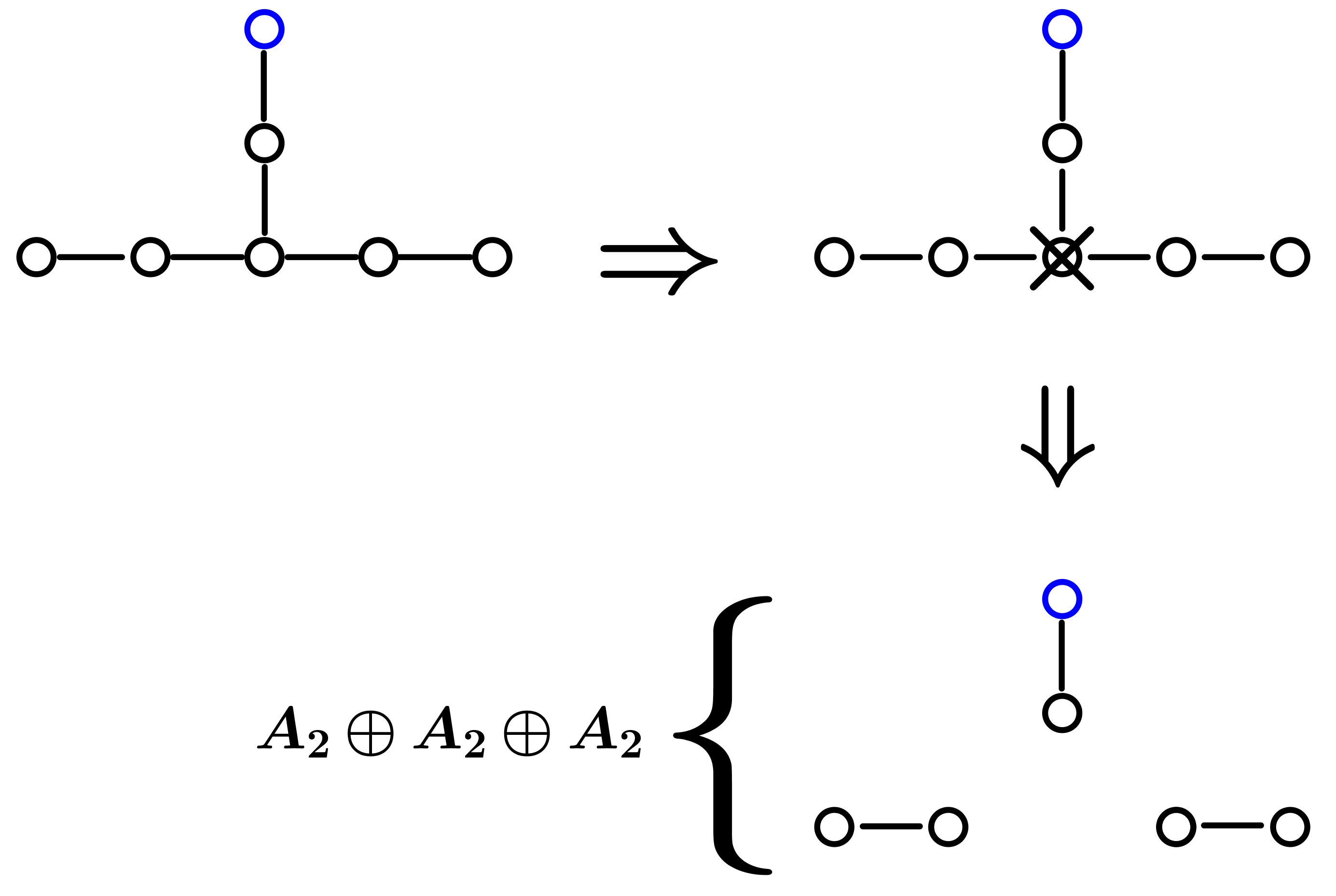}
    \caption{$A_2^{\oplus 3}$ subalgebra in the $N=4,8$ case.}
    \label{fig:E6branchingA3}
\end{figure}

\item {\bf For $N=6$ mod $12$} ($q_{\alpha}= 2$), the minimal degree for the non-zero partial Casimirs is $j=2$. According to the system \eqref{sysLinEqmu}, we have to set the $\mu_2,\mu_6,\mu_8$ parameters to zero, as well as $\mu_{12} = w^6$. This requires four partial Casimirs of minimal degree 2. It turns out that there exists a unique subalgebra of $E_6$ doing the work, i.e. $A_1\oplus A_1\oplus A_1\oplus A_1$. We then have $\mathcal{M}=A_1^{\oplus 4}\oplus\mathcal{H}$, with $\mathcal{H}$ generated by the two external nodes in the Dynkin diagram of $E_6$. The Higgs field take values in the semi-simple part of $\mathcal{M}$. This choice yields the resolution of the two external nodes with Coxeter label 1 of the $E_6$ Dynkin diagram, and produces a $\mathbb{Z}_2$ discrete group in 5d (since $\mathcal{L}=D_4$ and $A_1^{\oplus 4}$ is its maximal subalgebra, see Section~\ref{sec:symmetry-group}), as depicted in Figure~\ref{fig:E6branching}.
\begin{figure}[H]
    \centering
    \includegraphics[scale=0.14]{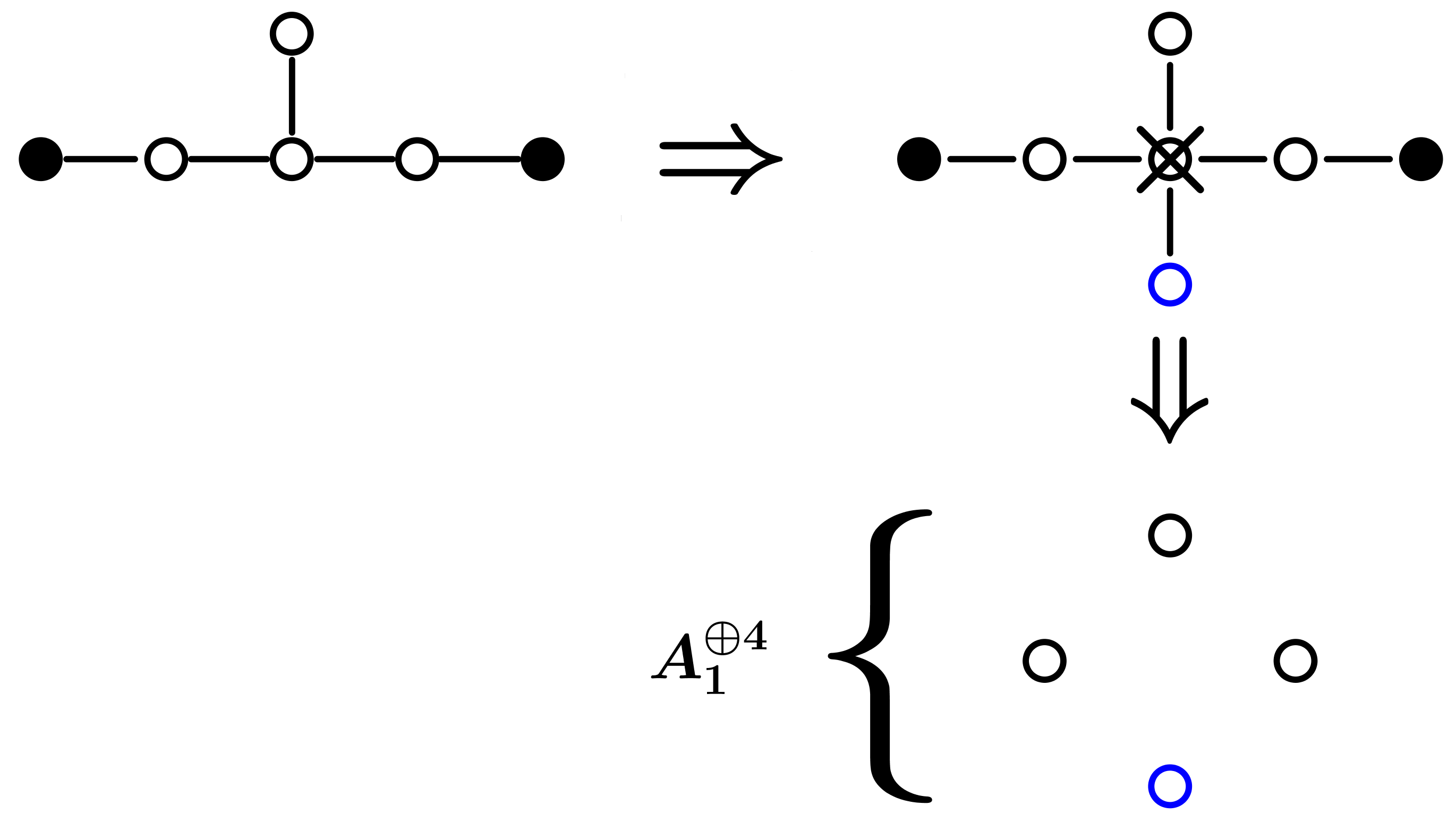}
    \caption{$A_1^{\oplus 4}$ subalgebra in the $N=6$ case.}
    \label{fig:E6branching}
\end{figure}

\item {\bf For $N=12$ mod $12$} ($q_{\alpha}= 1$), the minimal degree for the non-zero partial Casimirs is $j=1$. Then $\mathcal{M}$ is the Cartan subalgebra of $E_6$. As a result, all the simple roots of $E_6$ are blown up in the simultaneous resolution. The flavor charges of the 5d hypermultiplets are given, in some basis, by the root system of the $E_6$ algebra\footnote{In general, for all the completely resolvable cases of Tables \eqref{table modes 1} and \eqref{table modes 2}, the flavor charges are given by the roots of the corresponding algebra.}.

\end{description}

\subsubsection*{$\boldsymbol{E_7^{(14)}[N]}$ singularities}
The $E_7^{(14)}[N]$ singularities are expressed as deformed Du Val $E_7$ singularities:
\begin{equation}
    x^2+y^3+y z^3+z w^N=0\:.
\end{equation}
Notice that the only non-zero deformation parameter is
\begin{equation}\label{E7mu14}
    \mu_{14}(w) = w^N.
\end{equation}
The other (vanishing) deformation parameters are $\mu_{2},\mu_{6},\mu_{8},\mu_{10},\mu_{12},\mu_{18}$.
From \eqref{E7mu14}, we read $M=14$. Its divisors are:
\begin{equation}
  q_{\alpha} \in  \{1,2,7,14\}.
\end{equation}
With this in hand, we can start looking for the minimal degrees of candidate partial Casimirs, pinpointing the minimal subalgebra of $E_7$ containing $\Phi$ for a given $E_7^{(14)}[N]$. As in the previous section, we expect that the subalgebra corresponding to $E_7^{(14)}[N]$ is equal to the one of $E_7^{(14)}[N+14]$, given the degree 14 deformation parameter that is switched on.

\begin{description}
\item {\bf For $N=1,3,5,9,11,13$ mod $14$} ($q_{\alpha}= 14$), the minimal degree is $j=14$. Consequently, $\mathcal{M} = E_7$, with all Casimirs equal to zero except the maximal degree one.  This entails that no resolution is possible.

\item {\bf For $N=2,4,6,8,10,12$ mod $14$} ($q_{\alpha}= 7$), the minimal degree for the partial Casimirs is 7. In order to solve system \eqref{sysLinEqmu}, namely to fix $\mu_{14}$, we need only one partial Casimir. A degree 7 partial Casimir can be provided choosing $\mathcal{M} = A_6\oplus\langle\alpha_7^*\rangle$, which naturally lies inside $E_7$. This implies that a single node of $E_7$, with Coxeter label 2, gets inflated by the allowed resolution (see Figure~\ref{fig:E7A6branching}). This yields 5d hypers with charge 1 and 2, as well as uncharged hypers. The Higgs field $\Phi$ lives only in the semi-simple part of $\mathcal{M}$. See Figure~\ref{fig:E7A6branching}.

\begin{figure}[H]
    \centering
    \includegraphics[scale=0.25]{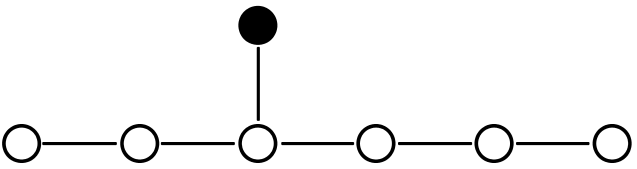}
    \caption{$A_6$ subalgebra in the $N=2,4,6,8,10,12$ case.}
    \label{fig:E7A6branching}
\end{figure}

\item {\bf For $N=7$ mod $14$} ($q_{\alpha}= 2$), the minimal degree for the partial Casimirs is 2. According to \eqref{sysLinEqmu}, we need seven distinct such Casimirs. It can be shown that indeed there exists a choice $\mathcal{M} =  A_1^{\oplus 7} \in E_7$, that yields seven partial Casimirs of degree 2.
This maximal subalgebra can be found noticing the chain of maximal subalgebras $E_7\supset A_1\oplus D_6 \supset A_1\oplus A_1^{\oplus 2} \oplus D_4 \supset A_1^7$, that is depicted in Figure~\ref{fig:E7xiebranching}.
\begin{figure}[H]
    \centering
    \includegraphics[scale=0.12]{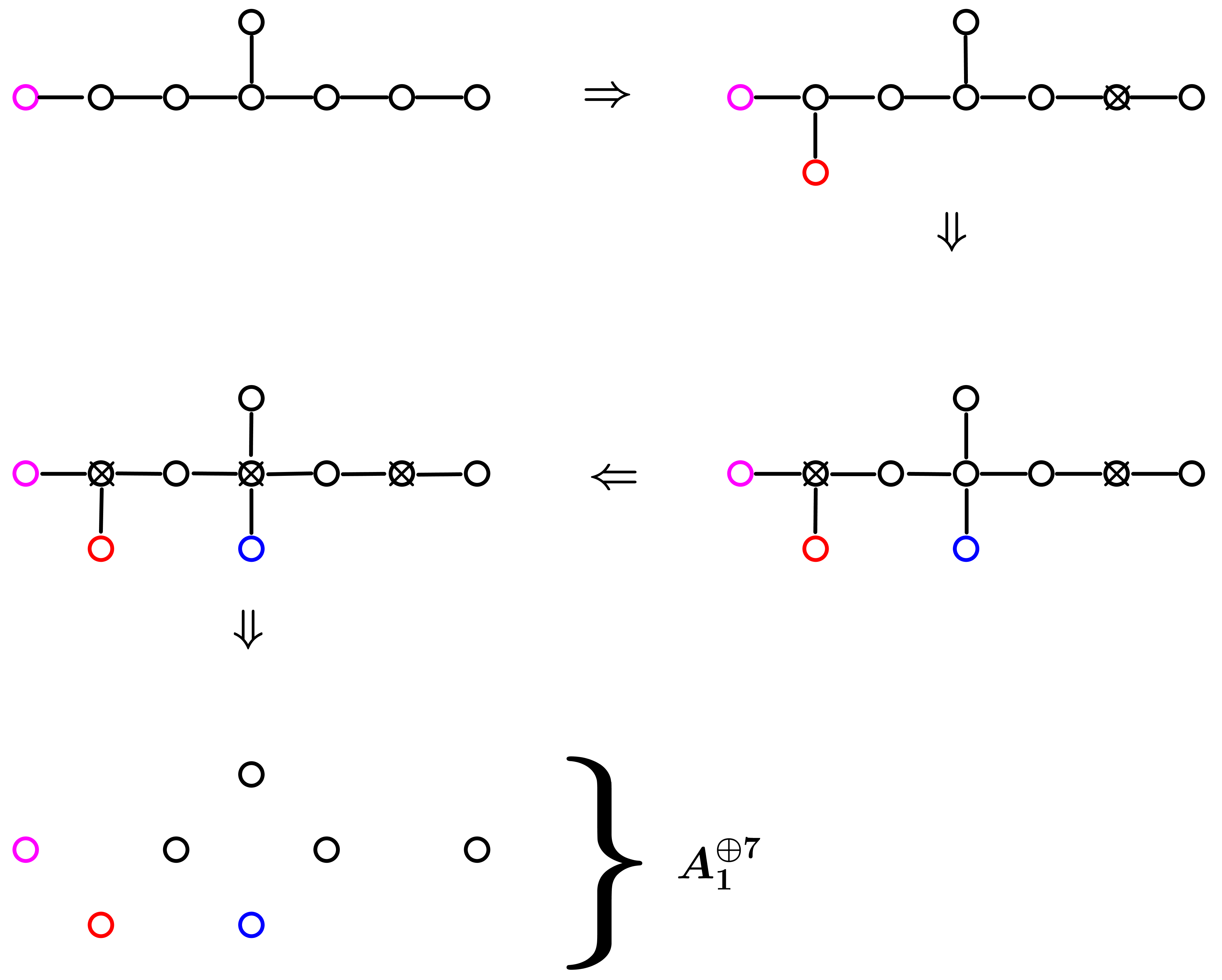}
    \caption{Maximal subalgebra in the $N=7$ case.}
    \label{fig:E7xiebranching}
\end{figure}
The three steps in obtaining the maximal subalgebra $A_1^{\oplus 7}$ of $E_7$, where nodes with Coxeter number equal to two are removed, tells us that  we have the non trivial discrete $\text{Stab}_{\mathcal G}(\Phi) = \mathbb Z_2^3$.

\item {\bf For $N=14$ mod $14$} ($q_{\alpha}= 1$), the minimal degree for the non-vanishing partial Casimirs is $j=1$. We need at least seven partial Casimirs to fix all the deformation parameters, and hence we can pick as partial Casimirs the Casimirs of the Cartan subalgebra of $E_7$. In this way, we see that all the simple roots of $E_7$ are blown-up in the simultaneous resolution. The flavor charges of the 5d hypers can be written as the root system of $E_7$.

\end{description}

\subsubsection*{Other quasi-homogeneous cDV singularities of type $E$}
Proceeding along the same path as the previous sections, we can readily find the minimal subalgebras containing the appropriate Higgs background $\Phi$ for each class of quasi-homogeneous cDV singularities arising from deformed $E_6,E_7,E_8$ singularities.\\
\indent We sum up our results in Table~\ref{table modes 1}, \ref{table modes 2}, \ref{table modes 3}, \ref{table discrete 1}, \ref{table discrete 2} and \ref{table discrete 3}. In particular, we list:
\begin{itemize}
    \item In the first column, the cDV singularity.
    \item In the second column, the maximal allowed simultaneous resolution (resolved nodes are in black). This fixes the  Levi subalgebra.
    \item In the third column, the minimal subalgebra $\mathcal{M} \subseteq \mathcal{L}$ containing $\Phi$. If it is non-trivial, this yields a discrete group in 5d.
    \item In the fourth column, the symmetry group preserved by $\Phi$. In general, it comprises both a continuous and a discrete factor.
    \item In the fifth column, the number of five-dimensional hypers localized in 5d, and their charges under the continuous and discrete symmetries. We also report the total number of hypers, to be compared with the number of normalizable complex structure deformations of the corresponding cDV singularity.
\end{itemize}


\newpage
\newgeometry{top=50pt,bottom=60pt,left=30pt,right=30pt}
\begin{table}[H]
\centering
$
\renewcommand{\arraystretch}{1.7}
\scalemath{0.7}{
\begin{array}{|C||P|L|U|D|}
\Xhline{4\arrayrulewidth}
\makecell{\large\textbf{Singularity}\\} & \large\textbf{Resolution pattern} & \large\makecell{\boldsymbol{\mathcal{M}}} & \large\textbf{Symmetry group} & 
\large\textbf{Hypers}\\
\hline
\hline
\textbf{Deformed }\boldsymbol{E_6} & \multicolumn{4}{c|}{} \\
\hline
\multirow{5}{*}{$(A_{N-1},E_6)$}& \makecell{N=12n:\includegraphics[scale=0.2]{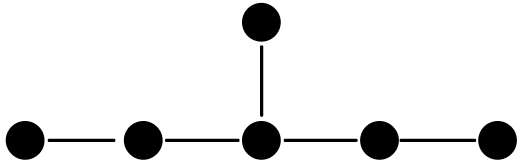}} & \mathfrak{t} & U(1)^6 &\makecell{\text{Charges: \textbf{root system} of } E_6  \\
\textbf{TOT: }\boldsymbol{3N } } \\ 
\cline{2-5}
 & \makecell{\\ \begin{array}{c}
 N=6n \\
 n\neq 2j \\
 \end{array}: \includegraphics[scale=0.2]{E6resolutionv1.png}\\} & A_1^{\oplus 4}\oplus \langle \alpha_1^*,\alpha_5^*\rangle&  U(1)_a\times U(1)_b\times \mathbb{Z}_2 & \renewcommand{\arraystretch}{1.5}\begin{array}{c}
 (q_a,q_b,q_{\mathbb{Z_2}}) =(1,0,1): 2n\\
  (q_a,q_b,q_{\mathbb{Z_2}}) =(1,0,0): 2n\\
  (q_a,q_b,q_{\mathbb{Z_2}}) =(0,1,1): 2n \\
    (q_a,q_b,q_{\mathbb{Z_2}}) =(0,1,0): 2n \\
 (q_a,q_b,q_{\mathbb{Z_2}}) =(1,1,1): 2n \\
  (q_a,q_b,q_{\mathbb{Z_2}}) =(1,1,0): 2n\\(q_a,q_b,q_{\mathbb{Z_2}}) =(0,0,1): 6n-2\\
  \textbf{TOT: }\boldsymbol{3N-2} \\
 \end{array} \\
\cline{2-5}
& \makecell{\\ \begin{array}{c}
 N=3n \\
 n\neq 2j \\
 \end{array}: \includegraphics[scale=0.2]{E6resolutionv1.png}} & D_4\oplus \langle \alpha_1^*,\alpha_5^*\rangle & U(1)_a\times U(1)_b &\renewcommand{\arraystretch}{1.5}\begin{array}{c}
 (q_a,q_b) =(1,0): 2n\\ (q_a,q_b) =(0,1): 2n \\
 (q_a,q_b) =(1,1): 2n \\ (q_a,q_b) =(0,0):3n-2\\
 \textbf{TOT: }\boldsymbol{3N-2 }
 \end{array}  \\
\cline{2-5}
& \makecell{\\ \begin{array}{c}
 N=4n \\
 n\neq 3j \\
 \end{array}: \includegraphics[scale=0.2]{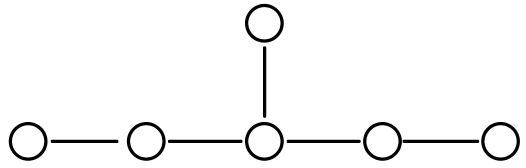}\\} & A_2^{\oplus 3}  & \mathbb{Z}_3 &\renewcommand{\arraystretch}{1.5}\begin{array}{c} q_{\mathbb{Z}_3} = 1: 9n \\
 q_{\mathbb{Z}_3} = 0: 3(n-1) \\
 \textbf{TOT: }\boldsymbol{3(N-1) }
 \end{array}\\
\cline{2-5}
 & \makecell{N \neq 3n,4n:\includegraphics[scale=0.2]{E6nores.png}\\} & E_6 & \emptyset & \textbf{TOT: }\boldsymbol{3(N-1) }\\
\hline
\hline
\end{array}
}
$
\caption{Higgs branch data for quasi-homogeneous cDV singularities of $(A_{N-1},E_6)$ type.}
\label{table modes 1}
\end{table}

\newpage
\begin{table}[H]
\centering
$
\renewcommand{\arraystretch}{1.7}
\scalemath{0.7}{
\begin{array}{|C||P|L|U|D|}
\Xhline{4\arrayrulewidth}
\makecell{\large\textbf{Singularity}\\} & \large\textbf{Resolution pattern} & \large\makecell{\boldsymbol{\mathcal{M}}} & \large\textbf{Symmetry group} &
\large\textbf{Hypers}\\
\hline
\hline
\multirow{4}{*}{$E_6^{(8)}[N]$} & \makecell{N=8n: \includegraphics[scale=0.2]{E6completeres.png}\\} & \mathfrak{t} & U(1)^6  &\makecell{\text{Charges: \textbf{root system} of } E_6  \\
\textbf{TOT: }\boldsymbol{\frac{9N}{2} } } \\ 
\cline{2-5}
 &  \makecell{\\ \begin{array}{c}
 N=4n \\
 n\neq 2j \\
 \end{array}:\includegraphics[scale=0.2]{E6resolutionv1.png} \\} & A_1^{\oplus 4}\oplus \langle \alpha_1^*,\alpha_5^*\rangle& U(1)_a\times U(1)_b\times \mathbb{Z}_2 &\renewcommand{\arraystretch}{1.5}\begin{array}{c}
 (q_a,q_b,q_{\mathbb{Z_2}}) =(1,0,1): 2n\\
  (q_a,q_b,q_{\mathbb{Z_2}}) =(1,0,0): 2n\\
  (q_a,q_b,q_{\mathbb{Z_2}}) =(0,1,1): 2n \\
    (q_a,q_b,q_{\mathbb{Z_2}}) =(0,1,0): 2n\\
 (q_a,q_b,q_{\mathbb{Z_2}}) =(1,1,1): 2n\\
  (q_a,q_b,q_{\mathbb{Z_2}}) =(1,1,0): 2n \\(q_a,q_b,q_{\mathbb{Z_2}}) =(0,0,1): 6n-2\\
   \textbf{TOT: }\boldsymbol{\frac{9N}{2}-2 }\\
 \end{array} \\
\cline{2-5}
 &\makecell{\\ \begin{array}{c}
 N=2n \\
 n\neq 2j \\
 \end{array}:\includegraphics[scale=0.2]{E6resolutionv1.png} \\} & D_4\oplus \langle \alpha_1^*,\alpha_5^*\rangle & U(1)_a\times U(1)_b & \renewcommand{\arraystretch}{1.5}\begin{array}{c}
 (q_a,q_b) =(1,0): 2n\\ (q_a,q_b) =(0,1): 2n \\
 (q_a,q_b) =(1,1):2n \\ (q_a,q_b) =(0,0): 3n-2\\
 \textbf{TOT: }\boldsymbol{\frac{9N}{2}-2  }\\
 \end{array}\\
\cline{2-5}
& \makecell{
 N=2n+1:\includegraphics[scale=0.2]{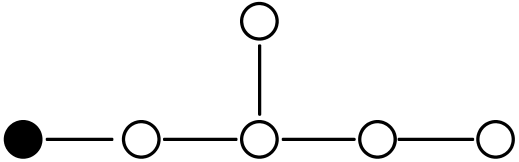} \\} & D_5\oplus \langle \alpha_1^*\rangle & U(1) &  \renewcommand{\arraystretch}{1.5}\begin{array}{c}
  q =1: 4n+2 \\
  q=0: 5n\\
   \textbf{TOT: }\boldsymbol{\frac{9N-5}{2} }\\
  \end{array}  \\
\hline
\hline
\multirow{3}{*}{$E_6^{(9)}[N]$} & \makecell{N=9n: \includegraphics[scale=0.2]{E6completeres.png}\\} & \mathfrak{t} & U(1)^6  & \makecell{\text{Charges: \textbf{root system} of } E_6  \\
\textbf{TOT: }\boldsymbol{4N } } \\ 
\cline{2-5}
&  \makecell{\\ \begin{array}{c}
 N=3n \\
 n\neq 3j \\
 \end{array}: \includegraphics[scale=0.2]{E6nores.png}\\} & A_2^{\oplus 3} & \mathbb{Z}_3 & \renewcommand{\arraystretch}{1.5}\begin{array}{c} q_{\mathbb{Z}_3} = 1: 9n \\
 q_{\mathbb{Z}_3} = 0:3(n-1) \\
  \textbf{TOT: }\boldsymbol{4N-3}\\
 \end{array}\\
\cline{2-5}
 & \makecell{N \neq 3n:\includegraphics[scale=0.2]{E6nores.png}\\} & E_6 & \emptyset &  \textbf{TOT: }\boldsymbol{4N-3 }\\
 \hline
\end{array}
}
$
\caption{Higgs branch data for quasi-homogeneous cDV singularities of $E_6^{(8)}[N]$ and $E_6^{(9)}[N]$ type.}
  \label{table modes 2}
\end{table}

\newpage
\begin{table}[H]
\centering
$
\renewcommand{\arraystretch}{1.7}
\scalemath{0.7}{
\begin{array}{|C||P|L|U|D|}
\Xhline{4\arrayrulewidth}
\makecell{\large\textbf{Singularity}\\} & \large\textbf{Resolution pattern} & \large\makecell{\boldsymbol{\mathcal{M}}} & \large\textbf{Symmetry group} &
\large\textbf{Hypers}\\
\hline
\hline
\textbf{Deformed }\boldsymbol{E_7} & \multicolumn{4}{c|}{} \\
\hline
\multirow{5}{*}{$(A_{N-1},E_7)$} & \makecell{N=18n: \includegraphics[scale=0.2]{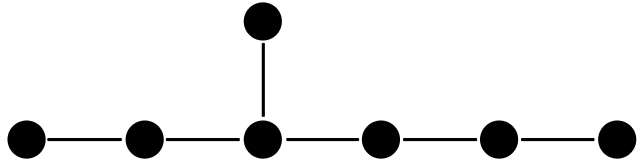}\\}& \mathfrak{t} & U(1)^7 &\makecell{\text{Charges: \textbf{root system} of } E_7  \\
\textbf{TOT: }\boldsymbol{\frac{7N}{2} } } \\ 
\cline{2-5}
& \makecell{\begin{array}{c}
 N=9n \\
 n\neq 2j \\
 \end{array}: \includegraphics[scale=0.2]{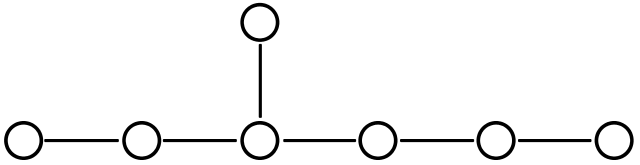}\\}&A_1^{\oplus 7} & \mathbb{Z}_2^3 &
 \renewcommand{\arraystretch}{1.5}\begin{array}{c}
  (q_{\mathbb{Z}_2}^{(a)},q_{\mathbb{Z}_2}^{(b)},q_{\mathbb{Z}_2}^{(c)})=(0,0,0): \frac{7}{2}(n-1) \\
 (q_{\mathbb{Z}_2}^{(a)},q_{\mathbb{Z}_2}^{(b)},q_{\mathbb{Z}_2}^{(c)})=(1,0,0): 4n \\
 (q_{\mathbb{Z}_2}^{(a)},q_{\mathbb{Z}_2}^{(b)},q_{\mathbb{Z}_2}^{(c)})=(0,1,0): 4n \\
 (q_{\mathbb{Z}_2}^{(a)},q_{\mathbb{Z}_2}^{(b)},q_{\mathbb{Z}_2}^{(c)})=(0,0,1): 4n  \\
 (q_{\mathbb{Z}_2}^{(a)},q_{\mathbb{Z}_2}^{(b)},q_{\mathbb{Z}_2}^{(c)})=(0,1,1): 4n  \\
 (q_{\mathbb{Z}_2}^{(a)},q_{\mathbb{Z}_2}^{(b)},q_{\mathbb{Z}_2}^{(c)})=(1,1,0):4n   \\
 (q_{\mathbb{Z}_2}^{(a)},q_{\mathbb{Z}_2}^{(b)},q_{\mathbb{Z}_2}^{(c)})=(1,0,1):4n   \\
 (q_{\mathbb{Z}_2}^{(a)},q_{\mathbb{Z}_2}^{(b)},q_{\mathbb{Z}_2}^{(c)})=(1,1,1):4n \\
 \textbf{TOT: }\boldsymbol{\frac{7(N-1)}{2}}\\
 \end{array}\\
\cline{2-5}
&\makecell{\begin{array}{c}
 N=6n \\
 n\neq 3j \\
 \end{array}\includegraphics[scale=0.2]{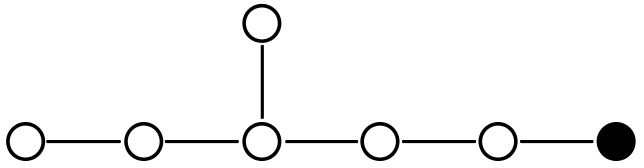} \\} & A_2^{\oplus 3}\oplus\langle \alpha_6^*\rangle & U(1) \times \mathbb{Z}_3  & \renewcommand{\arraystretch}{1.1}\begin{array}{c}
(q,q_{\mathbb{Z}_3})=(1,0): 3n \\
(q,q_{\mathbb{Z}_3})=(1,1): 3n \\
(q,q_{\mathbb{Z}_3})=(1,2): 3n \\
(q,q_{\mathbb{Z}_3})=(0,0): 3(n-1) \\
(q,q_{\mathbb{Z}_3})=(0,1):9n \\
  \textbf{TOT: }\boldsymbol{\frac{7N}{2}-3}\\
 \end{array}\\
\cline{2-5}
& \makecell{\begin{array}{c}
 N=2n+1 \\
 2n\neq 9j-1 \\
 \end{array}:\includegraphics[scale=0.2]{E7nores.png}\\} & E_7 & \emptyset & \textbf{TOT: }\boldsymbol{\frac{7(N-1)}{2}}\\
\cline{2-5}
&\makecell{\begin{array}{c}
 N=2n \\
 n\neq 3j \\
 \end{array}:\includegraphics[scale=0.2]{E7resolutionv1.png} \\} & E_6\oplus\langle \alpha_6^*\rangle &  U(1) & \renewcommand{\arraystretch}{1.1}\begin{array}{c}
q=1: 3n \\
 q=0: 4n-3 \\
 \textbf{TOT: }\boldsymbol{\frac{7N}{2}-3}\\
 \end{array}\\
\hline
\hline
\multirow{5}{*}{$E_7^{(14)}[N]$}& \makecell{N=14n: \includegraphics[scale=0.2]{E7completeres.png}\\} & \mathfrak{t} & U(1)^7 & \makecell{\text{Charges: \textbf{root system} of } E_7  \\
\textbf{TOT: }\boldsymbol{\frac{9N}{2} } } \\ 
\cline{2-5}
&  \makecell{\begin{array}{c}
 N=7n \\
 n\neq 2j \\
 \end{array}:\includegraphics[scale=0.2]{E7nores.png} \\}& A_1^{\oplus 7}& \mathbb{Z}_2^3 & \renewcommand{\arraystretch}{1.5}\begin{array}{c}
  (q_{\mathbb{Z}_2}^{(a)},q_{\mathbb{Z}_2}^{(b)},q_{\mathbb{Z}_2}^{(c)})=(0,0,0): \frac{7}{2}(n-1) \\
 (q_{\mathbb{Z}_2}^{(a)},q_{\mathbb{Z}_2}^{(b)},q_{\mathbb{Z}_2}^{(c)})=(1,0,0): 4n \\
 (q_{\mathbb{Z}_2}^{(a)},q_{\mathbb{Z}_2}^{(b)},q_{\mathbb{Z}_2}^{(c)})=(0,1,0): 4n \\
 (q_{\mathbb{Z}_2}^{(a)},q_{\mathbb{Z}_2}^{(b)},q_{\mathbb{Z}_2}^{(c)})=(0,0,1): 4n  \\
 (q_{\mathbb{Z}_2}^{(a)},q_{\mathbb{Z}_2}^{(b)},q_{\mathbb{Z}_2}^{(c)})=(0,1,1): 4n  \\
 (q_{\mathbb{Z}_2}^{(a)},q_{\mathbb{Z}_2}^{(b)},q_{\mathbb{Z}_2}^{(c)})=(1,1,0):4n   \\
 (q_{\mathbb{Z}_2}^{(a)},q_{\mathbb{Z}_2}^{(b)},q_{\mathbb{Z}_2}^{(c)})=(1,0,1):4n   \\
 (q_{\mathbb{Z}_2}^{(a)},q_{\mathbb{Z}_2}^{(b)},q_{\mathbb{Z}_2}^{(c)})=(1,1,1):4n \\
  \textbf{TOT: }\boldsymbol{\frac{9N-7}{2}}\\
 \end{array}\\
\cline{2-5}
& \makecell{\begin{array}{c}
 N=2n+1 \\
 2n\neq 7j-1 \\
 \end{array}: \includegraphics[scale=0.2]{E7nores.png}\\} &E_7& \emptyset & \textbf{TOT: }\boldsymbol{\frac{9N-7}{2}}\\
\cline{2-5}
& \makecell{\begin{array}{c}
 N=2n \\
 n\neq 7j \\
 \end{array}:\includegraphics[scale=0.2]{E7resolutionv2.png} \\} & A_6\oplus\langle \alpha_7^*\rangle & U(1) & \renewcommand{\arraystretch}{1.1}\begin{array}{c}
q=2: n\\
q=1: 5n\\
q = 0: 3(n-1) \\
\textbf{TOT: }\boldsymbol{\frac{9N}{2}-3}\\
\end{array}\\
\hline
\end{array}
}
$
  \caption{Higgs branch data for quasi-homogeneous cDV singularities of $(A_{N-1},E_7)$ and $E_7^{(14)}[N]$ type.}
  \label{table modes 3}
\end{table}

\begin{table}[H]
\centering
$
\scalemath{0.7}{
\renewcommand{\arraystretch}{1.7}
\begin{array}{|C||P|L|U|D|}
\Xhline{4\arrayrulewidth}
\makecell{\large\textbf{Singularity}\\} & \large\textbf{Resolution pattern} & \large\makecell{\boldsymbol{\mathcal{M}}} & \large\textbf{Symmetry group} &
\large\textbf{Hypers}\\
\hline
\hline
\textbf{Deformed }\boldsymbol{E_8} & \multicolumn{4}{c|}{} \\
\hline
\multirow{5}{*}{$(A_{N-1},E_8)$}
 &\makecell{N=30n: \includegraphics[scale=0.2]{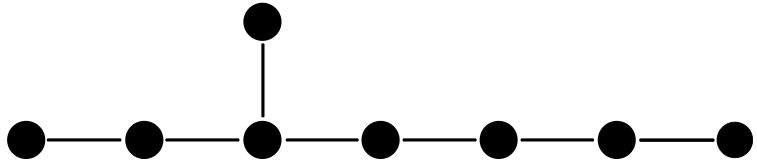} \\} & \mathfrak{t} & U(1)^8 &\makecell{\text{Charges: \textbf{root system} of } E_8  \\
\textbf{TOT: }\boldsymbol{4N } } \\ 
 \cline{2-5}
  & \makecell{\\ \begin{array}{c}
 N=6n \\
 n\neq 5j \\
 \end{array}: \includegraphics[scale=0.2]{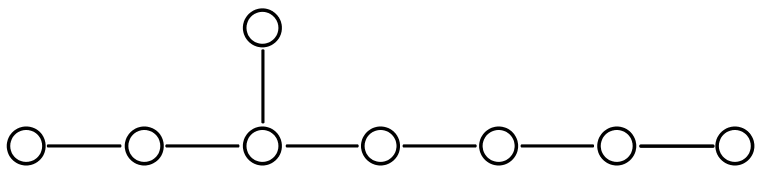} \\}& A_4 \oplus A_4  & \mathbb{Z}_5 & \renewcommand{\arraystretch}{1.1}\begin{array}{c}
 q_{\mathbb{Z}_5} = 2: 10n\\
 q_{\mathbb{Z}_5} =1: 10n\\
 q_{\mathbb{Z}_5} = 0: 4(n-1)\\
 \textbf{TOT: }\boldsymbol{4(N-1)} \\
 \end{array}\\
\cline{2-5}
 & \makecell{\\ \begin{array}{c}
 N=10n \\
 n\neq 3j \\
 \end{array}: \includegraphics[scale=0.2]{E8nores.png} \\} & A_2^{\oplus 4} & \mathbb{Z}_3^2 & \renewcommand{\arraystretch}{1.1}\begin{array}{c}
( q_{\mathbb{Z}_3}^{(a)}, q_{\mathbb{Z}_3}^{(b)}) = (1,0): 9n\\
( q_{\mathbb{Z}_3}^{(a)}, q_{\mathbb{Z}_3}^{(b)}) = (0,1): 9n\\
( q_{\mathbb{Z}_3}^{(a)}, q_{\mathbb{Z}_3}^{(b)}) = (1,1): 9n\\
( q_{\mathbb{Z}_3}^{(a)}, q_{\mathbb{Z}_3}^{(b)}) = (1,2): 9n\\
( q_{\mathbb{Z}_3}^{(a)}, q_{\mathbb{Z}_3}^{(b)}) = (0,0): 4(n-1)\\
 \textbf{TOT: }\boldsymbol{4(N-1)} \\
\end{array}\\
\cline{2-5}
 &\makecell{\\ \begin{array}{c}
 N=15n \\
 n\neq 2j \\
 \end{array}: \includegraphics[scale=0.2]{E8nores.png} \\} & A_1^{\oplus 8} & \mathbb{Z}_2^4 &
\renewcommand{\arraystretch}{1.1}\begin{array}{c}
(q_{\mathbb{Z}_2}^{(a)},q_{\mathbb{Z}_2}^{(b)},q_{\mathbb{Z}_2}^{(c)},q_{\mathbb{Z}_2}^{(d)})= (0,0,0,0): 4(n-1)\\
(q_{\mathbb{Z}_2}^{(a)},q_{\mathbb{Z}_2}^{(b)},q_{\mathbb{Z}_2}^{(c)},q_{\mathbb{Z}_2}^{(d)})= (1,0,0,0): 4n\\
(q_{\mathbb{Z}_2}^{(a)},q_{\mathbb{Z}_2}^{(b)},q_{\mathbb{Z}_2}^{(c)},q_{\mathbb{Z}_2}^{(d)})= (0,1,0,0): 4n\\
(q_{\mathbb{Z}_2}^{(a)},q_{\mathbb{Z}_2}^{(b)},q_{\mathbb{Z}_2}^{(c)},q_{\mathbb{Z}_2}^{(d)})= (0,0,1,0): 4n\\
(q_{\mathbb{Z}_2}^{(a)},q_{\mathbb{Z}_2}^{(b)},q_{\mathbb{Z}_2}^{(c)},q_{\mathbb{Z}_2}^{(d)})= (0,0,0,1): 4n\\
(q_{\mathbb{Z}_2}^{(a)},q_{\mathbb{Z}_2}^{(b)},q_{\mathbb{Z}_2}^{(c)},q_{\mathbb{Z}_2}^{(d)})= (1,1,0,0): 4n\\
(q_{\mathbb{Z}_2}^{(a)},q_{\mathbb{Z}_2}^{(b)},q_{\mathbb{Z}_2}^{(c)},q_{\mathbb{Z}_2}^{(d)})= (1,0,1,0): 4n\\
(q_{\mathbb{Z}_2}^{(a)},q_{\mathbb{Z}_2}^{(b)},q_{\mathbb{Z}_2}^{(c)},q_{\mathbb{Z}_2}^{(d)})= (1,0,0,1): 4n\\
(q_{\mathbb{Z}_2}^{(a)},q_{\mathbb{Z}_2}^{(b)},q_{\mathbb{Z}_2}^{(c)},q_{\mathbb{Z}_2}^{(d)})= (0,1,1,0): 4n\\
(q_{\mathbb{Z}_2}^{(a)},q_{\mathbb{Z}_2}^{(b)},q_{\mathbb{Z}_2}^{(c)},q_{\mathbb{Z}_2}^{(d)})= (0,1,0,1): 4n\\
(q_{\mathbb{Z}_2}^{(a)},q_{\mathbb{Z}_2}^{(b)},q_{\mathbb{Z}_2}^{(c)},q_{\mathbb{Z}_2}^{(d)})= (0,0,1,1): 4n\\
(q_{\mathbb{Z}_2}^{(a)},q_{\mathbb{Z}_2}^{(b)},q_{\mathbb{Z}_2}^{(c)},q_{\mathbb{Z}_2}^{(d)})= (1,1,1,0): 4n\\
(q_{\mathbb{Z}_2}^{(a)},q_{\mathbb{Z}_2}^{(b)},q_{\mathbb{Z}_2}^{(c)},q_{\mathbb{Z}_2}^{(d)})= (1,1,0,1): 4n\\
(q_{\mathbb{Z}_2}^{(a)},q_{\mathbb{Z}_2}^{(b)},q_{\mathbb{Z}_2}^{(c)},q_{\mathbb{Z}_2}^{(d)})= (1,0,1,1): 4n\\
(q_{\mathbb{Z}_2}^{(a)},q_{\mathbb{Z}_2}^{(b)},q_{\mathbb{Z}_2}^{(c)},q_{\mathbb{Z}_2}^{(d)})= (0,1,1,1): 4n\\
\textbf{TOT: }\boldsymbol{4(N-1)} \\
\end{array} \\
\cline{2-5}
 & \makecell{N \neq 6n,10n,15n:\includegraphics[scale=0.2]{E8nores.png} \\}& E_8 & \emptyset &  \textbf{TOT: }\boldsymbol{4(N-1)}\\
\hline
\hline
\end{array}
}
$
\caption{Higgs branch data for quasi-homogeneous cDV singularities of $(A_{N-1},E_8)$ type.}
\label{table discrete 1}
\end{table}

\newpage
\begin{table}[H]
\centering
$
\scalemath{0.7}{
\renewcommand{\arraystretch}{1.7}
\begin{array}{|C||P|L|U|D|}
\Xhline{4\arrayrulewidth}
\makecell{\large\textbf{Singularity}\\} & \large\textbf{Resolution pattern} & \large\makecell{\boldsymbol{\mathcal{M}}} & \large\textbf{Symmetry group} &
\large\textbf{Hypers}\\
\hline
\hline
 \multirow{6}{*}{$E_8^{(24)}[N]$} &\makecell{N=24n: \includegraphics[scale=0.2]{E8completeres.png} \\} & \mathfrak{t} & U(1)^8  &\makecell{\text{Charges: \textbf{root system} of } E_8  \\
\textbf{TOT: }\boldsymbol{5N } } \\ 
\cline{2-5}
 &\makecell{\\ \begin{array}{c}
 N=12n \\
 n\neq 2j \\
 \end{array}: \includegraphics[scale=0.2]{E8nores.png} \\} & A_1^{\oplus 8}& \mathbb{Z}_2^4 & \renewcommand{\arraystretch}{1.1}\begin{array}{c}
(q_{\mathbb{Z}_2}^{(a)},q_{\mathbb{Z}_2}^{(b)},q_{\mathbb{Z}_2}^{(c)},q_{\mathbb{Z}_2}^{(d)})= (0,0,0,0): 4(n-1)\\
(q_{\mathbb{Z}_2}^{(a)},q_{\mathbb{Z}_2}^{(b)},q_{\mathbb{Z}_2}^{(c)},q_{\mathbb{Z}_2}^{(d)})= (1,0,0,0): 4n\\
(q_{\mathbb{Z}_2}^{(a)},q_{\mathbb{Z}_2}^{(b)},q_{\mathbb{Z}_2}^{(c)},q_{\mathbb{Z}_2}^{(d)})= (0,1,0,0): 4n\\
(q_{\mathbb{Z}_2}^{(a)},q_{\mathbb{Z}_2}^{(b)},q_{\mathbb{Z}_2}^{(c)},q_{\mathbb{Z}_2}^{(d)})= (0,0,1,0): 4n\\
(q_{\mathbb{Z}_2}^{(a)},q_{\mathbb{Z}_2}^{(b)},q_{\mathbb{Z}_2}^{(c)},q_{\mathbb{Z}_2}^{(d)})= (0,0,0,1): 4n\\
(q_{\mathbb{Z}_2}^{(a)},q_{\mathbb{Z}_2}^{(b)},q_{\mathbb{Z}_2}^{(c)},q_{\mathbb{Z}_2}^{(d)})= (1,1,0,0): 4n\\
(q_{\mathbb{Z}_2}^{(a)},q_{\mathbb{Z}_2}^{(b)},q_{\mathbb{Z}_2}^{(c)},q_{\mathbb{Z}_2}^{(d)})= (1,0,1,0): 4n\\
(q_{\mathbb{Z}_2}^{(a)},q_{\mathbb{Z}_2}^{(b)},q_{\mathbb{Z}_2}^{(c)},q_{\mathbb{Z}_2}^{(d)})= (1,0,0,1): 4n\\
(q_{\mathbb{Z}_2}^{(a)},q_{\mathbb{Z}_2}^{(b)},q_{\mathbb{Z}_2}^{(c)},q_{\mathbb{Z}_2}^{(d)})= (0,1,1,0): 4n\\
(q_{\mathbb{Z}_2}^{(a)},q_{\mathbb{Z}_2}^{(b)},q_{\mathbb{Z}_2}^{(c)},q_{\mathbb{Z}_2}^{(d)})= (0,1,0,1): 4n\\
(q_{\mathbb{Z}_2}^{(a)},q_{\mathbb{Z}_2}^{(b)},q_{\mathbb{Z}_2}^{(c)},q_{\mathbb{Z}_2}^{(d)})= (0,0,1,1): 4n\\
(q_{\mathbb{Z}_2}^{(a)},q_{\mathbb{Z}_2}^{(b)},q_{\mathbb{Z}_2}^{(c)},q_{\mathbb{Z}_2}^{(d)})= (1,1,1,0): 4n\\
(q_{\mathbb{Z}_2}^{(a)},q_{\mathbb{Z}_2}^{(b)},q_{\mathbb{Z}_2}^{(c)},q_{\mathbb{Z}_2}^{(d)})= (1,1,0,1): 4n\\
(q_{\mathbb{Z}_2}^{(a)},q_{\mathbb{Z}_2}^{(b)},q_{\mathbb{Z}_2}^{(c)},q_{\mathbb{Z}_2}^{(d)})= (1,0,1,1): 4n\\
(q_{\mathbb{Z}_2}^{(a)},q_{\mathbb{Z}_2}^{(b)},q_{\mathbb{Z}_2}^{(c)},q_{\mathbb{Z}_2}^{(d)})= (0,1,1,1): 4n\\
\textbf{TOT: }\boldsymbol{5N-4} \\
\end{array} \\
\cline{2-5}
 &\makecell{\\ \begin{array}{c}
 N=6n \\
 n\neq 2j \\
 \end{array}: \includegraphics[scale=0.2]{E8nores.png} \\} & D_4 \oplus D_4 & \mathbb{Z}_2^2 & 
 \renewcommand{\arraystretch}{1.1}\begin{array}{c}
( q_{\mathbb{Z}_2}^{(a)}, q_{\mathbb{Z}_2}^{(b)}) = (1,0): 8n\\
( q_{\mathbb{Z}_2}^{(a)}, q_{\mathbb{Z}_2}^{(b)}) = (0,1): 8n\\
( q_{\mathbb{Z}_2}^{(a)}, q_{\mathbb{Z}_2}^{(b)}) = (1,1): 8n\\
( q_{\mathbb{Z}_2}^{(a)}, q_{\mathbb{Z}_2}^{(b)}) = (0,0): 6n-4\\
\textbf{TOT: }\boldsymbol{5N-4} \\
\end{array}\\
\cline{2-5}
 &\makecell{\\ \begin{array}{c}
 N=3n \\
 n\neq 2j \\
 \end{array}: \includegraphics[scale=0.2]{E8nores.png} \\} & D_8 & \mathbb{Z}_2 &  \renewcommand{\arraystretch}{1.1}\begin{array}{c}
q_{\mathbb{Z}_2} = 1: 7n-4\\
q_{\mathbb{Z}_2} =0: 8n\\
\textbf{TOT: }\boldsymbol{5N-4} \\
\end{array}\\
\cline{2-5}
 &\makecell{\\ \begin{array}{c}
 N=8n \\
 n\neq 3j \\
 \end{array}: \includegraphics[scale=0.2]{E8nores.png} \\} & A_2^{\oplus 4} & \mathbb{Z}_3^2  &\renewcommand{\arraystretch}{1.1}\begin{array}{c}
( q_{\mathbb{Z}_3}^{(a)}, q_{\mathbb{Z}_3}^{(b)}) = (1,0): 9n\\
( q_{\mathbb{Z}_3}^{(a)}, q_{\mathbb{Z}_3}^{(b)}) = (0,1): 9n\\
( q_{\mathbb{Z}_3}^{(a)}, q_{\mathbb{Z}_3}^{(b)}) = (1,1): 9n\\
( q_{\mathbb{Z}_3}^{(a)}, q_{\mathbb{Z}_3}^{(b)}) = (1,2): 9n\\
( q_{\mathbb{Z}_3}^{(a)}, q_{\mathbb{Z}_3}^{(b)}) = (0,0): 4(n-1)\\
 \textbf{TOT: }\boldsymbol{5N-4} \\
\end{array}\\ 
 \cline{2-5}
 &\makecell{N \neq 3n,8n:\includegraphics[scale=0.2]{E8nores.png} \\} & E_8 & \emptyset &\textbf{TOT: }\boldsymbol{5N-4} \\
\hline
\hline
\end{array}
}
$
\caption{Higgs branch data for quasi-homogeneous cDV singularities of $E_8^{(24)}[N]$ type.}
\label{table discrete 2}
\end{table}

\newpage
\begin{table}[H]
\centering
$
\scalemath{0.7}{
\renewcommand{\arraystretch}{1.7}
\begin{array}{|C||P|L|U|D|}
\Xhline{4\arrayrulewidth}
\makecell{\large\textbf{Singularity}\\} & \large\textbf{Resolution pattern} & \large\makecell{\boldsymbol{\mathcal{M}}} & \large\textbf{Symmetry group} &
\large\textbf{Hypers}\\
\hline
\hline
 \multirow{5}{*}{$E_8^{(20)}[N]$} &\makecell{N=20n: \includegraphics[scale=0.2]{E8completeres.png} \\} & \mathfrak{t} & U(1)^8  & \makecell{\text{Charges: \textbf{root system} of } E_8  \\
\textbf{TOT: }\boldsymbol{6N } } \\ 
\cline{2-5}
& \makecell{\\ \begin{array}{c}
 N=10n \\
 n\neq 2j \\
 \end{array}: \includegraphics[scale=0.2]{E8nores.png}\\} & A_1^{\oplus 8} & \mathbb{Z}_2^4 &\renewcommand{\arraystretch}{1.1}\begin{array}{c}
(q_{\mathbb{Z}_2}^{(a)},q_{\mathbb{Z}_2}^{(b)},q_{\mathbb{Z}_2}^{(c)},q_{\mathbb{Z}_2}^{(d)})= (0,0,0,0): 4(n-1)\\
(q_{\mathbb{Z}_2}^{(a)},q_{\mathbb{Z}_2}^{(b)},q_{\mathbb{Z}_2}^{(c)},q_{\mathbb{Z}_2}^{(d)})= (1,0,0,0): 4n\\
(q_{\mathbb{Z}_2}^{(a)},q_{\mathbb{Z}_2}^{(b)},q_{\mathbb{Z}_2}^{(c)},q_{\mathbb{Z}_2}^{(d)})= (0,1,0,0): 4n\\
(q_{\mathbb{Z}_2}^{(a)},q_{\mathbb{Z}_2}^{(b)},q_{\mathbb{Z}_2}^{(c)},q_{\mathbb{Z}_2}^{(d)})= (0,0,1,0): 4n\\
(q_{\mathbb{Z}_2}^{(a)},q_{\mathbb{Z}_2}^{(b)},q_{\mathbb{Z}_2}^{(c)},q_{\mathbb{Z}_2}^{(d)})= (0,0,0,1): 4n\\
(q_{\mathbb{Z}_2}^{(a)},q_{\mathbb{Z}_2}^{(b)},q_{\mathbb{Z}_2}^{(c)},q_{\mathbb{Z}_2}^{(d)})= (1,1,0,0): 4n\\
(q_{\mathbb{Z}_2}^{(a)},q_{\mathbb{Z}_2}^{(b)},q_{\mathbb{Z}_2}^{(c)},q_{\mathbb{Z}_2}^{(d)})= (1,0,1,0): 4n\\
(q_{\mathbb{Z}_2}^{(a)},q_{\mathbb{Z}_2}^{(b)},q_{\mathbb{Z}_2}^{(c)},q_{\mathbb{Z}_2}^{(d)})= (1,0,0,1): 4n\\
(q_{\mathbb{Z}_2}^{(a)},q_{\mathbb{Z}_2}^{(b)},q_{\mathbb{Z}_2}^{(c)},q_{\mathbb{Z}_2}^{(d)})= (0,1,1,0): 4n\\
(q_{\mathbb{Z}_2}^{(a)},q_{\mathbb{Z}_2}^{(b)},q_{\mathbb{Z}_2}^{(c)},q_{\mathbb{Z}_2}^{(d)})= (0,1,0,1): 4n\\
(q_{\mathbb{Z}_2}^{(a)},q_{\mathbb{Z}_2}^{(b)},q_{\mathbb{Z}_2}^{(c)},q_{\mathbb{Z}_2}^{(d)})= (0,0,1,1): 4n\\
(q_{\mathbb{Z}_2}^{(a)},q_{\mathbb{Z}_2}^{(b)},q_{\mathbb{Z}_2}^{(c)},q_{\mathbb{Z}_2}^{(d)})= (1,1,1,0): 4n\\
(q_{\mathbb{Z}_2}^{(a)},q_{\mathbb{Z}_2}^{(b)},q_{\mathbb{Z}_2}^{(c)},q_{\mathbb{Z}_2}^{(d)})= (1,1,0,1): 4n\\
(q_{\mathbb{Z}_2}^{(a)},q_{\mathbb{Z}_2}^{(b)},q_{\mathbb{Z}_2}^{(c)},q_{\mathbb{Z}_2}^{(d)})= (1,0,1,1): 4n\\
(q_{\mathbb{Z}_2}^{(a)},q_{\mathbb{Z}_2}^{(b)},q_{\mathbb{Z}_2}^{(c)},q_{\mathbb{Z}_2}^{(d)})= (0,1,1,1): 4n\\
\textbf{TOT: }\boldsymbol{6N-4} \\
\end{array} \\
\cline{2-5}
& \makecell{\\ \begin{array}{c}
 N=5n \\
 n\neq 2j \\
 \end{array}: \includegraphics[scale=0.2]{E8nores.png} \\} & D_4 \oplus D_4  & \mathbb{Z}_2^2 &  \renewcommand{\arraystretch}{1.1}\begin{array}{c}
( q_{\mathbb{Z}_2}^{(a)}, q_{\mathbb{Z}_2}^{(b)}) = (1,0): 8n\\
( q_{\mathbb{Z}_2}^{(a)}, q_{\mathbb{Z}_2}^{(b)}) = (0,1): 8n\\
( q_{\mathbb{Z}_2}^{(a)}, q_{\mathbb{Z}_2}^{(b)}) = (1,1): 8n\\
( q_{\mathbb{Z}_2}^{(a)}, q_{\mathbb{Z}_2}^{(b)}) = (0,0): 6n-4\\
\textbf{TOT: }\boldsymbol{6N-4} \\
\end{array}\\
\cline{2-5}
& \makecell{\\ \begin{array}{c}
 N=4n \\
 n\neq 5j \\
 \end{array}: \includegraphics[scale=0.2]{E8nores.png} \\} & A_4 \oplus A_4 & \mathbb{Z}_5 & \renewcommand{\arraystretch}{1.1}\begin{array}{c}
 q_{\mathbb{Z}_5} = 2: 10n\\
 q_{\mathbb{Z}_5} =1: 10n\\
 q_{\mathbb{Z}_5} = 0: 4(n-1)\\
 \textbf{TOT: }\boldsymbol{6N-4} \\
 \end{array}\\
\cline{2-5}
& \makecell{N \neq 4n,5n: \includegraphics[scale=0.2]{E8nores.png} \\} & E_8  & \emptyset & \textbf{TOT: }\boldsymbol{6N-4} \\
\hline
\end{array}
}
$
\caption{Higgs branch data for quasi-homogeneous cDV singularities of $E_8^{(20)}[N]$ type.}
\label{table discrete 3}
\end{table}

\restoregeometry

\newpage

\section{T-branes}\label{Sec:T-branes}
The analysis performed in this work, other than allowing the classification of 5d theories from M-theory on quasi-homogeneous cDV singularities, further elucidates the pivotal role of T-branes for the physical description of such theories. In fact, in the preceding sections we have always searched for a Higgs background in some ADE Lie algebra $\mathcal{G}$, that \textit{maximizes} the number of hypermultiplets of the 5d theory, namely the dimension of the Higgs Branch, at the same time breaking the 7d gauge group in the \textit{least} brutal way. These requirements translate into imposing that the Higgs background $\Phi(w)$ lives in the \textit{minimal} subalgebra $\mathcal{M}$ of $\mathcal{G}$ that allows for a holomorphic dependence of its Casimir invariants on the deformation parameter $w$. In Section \ref{section higgs vev}, we have developed the machinery to satisfy this constraint for all the quasi-homogeneous cDV singularities.\\
\indent It must be stressed, though, that looking for the minimal subalgebra is a mere choice, enabling comparisons and checks with other existing methods to extract the 5d Higgs Branch, but that it is by no means unique, nor necessary from a M-theory point of view. Indeed, in general the Higgs background can be embedded into some larger subalgebra $\mathcal{M}_{\text{T-brane}} \supset \mathcal{M}$, while generating the same threefold equation. This may yield:
\begin{enumerate}
    \item Less localized modes and a smaller unbroken continuous symmetry in 5d.\footnote{
     In this case part of the resolution is obstructed, even though it would appear possible from the geometry.} \label{item 1}
     \item A smaller unbroken discrete symmetry in 5d. \label{item 2}
    \item A combination of \ref{item 1} and \ref{item 2}. \label{item 3}
\end{enumerate}
In this regard, the most trivial choice one can pick is:
\begin{equation}
    \Phi \in \mathcal{M}_{\text{T-brane}} = \mathcal{G},
\end{equation}
namely embedding the Higgs field in the whole algebra.\footnote{This can be naturally achieved recalling a theorem by Slodowy \cite{Slodowy}, that establishes a one-to-one correspondence between the Slodowy slice through the principal nilpotent orbit of $\mathcal{G}$ and the coordinates on $\mathfrak{t}/\mathcal{W}$. Employing this fact one can, for \textit{all} the quasi-homogeneous cDV singularities, pick as Higgs background an element in the Slodowy slice through the principal nilpotent orbit of the corresponding $\mathcal{G}$, with appropriate coefficients.} This completely breaks the 7d gauge group and does not produce any hypermultiplet in 5d.\\

\indent Let us consider a trivial example for the $(A_1,A_3)$ singularity. The Higgs background producing the maximal amount of modes, as well as the expected $U(1)$ flavor symmetry, lies in the algebra $\mathcal{M} = A_1 \oplus A_1 \oplus \langle \alpha_2^* \rangle \subset A_3$, and reads:
\begin{equation}
    \Phi_{\text{T-brane}} = \left(\begin{array}{cccc}
 0 & 1 & 0 & 0 \\
 w & 0 & 0 & 0 \\
 0 & 0 & 0 & 1 \\
0 & 0 & -w & 0 \\
    \end{array} \right).
\end{equation}
In this case, we could have also chosen the following Higgs background:
\begin{equation}
    \Phi_{\text{T-brane}} = \left(\begin{array}{cccc}
 0 & 1 & 0 & 0 \\
 0 & 0 & 1 & 0 \\
 0 & 0 & 0 & 1 \\
 w^2 & 0 & 0 & 0 \\
    \end{array} \right).
\end{equation}
This background obviously reproduces the defining equation of the $(A_1,A_3)$ singularity, via \eqref{threefolds}, but breaks \textit{all} the 7d gauge group (in contrast with a preserved $U(1)$ in the case of $\Phi$ in the minimal allowed subalgebra $\mathcal{M}$), and does \textit{not} localize any mode in 5d. This is an example of phenomenon \ref{item 1}.\\
\indent Furthermore, there can be T-brane cases preserving a smaller discrete group in 5d with respect to their counterpart obtained from $\Phi$ in the minimal allowed subalgebra $\mathcal{M}$. Let us take a look again at the $(A_2,D_4)$ example examined in Section \ref{sec:expl-exampl-a_2}, with Higgs background living in the minimal allowed subalgebra:
\begin{equation}
    \Phi \in \mathcal{M} = A_1^{\oplus 4}.
\end{equation}
This choice yields:
\begin{itemize}
    \item 4 hypers in 5d.
    \item A preserved $\mathbb{Z}_2$ discrete symmetry in 5d.
\end{itemize}
On the other hand, one could have also made the choice:
\begin{equation}
    \Phi \in \mathcal{M}_{\text{T-brane}} =D_4=\mathcal{G},
\end{equation}
that explicitly reads, in the basis convention of \cite{Collingwood}:
\begin{equation}
    \Phi_{\text{T-brane}} = \left(
\begin{array}{cccc|cccc}
 0 & 1 & 0 & 0 & 0 & 0 & 0 & 0 \\
 0 & 0 & w & 0 & 0 & 0 & 0 & 0 \\
 0 & 0 & 0 & 1 & 0 & 0 & 0 & 1 \\
 0 & 0 & 0 & 0 & 0 & 0 & -1 & 0 \\
 \hline
 0 & -\frac{w}{4} & 0 & 0 & 0 & 0 & 0 & 0 \\
 \frac{w}{4} & 0 & 0 & 0 & -1 & 0 & 0 & 0 \\
 0 & 0 & 0 & 0 & 0 & -w & 0 & 0 \\
 0 & 0 & 0 & 0 & 0 & 0 & -1 & 0 \\
\end{array}
\right).
\end{equation}
It is then easy to check that $\Phi_{\text{T-brane}}$ produces:
\begin{itemize}
    \item 4 hypers in 5d.
    \item No preserved discrete symmetry in 5d.
\end{itemize}
The dimension of the Higgs Branch is unaffected, but the discrete symmetry is broken: this is the most simple example of phenomenon \ref{item 2}.\\

\begin{figure}[H]
    \centering
    \includegraphics[scale=0.28]{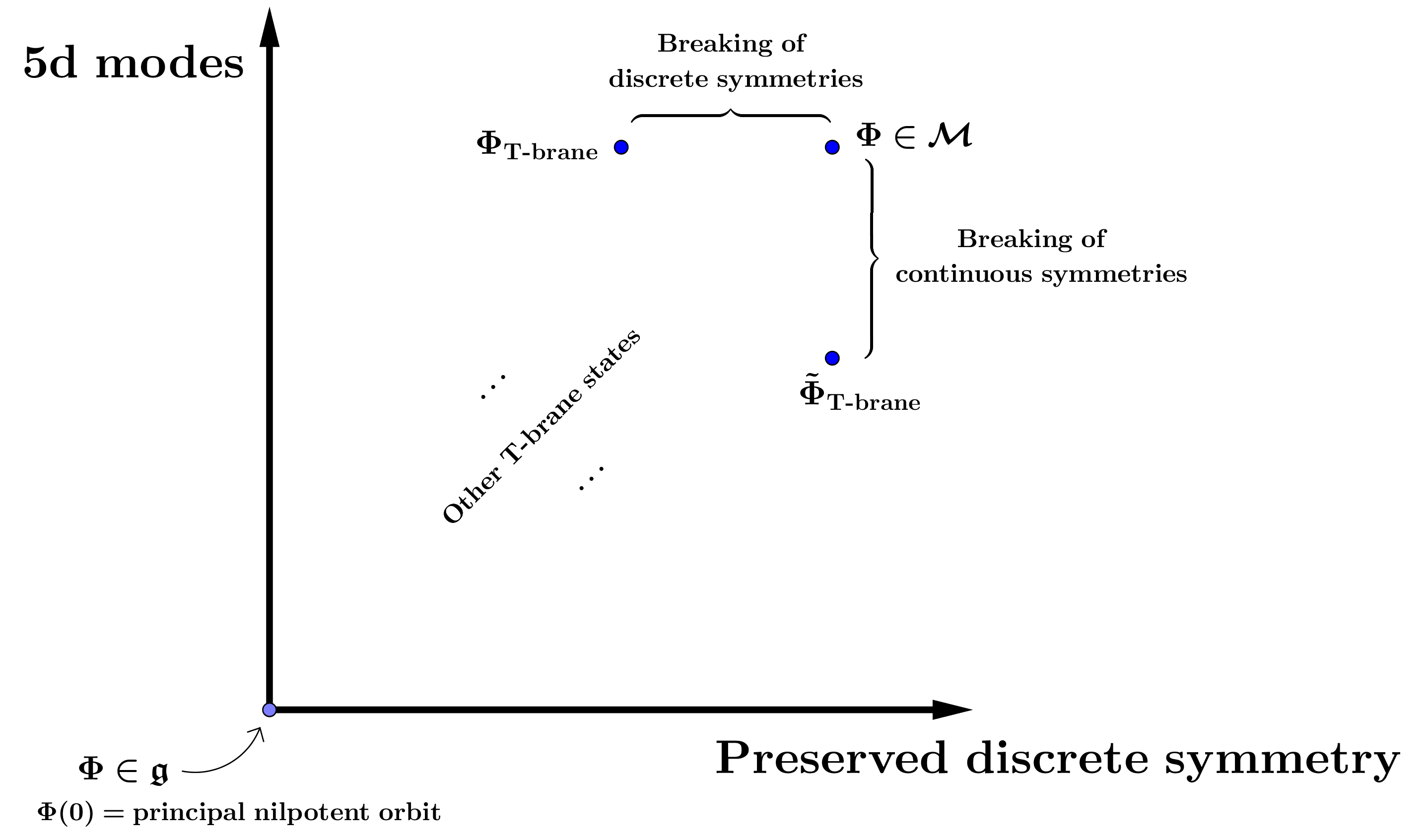}
    \caption{Allowed 5d theories from T-branes.}
    \label{T-brane hierarchy}
\end{figure}

\indent In full generality, one can easily construct Higgs backgrounds living in some subalgebra $\mathcal{M}_{\text{T-brane}} \supset \mathcal{M}$ such that both phenomenon \ref{item 1} and \ref{item 2} arise. This fact entails that, given a quasi-homogeneous cDV singularity\footnote{We remark that T-brane states such as the ones described in the text may appear in \textit{all} one-parameter deformed ADE singularities.}, a plethora of consistent 5d theories, with varying dimension of the Higgs Branch, as well as diverse flavor and discrete symmetries, are possible. $\Phi \in \mathcal{M}$ is the choice producing the largest Higgs Branch dimension, as well as the smallest breaking of the 7d gauge group.
This is another manifestation of the fact that the geometry of the M-theory background does not uniquely fix the effective low dimensional theory \cite{Cecotti:2010bp,Anderson:2013rka,Collinucci:2014qfa,Collinucci:2014taa,Collinucci:2016hpz,Collinucci:2017bwv,Anderson:2017rpr}. 
Intuitively one faces the possibilities depicted in Figure \ref{T-brane hierarchy}.\\
It would be extremely interesting to understand the counterpart of the 5d theories arising from T-brane backgrounds in complementary approaches, such as the techniques relying on magnetic quivers.

\section{Conclusions}\label{conclusion}
In the present work, we have completely classified the Higgs Branches of the 5d SCFTs arising from M-theory on quasi-homogeneous cDV singularities, explicitly computing their dimension, the flavor and discrete symmetries, as well as the charges of the hypermultiplets under such symmetries. This has been possible thanks to our novel method, allowing to associate with every quasi-homogeneous cDV singularity an explicit Higgs background (and viceversa) encoding the physics of the 5d SCFT, in a completely systematic fashion.
In general, we observe that the Higgs Branches of 5d theories arising from M-theory on quasi-homogeneous cDV singularities consist of free hypers, up to discrete gauging.

In the cases overlapping with the work of \cite{Carta:2022spy}, namely the $(A,E)$ singularities, we find complete agreement between our perspective and their Type IIB approach, as regards the HB dimension, the symmetry group and the charges of the hypers under such group. 

Furthermore, as preliminarily explored in \cite{DeMarco:2021try}, we notice a one-to-one correspondence between the discrete 0-form symmetries enjoyed by the 5d SCFTs from M-theory on quasi-homogeneous cDV singularities (that are nothing by the discrete symmetries preserved by our Higgs background), and the discrete 1-form symmetries of Type IIB on the same singularities (that can be extracted e.g.\ computing the torsion of the singular threefolds at infinity). This is consistent with the analysis of \cite{Closset:2020afy,Closset:2020scj,Closset:2021lwy}: compactifying the 4d SCFT on a circle, the line operators charged under the 1-form symmetry and wrapping the circle become point-like operators in 3d and, correspondingly, the one-form symmetry becomes a zero-form symmetry acting on the magnetic quiver Coulomb branch; one then ends up with a 0-form symmetry acting on the 5d Higgs Branch.

Finally, we have highlighted the crucial role of T-brane backgrounds in determining the physical content of the 5d theories: chosen a quasi-homogeneous cDV singularity, a wide range of different Higgs Branches, corresponding to different Higgs fields $\Phi$ and varying both in dimension and in flavor and discrete symmetries, are compatible with the same geometry. Among them, only the HB with maximal dimension and largest symmetry group  matches the results known in the literature, such as the ones for the $(A,E)$ singularities examined by \cite{Carta:2022spy}.  
Such framework arises in a natural fashion from our approach based on a Higgs field. 
It would be interesting to elucidate the meaning of the T-brane related Higgs Branches when the 5d theories engineered from the studied singularities are obtained with different methods.

\section*{Acknowledgments}

We thank  A. Collinucci for insightful discussions. We thank
F. Carta, S. Giacomelli, N. Mekareeya, A. Mininno for useful remarks and exchanges on the results of the analysis. M.D.M. and A.S. thank S. Nadir Meynet, F. Mignosa and M. Sacchi for useful suggestions.
A.S. and R.V. acknowledge support by INFN Iniziativa Specifica ST\&FI. M.D.M. acknowledges support by INFN Iniziativa Specifica GAST. The work of A.S. has been supported by the "Fondazione Angelo Della Riccia". A.S. would like to thank Andr\'es Collinucci and the Theoretical and Mathematical Physics Department of the Universit\'e Libre de Bruxelles for stimulating discussions and hospitality during the completion of this work.

\appendix

\section{Example of zero mode computations}\label{appendix C}

In this appendix, we give an explicit example of zero mode computation in a case that is more involved than the conifold one presented in Section~\ref{5dmodes}. We consider the $(A_1,A_3)$ threefold. The Higgs field is in the  subalgebra ($\mathcal{M}=\mathcal{L}$)
\begin{equation}
\mathcal{M}= A_1^{(1)}\oplus A_1^{(3)}\oplus\mathcal{H} \subset A_3\,,\qquad\mbox{with }\, \mathcal{H}=\langle\alpha_2^*\rangle.
\end{equation}
The algebra $\mathcal{G}=A_3$ can be decomposed into representations of $\mathcal{M}$ as
\begin{eqnarray}\label{A3branchingA1A1}
A_3 &=& (3,1)_{0}\oplus(1,3)_{0}\oplus(1,1)_{0}  \oplus\left[ (2,2)_{1} \oplus c.c.\right]
\end{eqnarray}

The representations $ (3,1)_{0}\oplus(1,3)_{0}\oplus(1,1)_{0} $ support one 7d mode each, as can be checked by a simple computation. Let us focus on the most interesting representation, i.e. $(2,2)_{1}$. Let us write $\Phi$ in this four dimensional
representation:
\begin{equation}
\Phi_{(2,2)_{1}} = \left(
\begin{array}{cccc}
 0 & 1 & 0 & 0 \\
 0 & 0 & 1 & 0 \\
 0 & 0 & 0 & w \\
 -4 w & 0 & 0 & 0 \\
\end{array}
\right) 
\end{equation}
Since we fixed a basis of
$(2,2)_{1}$, then $\varphi\rvert_{(2,2)_{1}}$ and $g\rvert_{(2,2)_{1}}$ are  vectors with four entries, along the
basis elements of $(2,2)_1$ and, in particular, 
\begin{equation}
  \label{vector notation for modes}
  \left[\Phi,g\rvert_{(2,2)_{1}}\right] = \Phi_{(2,2)_{1}} \cdot \left(
\begin{array}{c}
 g_1 \\
 g_2 \\
 g_3 \\
 g_4 \\
\end{array}
\right)  = \left(
\begin{array}{c}
 g_2 \\
 g_3 \\
 g_4 w \\
 -4 g_1 w \\
\end{array}
\right),
\end{equation}
with $g_i$, with $i = 1,2,3,4$ holomorphic functions of $w$. We now perform
the gauge fixing: we have to solve (\ref{eqZeroModes}) inside $(2,2)_1$:
\begin{equation}
  \label{gauge fixing}
\left(
\begin{array}{c}
 \varphi_1 \\
 \varphi_2 \\
 \varphi_3 \\
 \varphi_4 \\
\end{array}
\right)+ \left(
\begin{array}{c}
g_2 \\
g_3 \\
g_4 w \\
 -4 g_1 w \\
\end{array}
\right)  = 0,
\end{equation}
where $\varphi_i$ are holomorphic functions of $w$. We see that we can pick $g_2
= - \varphi_1, g_3 = - \varphi_2$, that will completely gauge-fix to zero the first
two entries of $\varphi\rvert_{(2,2)_1}$. On the other hand, $\varphi_{3,4}\in \mathbb{C}[w]/(w)$, i.e.\ they give 5d modes localized at $w =0$, that are charged under the $U(1)$ flavor group. The complex conjugate representation analogously gives two modes with opposite charge. Hence in total we obtain two free hypermultiplets that are charged under $U(1)$.
With the same method, one can check that the other irreducible representations do not localize any 5d mode.

\section{Slodowy slices and nilpotent orbits}\label{appendix D}

In this appendix, we provide definitions for the \textit{Slodowy slices}, used in the main text to explicitly construct Higgs backgrounds $\Phi$, and show how nilpotent orbits influence the number of localized 5d modes.

Let us start giving a definition of Slodowy slices.
 Consider a nilpotent element~$\mathrm{x}\in\mathfrak{g}$  belonging to some nilpotent orbit $\mathcal{O}$: the Jacobson-Morozov theorem ensures that there exists a standard triple $\{\mathrm{x},\mathrm{y},\mathrm{h}\}$ of elements in $\mathfrak{g}$ satisfying the $\mathfrak{su}(2)$ algebra relations.\footnote{The triple related to $\mathrm{x}$ is unique up to conjugation.}
Now, we define the \emph{Slodowy slice} through the point $\mathrm{x}$ as those Lie algebra elements satisfying:
\begin{equation}\label{Slodowy definition}
\mathcal{S}_\mathrm{x}=\{\mathrm{z}\in \mathfrak{g}\hspace{0.2cm}|\hspace{0.2cm}[\mathrm{z}-\mathrm{x},\mathrm{y}]=0\}.
\end{equation}
In the main text, we allow ourselves to switch on Higgs backgrounds
\begin{equation}\label{higgs decomposition}
  \Phi(w) = \sum_h \Phi_h(w) + \sum_{a=1}^\ell \varrho_1^a(w) \alpha_a^*
\end{equation}
with $\Phi(w) \in \mathcal{M} \subset \mathcal{G}$, where $\mathcal{M} = \bigoplus_h
\mathcal M_h \oplus \mathcal H$ and for some $\mathcal{G} \in ADE$ depending on the examined cDV singularity, only along the Slodowy slices in $\mathcal{M}_h$ through the nilpotent $\Phi_h(0)$.\\

To pick the right nilpotent orbit at fixed $\mathcal M_h$ we can proceed as follows. First, we can compute the quasi-homogeneous weights of the coordinates of all the Slodowy slices associated to the nilpotent orbits of $\mathcal M_h$. Comparing them with the expression of the Casimirs of $\Phi_h(w)$, we exclude many Slodowy slices that can not host a holomorphic $\Phi_h(w)$ due to the quasi-homogeneous scaling. Then, we pick $\Phi_h(w)$ along the Slodowy slice, among the remaining ones, associated to the nilpotent orbit of largest codimension.\\

Actually, as in the analysis of quasi-homogeneous cDV singularities we need to switch on only selected Casimirs in the addends $\Phi_h(w)$ of decomposition \eqref{higgs decomposition}, we can explicitly state a ``canonical'' choice of the Slodowy slice element in $\mathcal{M}_h$ that we are turning on. Let us immediately give the recipe for the addends in $\mathcal{M}_h = A_n$, for some $n$. In these cases, in the main text we \textit{always} need to turn on the top degree Casimirs, and nothing else. Hence, we can pick as canonical form for $\Phi_h(w)$ the following element 
(its shape can be gleaned from the form of the Slodowy slice through the principal nilpotent orbit of $\mathcal{M}_h$, with only the top Casimir switched on):
\begin{equation}\label{canonical A higgs}
    \Phi_h(w) = c_1 e_{\alpha_1} + c_2 e_{\alpha_2} + \ldots + c_n e_{\alpha_n}+ c_{n+1} e_{-\alpha_1-\alpha_2-\ldots-\alpha_n},
\end{equation}\\
where the $c_i$, $i=1,\ldots,n+1$ can either be constant or depend on $w$ (though not all of them can be constant, otherwise we would realize a non-nilpotent $\Phi_h(0)$).\\
\indent The form \eqref{canonical A higgs} yields a non-vanishing top degree Casimir $\rho_{\text{top}} = \prod_{i=1}^{n+1}c_i$, and allows $\Phi_h(0)$ to belong to any nilpotent orbit in the $A$ algebra, by a careful choice of the coefficients\footnote{Namely, recalling that nilpotent orbits of the $A_n$ algebras are in correspondence with the allowed Jordan forms in a matrix representation of $sl_{n+1}$, we can set some of the $c_i$ to 1 and the rest to $\tilde{c}_i w$, with $\tilde{c}_i$ a constant, obtaining any desired Jordan form.}.\\
\indent A similar reasoning works for the $\mathcal{M}_h = D$ and the $\mathcal{M}_h = E$ cases, in which it is sufficient, for the purposes of the main text, to turn on only some of the possible Casimirs. More precisely, in the $D$ cases we might need either the top degree Casimir, or the Casimirs having the same degree as the Pfaffian (for the definition of the Casimirs, we refer to Table \ref{threefolds}). To turn on only these Casimirs, we can construct a ``canonical'' $\Phi_h(w)$ in a fashion similar to the $A$ cases: the only difference is that in general we have a choice between \textit{two} such canonical forms, one inspired by the Slodowy slice through the principal nilpotent orbit, and the other along the subregular nilpotent orbit\footnote{This happens because in the $D$ cases not all nilpotent orbits can be obtained from the principal nilpotent orbit by removing some algebra elements.}. An analogous story goes for the $E_n$ cases, in which we can pick as many canonical forms as the number of orbits with the ``$E_n$'' label (displayed in Tables in \cite{Collingwood}).\\

The choice of nilpotent orbit where $\Phi(0)$ lies strongly influences the physics of the underlying 5d theory.
In \cite{DeMarco:2021try} we introduced a formula to obtain preliminary information regarding the number of localized 5d modes without performing the zero mode computations illustrated in Section \ref{5dmodes}. We found that the number of Lie algebra $\mathcal{G}$ elements that support 5d zero modes (i.e.\ that they are neither completely gauge fixed to zero nor remain untouched by gauge transformations) is given by the following \emph{codimension formula}:
\begin{equation}\label{codimension formula}
    n_{\text{ind}} = \text{cod}_{\mathbb C}\Big(\mathcal{O}_{0}  \hookrightarrow \mathcal{N}\Big),
\end{equation}
where $\mathcal{O}_0$ is the nilpotent orbit in which the Higgs background $\Phi$ lives on the origin:
\begin{equation}
    \Phi(w)|_{w=0} = \Phi(0) \in \mathcal{O}_0,
\end{equation}
and $\mathcal{N}$ is the nilpotent cone of $\mathcal{G}$.\\
\indent This is of great practical relevance for our computations, as it is roughly telling us that the higher the codimension of $\mathcal{O}_0$, the higher is the number of 5d modes localized by $\Phi(w)$. Using this fact, we can give a prescription to find the correct Higgs background field in all cases where ambiguities might arise. More precisely, there can be cases in which multiple Higgs backgrounds reproduce the same threefold equation pertaining to some quasi-homogeneous cDV singularity, yet yielding different amount of 5d modes. This means that the Higgs backgrounds localizing \textit{less} modes are T-brane states. The way to immediately tell which is the correct Higgs is then implied by the codimension formula \eqref{codimension formula}:\\

\indent\textit{The Higgs field localizing the maximal amount of 5d modes satisfies: }
\begin{equation*}
    \Phi(w)|_{w=0} = \Phi(0) \in \mathcal{O}^{\textit{low}}_0,
\end{equation*}
\textit{with $\mathcal{O}^{\text{low}}_0$ the nilpotent orbit of lowest dimension (that is, biggest codimension) allowed by the compatibility with the threefold equation}.\\

This is equivalent to requiring that every addend $\Phi_h(0)$ in \eqref{higgs decomposition} lies in the smallest allowed nilpotent orbit of the corresponding subalgebra, compatibly with the threefold equation. 

\indent Let us give a trivial example. Given the $(A_2,A_4)$ cDV singularity, we construct the Higgs background using the canonical form in \eqref{canonical A higgs}. We could have (among other choices) two different Higgs backgrounds with linear coefficients in $w$:
\begin{equation}
\renewcommand{\arraystretch}{1}
   \Phi_1 =  \left(\begin{array}{ccccc}
0 & 1 &0 &0 &0 \\
0 &0 & w &0 & 0\\
0 &0 & 0& 1 &0 \\
0 &0 &0 & 0& w\\
w & 0&0 &0 & 0\\
    \end{array} \right), \quad\quad
    \Phi_2 = \left(\begin{array}{ccccc}
0 & 1 &0 &0 &0 \\
0 &0 & 1 &0 & 0\\
0 &0 & 0& w &0 \\
0 &0 &0 & 0& w\\
w & 0&0 &0 & 0\\
    \end{array} \right).
\end{equation}
The Higgs localizing the maximal amount modes is $\Phi_1$, because:
\begin{equation}
    \Phi_1(0) \in \mathcal{O}_{[2,2,1]}, \quad\quad \Phi_2(0) \in \mathcal{O}_{[3,1,1]},
\end{equation}
and the orbit\footnote{We have used the conventions labelling nilpotent orbits of \cite{Collingwood}.} $\mathcal{O}_{[2,2,1]}$ has a bigger codimension than $\mathcal{O}_{[3,1,1]}$.\\

\indent Thus, recalling the form \eqref{canonical A higgs} and its top Casimir $\rho_{\text{top}} = \prod_{i=1}^{n+1}c_i$, we can lay down the following general recipe to promptly construct the Higgs background $\Phi_h(w) \in \mathcal{M}_h = A_n$: if we require $\rho_{\text{top}} = w^k$, with $k<n$, the corresponding $\Phi_h(w)$ has the shape $\eqref{canonical A higgs}$, with $k$ parameters $c_i$ equal to $w$, and the rest equal to 1. The 1's are distributed in such a way that $\Phi_h(0)$ lies in the nilpotent orbit labelled by a partition of $k$ parts $[d_1,\ldots,d_k]$ with the largest codimension among the allowed ones.

\section{Casimirs for the $E_6,E_7,E_8$ families}\label{appendix E}
In this Appendix, we present the explicit expressions of the coefficients of the versal deformations of $E_6,E_7,E_8$ in terms of the Casimir invariants of the Higgs backgrounds $\Phi$. This allows to build a bridge between a given quasi-homogeneous cDV threefold arising from a deformation of a $\mathcal{G}=E_6,E_7,E_8$ singularity and a Higgs background $\Phi$.\\

The $E_r$ singularities possess $r$ deformation parameters:
\begin{equation}\label{En epsilon}
\begin{array}{c|cl}
\boldsymbol{E_6} & \mu_i & \text{for }i=2,5,6,8,9,12  \\
\boldsymbol{E_7} & \mu_i & \text{for }i=2,6,8,10,12,14,18\\
\boldsymbol{E_8} & \mu_i & \text{for }i=2,8,12,14,18,20,24,30 \\
\end{array}
\end{equation}
entering in the equation of the family as in \eqref{threefolds}.

Now fix the representations of $\mathcal{G}=E_6,E_7,E_8$ as follows: $\boldsymbol{27}$ for $E_6$, $\boldsymbol{133}$ for $E_7$ and $ \boldsymbol{248} $ for $E_8$.
One then defines the Casimirs of an element $\mathrm{g} \in \mathcal{G}$ in the respective representation as\cite{Dixmier,Okubo}:
\begin{equation}\label{En casimirs}
\begin{array}{c|cl}
\boldsymbol{E_6} & c_{k_i}(\mathrm{g}) = \text{Tr}(\mathrm{g}^{k_i}) & \text{for }k_i=2,5,6,8,9,12  \\
\boldsymbol{E_7} & \tilde{c}_{k_i}(\mathrm{g}) = \text{Tr}(\mathrm{g}^{k_i}) & \text{for }k_i=2,6,8,10,12,14,18\\
\boldsymbol{E_8} & \hat{c}_{k_i}(\mathrm{g}) = \text{Tr}(\mathrm{g}^{k_i}) & \text{for }k_i=2,8,12,14,18,20,24,30 \\
\end{array},
\end{equation}
and $i=1,...,r$.
In particular, we are interested in the relationship between the Casimirs of $\mathrm{g} = \Phi$ and the deformation parameters \eqref{En epsilon}.\\

The result for the $E_6$ case is:
\begin{equation}\label{E6 epsilon}
\scalemath{1}{
\begin{split}
& \mu_2 = -\frac{c_{2}}{24}\\
& \mu_5 = \frac{c_{5}}{60}\\
& \mu_6 = \frac{c_{2}^3}{13824}-\frac{c_{6}}{144}\\
& \mu_8 = -\frac{c_{2}^4}{110592}+\frac{13 c_{2} c_{6}}{8640}-\frac{c_{8}}{240}\\
& \mu_9 = \frac{c_{9}}{756}-\frac{c_{2}^2 c_{5}}{11520}\\
& \mu_{12} =-\frac{c_{12}}{3240}+\frac{109 c_{2}^6}{4299816960}-\frac{847 c_{2}^3 c_{6}}{134369280}+\frac{109 c_{2}^2 c_{8}}{3732480}+\frac{13 c_{2} c_{5}^2}{466560}+\frac{61 c_{6}^2}{933120}.\\
\end{split}}
\end{equation}

For the $E_7$ case:
\begin{equation}\label{E7 epsilon}
\scalemath{1}{
\setlength{\jot}{16pt}
\begin{split}
\mu_2& =\frac{\tilde{c}_{2}}{18} \\
\mu_6 &=\frac{\tilde{c}_{2}^3}{139968}-\frac{\tilde{c}_{6}}{72} \\
\mu_8& =-\frac{7 \tilde{c}_{2}^4}{25194240}+\frac{11 \tilde{c}_{2} \tilde{c}_{6}}{16200}-\frac{\tilde{c}_{8}}{300} \\
\mu_{10} &=-\frac{2 \tilde{c}_{10}}{315}+\frac{\tilde{c}_{2}^5}{151165440}-\frac{17 \tilde{c}_{2}^2 \tilde{c}_{6}}{583200}+\frac{\tilde{c}_{2} \tilde{c}_{8}}{1400} \\
\mu_{12}& =-\frac{16 \tilde{c}_{10} \tilde{c}_{2}}{1148175}+\frac{\tilde{c}_{12}}{12150}-\frac{149 \tilde{c}_{2}^6}{10579162152960}+\frac{167 \tilde{c}_{2}^3 \tilde{c}_{6}}{3401222400}+\frac{737 \tilde{c}_{2}^2 \tilde{c}_{8}}{881798400}-\frac{31 \tilde{c}_{6}^2}{437400} \\
\mu_{14} &=\frac{8303 \tilde{c}_{10} \tilde{c}_{2}^2}{14935460400}-\frac{2201 \tilde{c}_{12} \tilde{c}_{2}}{217314900}+\frac{4 \tilde{c}_{14}}{62601}+\frac{11083 \tilde{c}_{2}^7}{24082404724998144}-\frac{11609 \tilde{c}_{2}^4 \tilde{c}_{6}}{5530387622400}\\
 &-\frac{1289 \tilde{c}_{2}^3 \tilde{c}_{8}}{1433804198400}+\frac{353 \tilde{c}_{2} \tilde{c}_{6}^2}{142242480}-\frac{31 \tilde{c}_{6} \tilde{c}_{8}}{1463400} \\
\mu_{18}& = \frac{12182634587 \tilde{c}_{10} \tilde{c}_{2}^4}{77806514663884339200}-\frac{564449 \tilde{c}_{10} \tilde{c}_{2} \tilde{c}_{6}}{3418744644000}+\frac{1844 \tilde{c}_{10} \tilde{c}_{8}}{3956880375}-\frac{27233975 \tilde{c}_{12} \tilde{c}_{2}^3}{11321053720935552}\\
 & +\frac{301 \tilde{c}_{12} \tilde{c}_{6}}{452214900}+\frac{307855 \tilde{c}_{14} \tilde{c}_{2}^2}{13588370378352}-\frac{2 \tilde{c}_{18}}{1507383}-\frac{886993691 \tilde{c}_{2}^9}{313644160640867419847393280}\\
 &+\frac{4713945967 \tilde{c}_{2}^6 \tilde{c}_{6}}{72026602145995788288000}-\frac{14715122551 \tilde{c}_{2}^5 \tilde{c}_{8}}{2334195439916530176000}-\frac{579011753 \tilde{c}_{2}^3 \tilde{c}_{6}^2}{23156700792822720000}\\
 &+\frac{2313866297 \tilde{c}_{2}^2 \tilde{c}_{6} \tilde{c}_{8}}{222355151645760000}-\frac{77393 \tilde{c}_{2} \tilde{c}_{8}^2}{3376537920000}-\frac{15011 \tilde{c}_{6}^3}{97678418400}.\\
\end{split}}
\end{equation}
For the $E_8$ case:
\begin{equation}\label{E8 epsilon}
\scalemath{0.8}{
\setlength{\jot}{13pt}
\begin{split}
\mu_2& =\frac{\hat{c}_{2}}{120} \\
\mu_8& =\frac{13 \hat{c}_{2}^4}{24883200000}-\frac{\hat{c}_{8}}{5760} \\
\mu_{12}& =\frac{\hat{c}_{12}}{181440}+\frac{101 \hat{c}_{2}^6}{3224862720000000}-\frac{\hat{c}_{2}^2 \hat{c}_{8}}{64512000} \\
\mu_{14}& =-\frac{71 \hat{c}_{12} \hat{c}_{2}}{798336000}+\frac{\hat{c}_{14}}{1108800}-\frac{2531 \hat{c}_{2}^7}{9029615616000000000}+\frac{103 \hat{c}_{2}^3 \hat{c}_{8}}{696729600000} \\
\mu_{18}& =-\frac{4451 \hat{c}_{12} \hat{c}_{2}^3}{689762304000000}+\frac{1523 \hat{c}_{14} \hat{c}_{2}^2}{12454041600000}-\frac{\hat{c}_{18}}{47174400}-\frac{26399 \hat{c}_{2}^9}{2080423437926400000000000}\\
 &+\frac{4747 \hat{c}_{2}^5 \hat{c}_{8}}{722369249280000000}+\frac{331 \hat{c}_{2} \hat{c}_{8}^2}{1672151040000} \\
\mu_{20}& = \frac{191071 \hat{c}_{12} \hat{c}_{2}^4}{2121019084800000000}+\frac{127 \hat{c}_{12} \hat{c}_{8}}{174569472000}-\frac{1165063 \hat{c}_{14} \hat{c}_{2}^3}{612738846720000000}+\frac{236627 \hat{c}_{18} \hat{c}_{2}}{434023349760000}\\
 &+\frac{10249681 \hat{c}_{2}^{10}}{61414099887587328000000000000}-\frac{2994007 \hat{c}_{2}^6 \hat{c}_{8}}{35540567064576000000000}-\frac{323371 \hat{c}_{2}^2 \hat{c}_{8}^2}{82269831168000000}-\frac{\hat{c}_{20}}{220809600}\\
\mu_{24}& =-\frac{193 \hat{c}_{12}^2}{17793312768000}+\frac{228270563 \hat{c}_{12} \hat{c}_{2}^6}{29320967828275200000000000}+\frac{234189517 \hat{c}_{12} \hat{c}_{2}^2 \hat{c}_{8}}{945465467240448000000}\\
 &-\frac{9171869023 \hat{c}_{14} \hat{c}_{2}^5}{52675933174824960000000000}-\frac{23281 \hat{c}_{14} \hat{c}_{2} \hat{c}_{8}}{9150846566400000}+\frac{561557071 \hat{c}_{18} \hat{c}_{2}^3}{8291582073815040000000}\\
 &+\frac{8268193432181 \hat{c}_{2}^{12}}{580761207304971815485440000000000000000}-\frac{20976434911 \hat{c}_{2}^8 \hat{c}_{8}}{3055351469407469568000000000000}\\
 &-\frac{16935675593 \hat{c}_{2}^4 \hat{c}_{8}^2}{33005339947302912000000000}-\frac{666323 \hat{c}_{2}^2 \hat{c}_{20}}{721337268326400000}+\frac{\hat{c}_{24}}{10061694720}-\frac{593 \hat{c}_{8}^3}{887354818560000} \\
\mu_{30}& = -\frac{636328729 \hat{c}_{12}^2 \hat{c}_{2}^3}{367646783551116410880000000}-\frac{189107437 \hat{c}_{12} \hat{c}_{14} \hat{c}_{2}^2}{277976001893990400000000}+\frac{2521 \hat{c}_{12} \hat{c}_{18}}{31907254579200000}\\
 &+\frac{122785779721089347 \hat{c}_{12} \hat{c}_{2}^9}{5354576379380206927872000000000000000000}+\frac{374760114643099 \hat{c}_{12} \hat{c}_{2}^5 \hat{c}_{8}}{685159914799807856640000000000000}\\
 &-\frac{199931513 \hat{c}_{12} \hat{c}_{2} \hat{c}_{8}^2}{94458563710156800000000}+\frac{28501673 \hat{c}_{14}^2 \hat{c}_{2}}{3860777804083200000000}-\frac{1634513578407571229 \hat{c}_{14} \hat{c}_{2}^8}{3206548401263100769075200000000000000000}\\
 &-\frac{3442332938170993 \hat{c}_{14} \hat{c}_{2}^4 \hat{c}_{8}}{593805259493166809088000000000000}+\frac{1223 \hat{c}_{14} \hat{c}_{8}^2}{112201334784000000}+\frac{15587535288859801 \hat{c}_{18} \hat{c}_{2}^6}{76346390506264304025600000000000000}\\
 &-\frac{1051350791 \hat{c}_{18} \hat{c}_{2}^2 \hat{c}_{8}}{1243310844834938880000000}+\frac{38736013334814563129113 \hat{c}_{2}^{15}}{919171413254131073937239231692800000000000000000000000}\\
 &-\frac{966205043352894287 \hat{c}_{2}^{11} \hat{c}_{8}}{46497194159854305977303040000000000000000000}-\frac{53516928494297557 \hat{c}_{2}^7 \hat{c}_{8}^2}{42002885419922588958720000000000000000}\\
 &-\frac{2159242595767 \hat{c}_{2}^5 \hat{c}_{20}}{737984035215212544000000000000}+\frac{21328481 \hat{c}_{2}^3 \hat{c}_{24}}{58332071437516800000000}+\frac{225239997090599 \hat{c}_{2}^3 \hat{c}_{8}^3}{119591548765057371340800000000000}\\
 &+\frac{72667 \hat{c}_{2} \hat{c}_{20} \hat{c}_{8}}{4518107320320000000}-\frac{\hat{c}_{30}}{1978376400000}.\\
\end{split}}
\end{equation}
\section{Mathematica code for computing the zero modes}\label{Appendix B}

In this Appendix we are going to describe the ancillary Mathematica
code that we used to analyze the Higgs branches of M-theory on the
quasi-homogeneous cDV. The code can be found on the same \textit{arXiv} page of this
paper. On the  \textit{arXiv} page, the reader can find a zipped folder,
containing, together with the Mathematica notebook code "CodeHiggsBranchDatav2.nb", nine text files. The text files have to be placed in one of the folders of the variable  \texttt{\$Paths} of
Mathematica and contain the explicit matrix realization of the exceptional Lie algebras\footnote{We took the explicit matrix realization of the exceptional Lie algebras from \cite{Cacciatori1,Cacciatori2,Deppisch}.}.\\
\indent The notebook file is divided in two sections. The first
section (``Main Code'') contains the functions that extract the Higgs
branch data from the Higgs field $\Phi$. The second
section ``Examples'' contains various examples where we show how to use the
routines contained in the section ``Main Code''. The Mathematica code can
be used also to analyze singularities that \textit{are not} quasi-homogeneous.
\paragraph{HbData function}

The most important function contained in the notebook is
\begin{equation*}
  \label{eq:163}
  \footnotesize{\text{\texttt{HbData[ADE, rank, simsrts, listhiggs, coeffhiggs, cartanhiggs, coeffcartan]}}}.
\end{equation*}
The arguments of the function are
\begin{itemize}
\item \texttt{ADE}: is a \texttt{Symbol} to be picked among \texttt{"A, DD, E6, E7, E8"} and specifies
  the type of ADE algebra associated with the threefold.
\item \texttt{rank}: is a positive \texttt{Integer} that specifies the rank of the
  ADE algebra associated with the threefold.
\item   \texttt{simsrts}: is a \texttt{List} of \texttt{Lists}. Each sublists
  represents a root of the Lie algebra specified by \texttt{ADE} and
  \texttt{rank}. \textit{The roots contained in} \texttt{simsrts} \textit{are the simple roots of the subalgebra $\mathcal M$ where $\Phi$ resides}.\footnote{They give the Dynkin diagram of the subalgebra $\mathcal{M}$.}\\
  \indent The roots are described by their integer coefficients
  with respect to the basis of the simple roots of $\mathcal{G}$.
 We labelled the simple roots as in
 the Figure \ref{roots labelling}.
 \begin{figure}[H]
    \centering
    \includegraphics[scale=0.14]{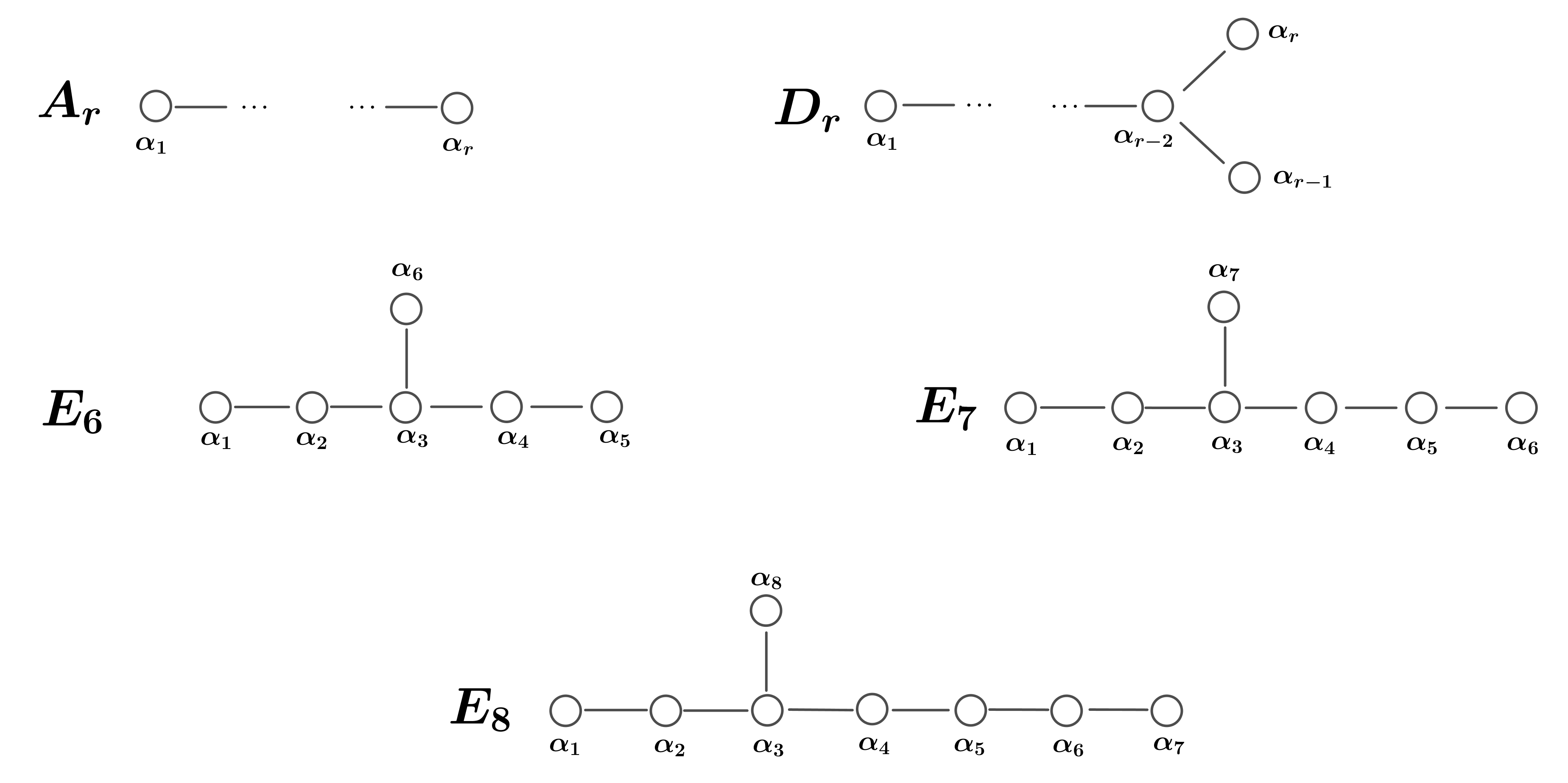}
    \caption{Roots labelling convention}
    \label{roots labelling}
  \end{figure}
 For example, the lowest
  root 
  of the $D_4$ Lie algebra is expressed  as
\begin{equation*}
  \label{eq:73app}
\left\{-1,-2,-1,-1\right\}.
\end{equation*}
  Concretely, considering the Higgs field in \eqref{A2D4 higgs first} as an example,
  we see that it lies in the subalgebra $\mathcal{M}=A_1^{4}$ of $D_4$. This subalgebra
  is generated by the three outer roots of the $D_4$ diagram and by the lowest
  root of $D_4$. In our notation, the corresponding input is
  \begin{eqnarray}
    \label{simsrt A2D4}
    \text{\texttt{simsrts}}= 
                                                         \left
                                                         \{
                                                         \left\{1,0,0,0\right\},
                                                         \left\{0,0,1,0\right\},
                                                         \left\{0,0,0,1\right\},
                                                         -\left\{1,2,1,1\right\}
\right\}.
  \end{eqnarray}
  
  The user can print on screen the roots system of $\mathcal G$ calling the  function 
  \begin{equation*}
      \text{\texttt{PrintRootSystem[ADE,rank]}},
  \end{equation*}
   the first argument being
  again the ADE type of $\mathcal{G}$, and the second argument its
  rank. \\
  The highest root of the root system can be obtained calling the
  function 
    \begin{equation*}
      \text{\texttt{PrintHighestRoot[ADE,rank]}}.
    \end{equation*}
    We will explain below how to prompt, using the function
    \texttt{PrintMatrix}, the explicit matrices representing, in the
    adjoint representation, the
    root vectors associated with the roots (as well as to the elements of  the basis of the
    Cartan subalgebra of $\mathcal{G}$).
\item \texttt{listhiggs}: is a \texttt{List} of \texttt{Lists}. Each sublist represents a root such
  that the Higgs field has a non-zero coefficient along the corresponding
  root-vector in $\mathcal{G}$. We input in this way all the components of the Higgs that \textit{do not} lie
  in the Cartan subalgebra; the elements in the Cartan subalgebra will be separately input with the
  variables \texttt{cartanhiggs} and \texttt{coeffcartan}. For the
  $(A_2,D_4)$ Higgs field \eqref{A2D4 higgs first} that we are taking as example, the variable
  \texttt{listhiggs} is
  \begin{eqnarray}
    \label{listhiggs A2D4}
    \text{\texttt{listhiggs}}= 
                                                         &&\Big\{          
                                                         \left\{1,0,0,0\right\},
                                                         \left\{0,0,1,0\right\},
                                                         \left\{0,0,0,1\right\},
                                                         -\left\{1,2,1,1\right\},
                                                            \nonumber \\
    &&\quad   -\left\{1,0,0,0\right\},
                                                        -\left\{0,0,1,0\right\},
                                                         -\left\{0,0,0,1\right\},
                                                         \left\{1,2,1,1\right\}
\Big\}.\nonumber \\
  \end{eqnarray}
\item \texttt{coeffhiggs}: is a \texttt{List} containing the coefficients
  corresponding to the elements of \texttt{listhiggs}. If we again consider
  the Higgs field of \eqref{A2D4 higgs first} we have 
  \begin{equation}
    \label{coeffhiggs A2D4}
    \text{\texttt{coeffhiggs}} = \left\{1,1,1,1,c_1 w,c_2 w,c_3 w,
        c_4 w\right\},
  \end{equation}
  where we lowered the index $h$ of the coefficients $c^h$ appearing in \eqref{A2D4 higgs first} for clarity of notation. 
\item \texttt{cartanhiggs}: is a \texttt{List} of positive \texttt{Integers} $n_i$, with
  $n_i = 1$, ..., \texttt{rank}, describing the elements of the Cartan
  subalgebra  of $\mathcal{G}$ along which the Higgs field has a
  non-zero coefficient. We chose the generators of the Cartan subalgebra to
  be the dual elements $\alpha_j^*$ of the simple roots.  For example,
  let's consider the $(A_{11},E_6)$ singularity. We saw 
  in table \ref{table modes 1} that its crepant resolution simultaneously resolves all the
  nodes of the $E_6$ Dynkin diagram. In terms of the Higgs fields, this
  means that $\Phi$ has to lie in the Cartan subalgebra of $E_6$. Inside
  the Cartan subalgebra, the Higgs field associated with $(A_{11},E_6)$  has
  non-zero component along all the $\alpha_i^*$, with $i = 1, ...,6$.
In order to pick a Higgs field with a non-zero component along all the $\alpha_i^*$
we input
  \begin{equation}
    \label{A11E6 input 1}
    \text{\texttt{cartanhiggs}}= \left\{1,2,3,4,5,6\right\}.
\end{equation}
If we initialize the variable \texttt{cartanhiggs} as in \eqref{A11E6 input
1}, we get an Higgs field with a (possibly)
non-zero component along all the $\alpha_i^*$. We note that, however, \texttt{cartanhiggs} does not specify the
precise value of the coefficients. The coefficients will be specified in the variable \texttt{coeffcartan}.
\item \texttt{coeffcartan}: is the \texttt{List} of the
  coefficients corresponding to the elements of \texttt{cartanhiggs}. For
  the $(A_{11},E_6)$ example, if we input
  \begin{eqnarray*}
     &&\text{\texttt{cartanhiggs}}=
        \left\{1,2,3,4,5,6\right\},  \nonumber \\
     &&  \text{\texttt{coeffcartan}}=
  \left\{w t_1,w t_2,w t_3,w t_4,w t_5,w t_6\right\} \nonumber 
\end{eqnarray*}
we picked the Higgs to have a coefficient $w t_i$ along the corresponding $\alpha_i^*$.
\end{itemize}
\paragraph{Output:} The function HbData has a void output and prints all the data that describe the
action of the flavor symmetries and discrete gauging symmetries on the
five-dimensional hypers. This permits to reconstruct the Higgs branch as
complex algebraic variety and the action of the flavor isometries on the
Higgs branch.

We remark here that the user can print on the screen the explicit matrix
(in the adjoint representation) 
associated with a certain value of the variables \texttt{listhiggs, coeffhiggs, cartanhiggs}, \texttt{coeffcartan} using
\begin{equation*}
  \footnotesize{\text{\texttt{PrintMatrix[ADE,rank,listhiggs,coeffhiggs,cartanhiggs,coeffcartan]}}}.
\end{equation*}
For example, if we want to visualize the matrix associated with the generator corresponding to the root
$\left\{0,1,0,0\right\}$ of $D_4$, we will input
\begin{equation*}
  \text{
    \texttt{
      PrintMatrix[DD,4,\{\{0,1,0,0\}\},\{1\},\{\},\{\}].
    }
    }
  \end{equation*}
  This permits the user to read off the explicit normalization we used for
  the generators of the Lie algebra\footnote{In particular, for classical Lie algebra, we followed
  the convention of \cite{Collingwood}.}.\\
\indent Summing up, to obtain the Higgs branch data of the Higgs field associated with the $(A_2,D_4)$ singularity, we will input the following data:
\begin{itemize}
\item \texttt{ADE = DD};
\item \texttt{rank = 4};
\item \texttt{simsrts}: The subalgebra $\mathcal M$ containing the Higgs field is $A_1^4
  \subset D_4$. The corresponding simple roots are $e_{\alpha_1}$,
  $e_{\alpha_3}$, $e_{\alpha_4}$ and the lowest root of $D_4$ (see figure \ref{D4coxeter}).  Consequently, we input \eqref{simsrt A2D4}.
\item Given \eqref{A2D4 higgs first}, recalling that $\alpha^h$, with $h =4$ is the
  lowest root of the $D_4$ Dynkin diagram, we input \eqref{listhiggs A2D4}, \eqref{coeffhiggs A2D4} as, respectively, \texttt{listhiggs} and \texttt{coeffhiggs}.
\item The subalgebra $\mathcal M = A_1^4$ has no 
  $\mathfrak{u}(1)$ factors. Consequently, the Higgs field can not have non-zero
  coefficients along the
  Cartan elements $\alpha_i^*$, with $i = 1, ..., 4$, dual to the simple roots of $D_4$. Then,
  we input\footnote{We can also choose \texttt{cartanhiggs} as a non-void list and set to zero
  the corresponding coefficients inside \texttt{coeffcartan}. For example,
  we can input
    \begin{equation*}
    \label{a2d4 cartan input 2}
    \text{\texttt{cartanhiggs}} = \left\{1,3,4\right\}, \quad \text{\texttt{coeffcartan}} = \left\{0,0,0\right\}.
  \end{equation*}}
  \begin{equation*}
    \label{a2d4 cartan input 1}
    \text{\texttt{cartanhiggs}} = \left\{\right\}, \quad \text{\texttt{coeffcartan}} = \left\{\right\}.
  \end{equation*}
\end{itemize}
We
report here a part of the output for the $(A_2,D_4)$ case\footnote{For the full
output please check the subsection "A2D4" inside the section ``Examples'' of the ancillary Mathematica
file.}.  The first part of the output is
\begin{center}
\includegraphics[scale=0.8]{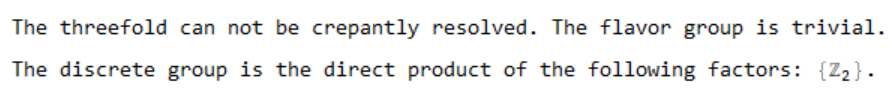}
\label{firstout}
\end{center}
The first line is telling us that the considered threefold does not admit any
small crepant resolution. The second line 
is telling us that
the discrete gauging group is non trivial. The discrete gauging group is
the direct product of the factors appearing between curly brackets in the
second line of the output. In this case, we have only one such factor, and
the discrete gauging group is isomorphic to $\mathbb Z_2$.\\
\indent The second part of the output consists of many blocks (one for each irreducible representation of the branching of
$\mathcal{G} = D_4$ with respect to the $A_1^4 \subset D_4$ subalgebra) of the
following type:
\begin{center}
\includegraphics[scale=0.8]{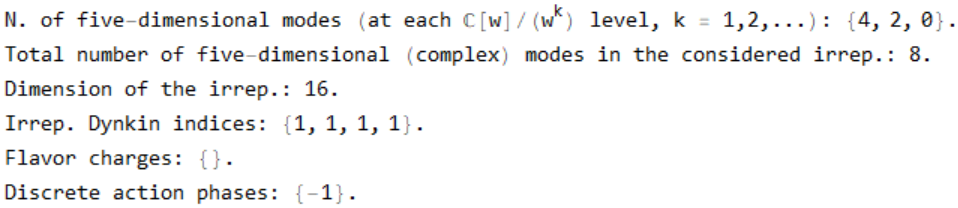}
\end{center}
From the first three lines we can reconstruct the number of five-dimensional modes localized in
the considered irreducible representation. In the first line, we read the \texttt{List}
$\left\{4,2,0\right\}$, this means that we have four modes localized in
$\mathbb C[w]/(w)$, two modes localized in $\mathbb C[w]/(w^2)$ and zero in
$\mathbb C[w]/(w^k)$ with $k > 2$. The
overall number of complex-valued modes is, hence, $4 * 1 + 2 * 2 = 8$. 
The last four lines tell us, respectively:
\begin{itemize}
\item  \textit{The complex dimension of the considered irreducible
    representation.} In the example, it is $16$.
\item \textit{The Dynkin indices of the highest weight state of the representation.} In the example, we read
  \begin{equation}
    \label{Dynkin indeces}
    \left\{1,1,1,1\right\}.
  \end{equation}
  Each of the number appearing in \eqref{Dynkin indeces} tells us the weight
  of the highest state of the considered irreducible representation with
  respect to the roots contained in  \texttt{simsrts}. In other words, the first number of
  \eqref{Dynkin indeces} is the Dynkin index of the highest weight state
  with respect to the first root appearing in \texttt{simsrts} (in this case
  $e_{\alpha_1} = \left\{1,0,0,0\right\}$) and so on. These data permit us to completely reconstruct the representation: in this case, \eqref{Dynkin indeces} tells us that we are considering the tensor product of all the fundamental representations of the four $A_1$ factors (that has dimension $2^4 = 16$).
\item \textit{The charges,  with respect to the flavor group
  generators, of the modes localized in the representation.} The generators of the flavor group are the Cartan elements $\alpha_i^*$  that are dual to the roots
  that get resolved. In this case, the flavor group is trivial (since no
  $\mathbb P^1$ can be simultaneosly resolved) and the list is void.
\item \textit{The action of the discrete gauging group on the considered
    irreducible representation}. As we just learned, for the $(A_2,D_4)$ case the
  discrete gauging group is $\mathbb Z_2$. We saw that the discrete gauging
  groups are generated by diagonal matrices that respect the branching of $\mathcal{G}$ with respect to $\mathcal{M}$. Hence, their
  generators act
  multiplying by the same phase all the elements of the considered
  irreducible representation. In this case, the output is telling us that the generator of the $\mathbb
  Z_2$ group acts multiplying all the elements of the considered
  irreducible representation by $-1$. In general, the list will contain as many phases as the factors of the discrete gauging group.
\end{itemize}
\paragraph{Overloaded version of HbData}

The function \texttt{HbData} is overloaded as
\begin{equation*}
\text{\texttt{HbData[ADE, rank, simsrts, higgs]}}.  
\end{equation*}
The overloaded version of \texttt{HbData} can be used to analyze, in the language of this paper, the
Higgs fields we presented in \cite{Collinucci:2021ofd,DeMarco:2021try}.
The first three arguments are exactly the same of the  version of the \texttt{HbData} function presented in the previous pages. The fourth argument is
a matrix representing the Higgs field. The Higgs field has to be input
\begin{itemize}
\item in the fundamental representations for the $A_r$, $D_r$ cases
  (following the notations in \cite{Collingwood});
\item in the $\textbf{27}$ representation for the $E_6$ case;
\item in the adjoint representation for the $E_7$, $E_8$ cases.
\end{itemize}

The output is exactly analogous to the one of \texttt{HbData[ADE, rank,
  simsrts, listhiggs, coeffhiggs, cartanhiggs, coeffcartan]}, and contains
the data of the Higgs branch of the five-dimensional SCFT associated with the
Higgs field profile \texttt{higgs} that we input in \texttt{HbData[ADE,
  rank, simsrts, higgs]}.


\providecommand{\href}[2]{#2}

\end{document}